\documentclass[12pt,a4paper]{article}
\pdfoutput=1
\usepackage{graphicx}
\usepackage{amssymb}
\usepackage{amsmath}              
\usepackage{bm}
\usepackage{color}
\usepackage{cite}
\usepackage{colortbl}
\usepackage{longtable}
\usepackage{subfigure}

\definecolor{Orange}{cmyk}{0,0.61,0.87,0}
\definecolor{JungleGreen}{cmyk}{0.99,0,0.52,0}
\definecolor{OliveGreen}{cmyk}{0.64,0,0.95,0.40}
\definecolor{Brown}{cmyk}{0,0.70,1,0.40}
\definecolor{RoyalBlue}{cmyk}{0.71,0.53,0,0.12}
\definecolor{Gray}{cmyk}{0,0,0,0.40}
\definecolor{LightPink}{cmyk}{0.0,0.25,0,0}
\definecolor{LLightPink}{cmyk}{0.0,0.10,0,0}
\definecolor{LightBlue}{cmyk}{0.25,0,0,0}
\definecolor{LightGray}{cmyk}{0,0,0,0.2}

\newcommand{\Slash}[1]{{\ooalign{\hfil/\hfil\crcr$#1$}}}

\usepackage[colorlinks=true, linkcolor=black, citecolor=black,
urlcolor=black]{hyperref}

\begin{document}

\begin{titlepage}

\begin{flushright}

LPT-Orsay-15-16 \\
FTPI--MINN--15/06 \\
UMN--TH--3420/15 \\
IPMU15-0020 \\
KCL-PH-TH/2015-09 \\

\end{flushright}

\vskip 1cm
\begin{center}

{\Large
{\bf 
Dark Matter and Gauge Coupling Unification in \\[5pt] Non-supersymmetric 
SO(10) Grand Unified Models 
}
}

\vskip 1cm

Yann Mambrini$^a$, 
Natsumi Nagata$^{b,c}$, 
Keith A. Olive$^b$, \\
J\'{e}r\'{e}mie Quevillon$^d$,
and
Jiaming Zheng$^b$

\vskip 0.5cm

{\it $^{a}$Laboratoire de Physique Th\'{e}orique Universit\'{e}
 Paris-Sud, F-91405 Orsay, France} \\[2pt]
{\it $^{b}$William I. Fine Theoretical Physics Institute, School of
 Physics and Astronomy, University of Minnesota, Minneapolis, MN 55455,
 USA}\\[2pt]
{\it $^{c}$Kavli Institute for the Physics and Mathematics of the Universe
 (WPI), Todai Institutes for Advanced Study, the University of Tokyo,
 Kashiwa 277-8568, Japan}\\[2pt]
{\it $^d$Theoretical Particle Physics \& Cosmology, Department of
 Physics, King's College London, London, WC2R 2LS, United Kingdom}
\date{\today}

\vskip 1.5cm

\begin{abstract} 

 Unlike minimal SU(5), SO(10) provides a straightforward path towards
 gauge coupling unification by modifying the renormalization group
 evolution of the gauge couplings above some intermediate scale which
 may also be related to the seesaw mechanism for neutrino
 masses. Unification can be achieved for several different choices of
 the intermediate gauge group below the SO(10) breaking scale. 
 In this work, we consider in detail the
 possibility that SO(10) unification may also provide a natural dark matter
 candidate, stability being guaranteed by a leftover $\mathbb{Z}_2$
 symmetry. We systematically examine the possible intermediate gauge
 groups which allow a non-degenerate, fermionic, Standard Model singlet
 dark matter candidate while at the same time respecting gauge coupling
 unification. Our analysis is done at the two-loop level. Surprisingly,
 despite the richness of SO(10), we find that only two models survive
 the analysis of phenomenological constraints, which include suitable
 neutrino masses, proton decay, and reheating.

\end{abstract}

\end{center}
\end{titlepage}

\section{Introduction}
\label{sec:introduction}

One of the often quoted motivations for supersymmetry (SUSY) is its ability to
improve the possibility for gauge coupling unification at the grand
unified (GUT) scale, which is not possible in minimal SU(5)
\cite{susy}. However, SO(10) has the built-in possibility for achieving
gauge coupling unification through several potential intermediate-scale
gauge groups \cite{Rajpoot:1980xy, Fukugita:1993fr,
Mambrini:2013iaa}. Of course, low-energy SUSY has many other
motivations including the presence of a dark matter (DM) candidate \cite{EHNOS}
whose stability is insured  if $R$-parity is conserved. However, under
very generic conditions, non-SUSY SO(10) models also possess a
remnant $\mathbb{Z}_2$ symmetry when an intermediate-scale U(1) symmetry
is broken \cite{Kibble:1982ae, Krauss:1988zc, Ibanez:1991hv,
Martin:1992mq,Kadastik:2009dj,Frigerio:2009wf}. Thus, several modest
extensions of minimal SO(10) may also allow for the possibility of DM.  

In building a successful SO(10), we must also require that the GUT and
intermediate mass scales be sufficiently large so as to ensure a proton
lifetime and neutrino masses compatible with experiment. Unfortunately,
these requirements are not realized for every choice of
intermediate-scale gauge group. The addition of a new SO(10) multiplet
containing a 
DM candidate will, however, affect the running of the gauge
couplings and can improve the desired unification of the gauge
couplings. For this reason, we suppose that the DM candidate be charged
under the intermediate gauge symmetries. The cosmological production of
DM could occur, for example, out of equilibrium from the thermal bath
(nonequilibrium thermal DM or NETDM  \cite{Mambrini:2013iaa}) in a
manner reminiscent of freeze-in scenarios \cite{freezein}. This
mechanism works with a stable particle which has no interaction with the SM
particles. Thus, we focus on singlet DM candidates. Further, as scalar
DM would most assuredly couple to the Standard Model (SM) Higgs, we limit our
attention here to fermionic DM. 

SO(10) grand unification is, of course, a general moniker for many candidate
theories of unification, as there are several possible intermediate gauge
groups and several possible choices for representations $R_1$ of Higgs
fields which break SO(10) to the intermediate gauge group,
$G_{\text{int}}$, and then again, several possible choices of
representations $R_2$ for the Higgs fields which break $G_{\text{int}}$
down to the SM. Furthermore, there are several possible choices
for the representation which contains DM. Thus, it may seem that
DM in SO(10) models is a rather robust and generic
feature. However, if we insist on maintaining gauge coupling unification
at a suitably high scale to guarantee proton stability, the number of
models is dramatically reduced. In fact, by limiting the dimension of the
representation containing DM to be no larger than a {\bf 210}, we
find that only two models survive. 

In this paper, we will systematically examine the possibility for
fermionic NETDM in SO(10) models, though our conclusions are more general
than the specific NETDM model. We will discuss the various possible
intermediate gauge groups and Higgs representations which allow for
gauge coupling unification, and we will demonstrate the effect of
including two-loop running of the renormalization group equations (RGEs). The
DM representation needs to be split so that only fermions with
the appropriate gauge quantum numbers survive at low energy. This
requires fine-tuning similar to the doublet-triplet separation problem
in GUTs. We also systematically consider viable DM
representation and their effect on the running of the gauge
couplings. In all but two distinct models, the presence of DM
spoils the desired unification of the gauge couplings. 

In the following, we begin by discussing the origin of a discrete
symmetry in a variety of models with different intermediate gauge groups
and the possible representations for DM and the splitting of
the DM multiplet. In Sec.~\ref{sec:gcu}, we first
demonstrate gauge coupling unification in these models (without DM) and
show the effect of including the two-loop functions in the RGE running
and one-loop threshold effects. We next consider the question of gauge
coupling unification in the presence of a DM multiplet. In
Sec.~\ref{sec:models}, we discuss the criteria which select only two
possible models in a specific example of the NETDM scenario
\cite{Mambrini:2013iaa}. The phenomenological aspects of these models
including neutrino masses, proton decay, and the production of DM through
reheating after inflation will be discussed in
Sec.~\ref{sec:phenomenology}. We also consider the case where the DM
field is a singlet under the intermediate gauge groups in
Sec.~\ref{sec:singletDM}. Our conclusions will be given in
Sec.~\ref{sec:conclusion}.

\section{Candidates}
\label{sec:candidates}

We assume that the SO(10) gauge group is 
spontaneously broken to an intermediate subgroup
$G_{\text{int}}$ at the GUT scale $M_{\text{GUT}}$, and
subsequently  broken to the SM gauge group $G_{\text{SM}}$ at an
intermediate scale $M_{\text{int}}$:  
\begin{equation}
 \text{SO}(10)\longrightarrow G_{\text{int}}\longrightarrow
 G_{\text{SM}}\otimes \mathbb{Z}_N ~,
\label{eq:decaypattern}
\end{equation}
with
$G_{\text{SM}}\equiv\text{SU(3)}_C\otimes \text{SU(2)}_L\otimes
\text{U(1)}_Y$. The Higgs multiplets which break SO(10) and
$G_{\text{int}}$ are called $R_1$ and $R_2$, respectively. In addition,
we require that there be a remnant discrete symmetry $\mathbb{Z}_N$ that
is capable of rendering a SM singlet field to be stable and hence
account for the DM in the Universe \cite{Kadastik:2009dj, Frigerio:2009wf}. 
The mechanism  for ensuring a remnant $\mathbb{Z}_N$ is discussed in detail in
Sec.~\ref{sec:discsym}, and the possible intermediate gauge groups that
accommodate the condition are summarized in Sec.~\ref{sec:intgauge}. 

If, moreover, the DM couplings are such that the candidate is not in thermal
equilibrium at early times, as in the NETDM scenario, we obtain
stringent constraints on the model structure. We will consider this subject in
Sec.~\ref{sec:degenprob}.

\subsection{Discrete symmetry in SO(10)}
\label{sec:discsym}

SO(10) is a rank-five group and has an extra U(1) symmetry beyond
U(1)$_Y$ in the SM gauge group. The U(1) charge assignment for fields in
an SO(10) multiplet is determined uniquely up to an overall factor. We
define the normalization factor such that all of the fields $\phi_i$ in
a given model have integer charges $Q_i$ with a minimum non-zero value
of $|Q_i|$ equal to $+1$. Now, let us suppose that a
Higgs field $\phi_H$ has a non-zero charge $Q_H$. Then, if $Q_H =0$
(mod.~$N$) with $N\geq 2$ an integer, the U(1) symmetry is broken to a
$\mathbb{Z}_N$ symmetry after the Higgs field obtains a vacuum
expectation value (VEV)
\cite{Krauss:1988zc,Ibanez:1991hv,Martin:1992mq}. One can easily show
this by noting that both the Lagrangian and the VEV $\langle
\phi_H\rangle$ are invariant under the following transformations: 
\begin{equation}
 \phi_i \to \exp\biggl(\frac{i2\pi Q_i}{N}\biggr)\phi_i ~,
~~~~~~
\langle \phi_H\rangle \to \exp\biggl(\frac{i2\pi Q_H}{N}\biggr)\langle \phi_H
\rangle =\langle \phi_H \rangle ~. 
\end{equation}
Thus, an SO(10) GUT may account for the stability of DM in terms of the
remnant $\mathbb{Z}_N$ symmetry originating from the extra U(1) gauge
symmetry.

The next task is to determine which type of irreducible
representations for the Higgs field $\phi_H$ can be exploited to realize
the discrete symmetry. To that end, we follow the discussion presented
in Ref.~\cite{DeMontigny:1993gy}. 
The discussion is based on the Dynkin formalism of the Lie
algebra \cite{DYNKINE.B.:1957}.\footnote{For a
review and references, see Refs.~\cite{Slansky:1981yr,Georgi:1982jb}. We
follow the convention of Ref.~\cite{Slansky:1981yr} in this paper. }
Since the rank of SO(10) is five, we have five independent 
generators which can be diagonalized simultaneously. We denote them by
$H_i$ ($i=1,\dots, 5$). They form the Cartan subalgebra of SO(10). 
Each component of a multiplet is characterized by a set of eigenvalues
of the generators, $\mu_i$ $(i=1,\dots, 5)$, called weights. We also
define the weight vector $\bm{\mu}
\equiv (\mu_1, \dots, \mu_5)$. The weights in
the adjoint representation are called roots $\alpha_i$, with
$\bm{\alpha} = (\alpha_1,\dots, \alpha_5)$ the root vector. Among the
root vectors, a set of five linearly independent vectors play an
important role. They are called simple roots, $\bm{\alpha}_i$ ($i=1,\dots
5$), and expressed by the Dynkin diagrams. In what follows, we consider
the weight and root vectors in the so-called Dynkin basis. In this
particularly useful basis, a weight vector $\bm{\mu}$ is expressed in terms
of a set of Dynkin labels given by
\begin{equation}
 \widetilde{\mu}_i = \frac{2\bm{\alpha}_i \cdot \bm{\mu}}{|\bm{\alpha}_i|^2} ~.
\end{equation}
It turns out that the Dynkin labels are always integers. For example,
the highest weight of the ${\bf 16}$ in SO(10) is expressed as
$(0~0~0~0~1)$, while that of the ${\bf 10}$ is given by $(1~0~0~0~0)$. 

On the other hand, it is convenient to express the Cartan generators
$H_i$ in the dual basis, where they are expressed in terms of
five-dimensional vectors $[\bar{h}_{i1},\dots, \bar{h}_{i5}]$ such that
their eigenvalues for a state corresponding to the weight $\bm{\mu}$ are
given by
\begin{equation}
 H_i(\bm{\mu}) = \sum_{j=1}^{5} \bar{h}_{ij} \widetilde{\mu}_j ~.
\end{equation}
We choose the five linearly independent Cartan generators as follows:
\begin{align}
 H_1 &= \frac{1}{2}[1~2~2~1~1] ~, \nonumber \\
 H_2 &= \frac{1}{2\sqrt{3}}[1~0~0~-1~1] ~, \nonumber \\
 H_3 &= \frac{1}{2}[0~0~1~1~1] ~, \nonumber \\
 H_4 &= \frac{1}{6}[-2~0~3~-1~1] ~, \nonumber \\
 H_5 &= [2~0~2~1~-1] ~.
\end{align}
Here, $H_1$ and $H_2$ correspond to the SU(3)$_C$ Cartan
generators $\lambda_3/2$ and $\lambda_8/2$, respectively, where
$\lambda_A$ $(A=1,\dots,8)$ are the Gell-Mann matrices; $H_3$ and $H_4$
are the weak isospin and hypercharge, $T_{3L}$ and $Y$,
respectively.\footnote{In the case of the flipped SU(5) scenario
\cite{Barr:1981qv,Antoniadis:1987dx},
the weak hypercharge is given by $Y=-\frac{1}{5}(H_4+H_5)$. } 
$H_5$ is related to the $B-L$ charge as $H_5=-5(B-L)+4Y$.
The additional U(1) symmetry required to generate a discrete symmetry is
provided by a linear combination of the Cartan generators containing
$H_5$. Following Ref.~\cite{DeMontigny:1993gy} (see also
Ref.~\cite{Ibanez:1991hv}), we define the extra U(1)
charge $Q_1$ by
\begin{equation}
 Q_1 =  -\frac{6}{5} H_4 -\frac{1}{5} H_5 
= [0~0~-1~0~0]
~.
\end{equation}
This U(1) charge can be also written as $Q_1 = (B-L)-2Y$. One can
readily find that all of the components in ${\bf 10}$ and ${\bf 16}$
have the U(1) charges of either $0$ or $\pm 1$.

Now we consider possible representations, $R_2$, for $\phi_H$ discussed
above. First, let us determine the possible weight vectors corresponding
to the component of $\phi_H$ that can have a VEV without breaking the SM
gauge group. Namely, such a component has a zero eigenvalue for $H_i$
($i=1,\dots, 4$). This condition tells us that the corresponding weight
vectors have the following form:
\begin{equation}
 \bm{\mu}_{N} = (-N~N~-N~0~N) ~.
\end{equation}
The $Q_1$ charges of the vectors are then given by 
\begin{equation}
 Q_1(\bm{\mu}_N) = N ~.
\end{equation}
It is found that the smallest irreducible representation that contains
the weight vector $\bm{\mu}_{N}$ has the highest weight\footnote{In
fact, we obtain $\bm{\mu}_N$ by subtracting the root vector
$(1~-1~1~0~0)$ from $\bm{\Lambda}_N$ $N$ times.} 
\begin{equation}
 \bm{\Lambda}_N = (0~0~0~0~N) ~.
\end{equation}
Its dimension is ${\bf 16}$, ${\bf 126}$, ${\bf
672}$, $\dots$ for $N=1,2,3,\dots$, respectively.\footnote{
The dimension of $\bm{\Lambda}_N$ for any $N$ is given by
\begin{equation}
 \text{dim}(\bm{\Lambda}_N) = 
(1+N)
\biggl(1+\frac{N}{2}\biggr)
\biggl(1+\frac{N}{3}\biggr)^2
\biggl(1+\frac{N}{4}\biggr)^2
\biggl(1+\frac{N}{5}\biggr)^2
\biggl(1+\frac{N}{6}\biggr)
\biggl(1+\frac{N}{7}\biggr)~.
\end{equation}
} 
To obtain a $\mathbb{Z}_N$ symmetry, $N\geq 2$ is required. Thus,
as long as we consider relatively small representations (such as those
with dimensions not exceeding 210), ${\bf 126}$ is the only
candidate\footnote{The next-to-smallest representation including
$\bm{\mu}_2$ is ${\bf 1728}$ with the highest weight (1~0~0~1~1).} for
the representation of $\phi_H$. In this case, the remnant discrete
symmetry is $\mathbb{Z}_2$.\footnote{For earlier work on the
remnant $\mathbb{Z}_2$ symmetry in SO(10), see Ref.~\cite{Kibble:1982ae}.}

Under the $\mathbb{Z}_2$ symmetry, the SM left-handed fermions are even,
while the SM right-handed fermions as well as the Higgs field are
odd. One can easily show that this symmetry is related to the product of
matter parity $P_M = (-1)^{3(B-L)}$ \cite{Farrar:1978xj} and the U(1)$_Y$
rotation by $6\pi$, $e^{6i\pi Y}$. Thus, if a
SM-singlet fermion (boson) has an even (odd) parity, the remnant
$\mathbb{Z}_2$ symmetry makes the particle stable. In
Table~\ref{tab:irrp}, we summarize irreducible representations that
contain $\bm{\mu}_N$. We only show those that have dimensions
less than or equal to 210. From the table, we find that a singlet fermion in a
${\bf 45}$, ${\bf 54}$, ${\bf 126}$, or ${\bf 210}$ representation, or a
singlet scalar boson in a ${\bf 16}$ or ${\bf 144}$ representation, can
be a DM candidate.

\begin{table}[t]
 \begin{center}
\caption{\it Irreducible representations containing $\bm{\mu}_N$.}
\label{tab:irrp}
\vspace{5pt}
\begin{tabular}{cccc}
\hline
\hline
&Representation & Highest weight & $\mathbb{Z}_2$ \\ 
\hline
$\bm{\mu}_0$ & ${\bf 45}$ & $(0~1~0~0~0)$ & $+$ \\
  & ${\bf 54}$ & $(2~0~0~0~0)$ & $+$ \\
  & ${\bf 210}$ & $(0~0~0~1~1)$ & $+$ \\
$\bm{\mu}_1$ & ${\bf 16}$ & $(0~0~0~0~1)$ & $-$ \\
 & ${\bf 144}$ & $(1~0~0~1~0)$ & $-$ \\
$\bm{\mu}_2$ & ${\bf 126}$ & $(0~0~0~0~2)$ & $+$ \\
\hline
\hline
\end{tabular}
 \end{center}
\end{table}

Note that although we need a ${\bf 126}$ Higgs field to break the
extra U(1) symmetry and produce a remnant $\mathbb{Z}_2$ symmetry, other
$\mathbb{Z}_2$-even singlet fields, ${\bf 45}$, ${\bf 54}$, ${\bf 210}$,
{\it etc}., can have VEVs simultaneously without breaking the
$\mathbb{Z}_2$ symmetry. While the latter do not break the 
$\mathbb{Z}_2$ symmetry, as discussed above, they are not capable 
of producing it, thus requiring the {\bf 126}.  We will use such fields
to obtain an  adequate mass spectrum and a non-degenerate DM candidate,
as discussed in Sec.~\ref{sec:degenprob} and
Sec.~\ref{sec:models}. $R_2$ will therefore refer to all representations
at the intermediate-scale which are responsible for either symmetry
breaking or intermediate scale masses and may be a combination of the
${\bf 126}$ and other representations listed in Table~\ref{tab:irrp}
with positive $\mathbb{Z}_2$ charge.  

\subsection{Intermediate gauge group}
\label{sec:intgauge}

As shown in Eq.~\eqref{eq:decaypattern}, the extra U(1) symmetry is
assumed to be broken at the intermediate scale, \textit{i.e.}, the ${\bf
126}$ Higgs field acquires a VEV of the order of $M_{\text{int}}$. Thus,
the intermediate gauge group $G_{\text{int}}$ should be of rank five. In
Table~\ref{tab:intgauge}, we summarize the rank-five subgroups of SO(10)
and the Higgs multiplets $R_1$ whose VEVs break SO(10) into the
subgroups. Again, we only consider the representations whose dimensions
are less than or equal to {\bf 210}. Here $D$ denotes the so-called $D$-parity
\cite{Kuzmin:1980yp}, that is, a $\mathbb{Z}_2$ symmetry with respect to
the exchange of SU(2)$_L\leftrightarrow \text{SU}(2)_R$. $D$-parity
can be related to an element of SO(10) \cite{Kuzmin:1980yp} under which
a fermion field transforms into its charge conjugate. In cases where the
$D$-parity is not broken by $R_1$, it is subsequently broken by $R_2$ at
the scale of $M_{\text{int}}$. In the NETDM scenario, the reheating
temperature is always below $M_{\text{int}}$, and therefore any
cosmological relics \cite{Kibble:1982ae} due to the breaking of
$D$-parity will be harmless. Note that the VEVs of the $R_1$ 
Higgs fields are even under the  
$\mathbb{Z}_2$ symmetry considered in Sec.~\ref{sec:discsym}. Thus,
there is no danger for this $\mathbb{Z}_2$ symmetry to be spontaneously
broken by the $R_1$ Higgs fields. 

\begin{table}[t]
 \begin{center}
\caption{\it Candidates for the intermediate gauge group $G_{\text{int}}$.}
\label{tab:intgauge}
\vspace{5pt}
\begin{tabular}{ll}
\hline
\hline
$G_{\text{int}}$ & $R_1$ \\
\hline
$\text{SU}(4)_C\otimes \text{SU}(2)_L \otimes \text{SU}(2)_R$& {\bf 210}\\
$\text{SU}(4)_C\otimes \text{SU}(2)_L \otimes \text{SU}(2)_R\otimes {D}$& {\bf
     54}\\
$\text{SU}(4)_C\otimes \text{SU}(2)_L \otimes \text{U}(1)_R$ & {\bf 45}\\
$\text{SU(3)}_C\otimes \text{SU}(2)_L \otimes \text{SU}(2)_R
 \otimes \text{U}(1)_{B-L}$ &{\bf 45}\\
$\text{SU(3)}_C\otimes \text{SU}(2)_L \otimes \text{SU}(2)_R
 \otimes \text{U}(1)_{B-L} \otimes D$ & {\bf 210}\\
$\text{SU(3)}_C\otimes \text{SU}(2)_L \otimes \text{U}(1)_R 
 \otimes \text{U}(1)_{B-L}$ & {\bf 45}, {\bf 210}\\
\hline
$\text{SU}(5) \otimes \text{U}(1)$ & {\bf 45}, {\bf 210}\\
$\text{Flipped}~ \text{SU}(5) \otimes \text{U}(1)$ & {\bf 45}, {\bf 210}\\
\hline
\hline
\end{tabular}
 \end{center}
\end{table}

\subsection{Fermion dark matter and degeneracy problem}
\label{sec:degenprob}

In the NETDM scenario, the DM should not be in thermal equilibrium. This
requirement disfavors scalar DM candidates, since a scalar, $\phi$, can
always have a quartic coupling with the SM Higgs field $H$: $\lambda_{\phi
H}|\phi|^2 |H|^2$. Unless $|\lambda_{\phi H}|$ is extremely small for
some reason, this coupling keeps scalar DM in thermal equilibrium
even when the temperature of the Universe becomes much lower than the
reheating temperature. Therefore, we focus on fermionic DM
in this paper. Following the discussion in Sec.~\ref{sec:discsym},
the DM candidate should be contained in either a ${\bf 45}$, ${\bf 54}$,
${\bf 126}$, or ${\bf 210}$ representation. 

\begin{table}[t]
 \begin{center}
\caption{\it Candidates for the NETDM.}
\label{tab:NETDM}
\vspace{5pt}
\begin{tabular}{lll}
\hline
\hline
$G_{\text{int}}$ & $R_{\text{DM}}$ & SO(10) \\
\hline
$\text{SU}(4)_C\otimes \text{SU}(2)_L \otimes \text{SU}(2)_R$&
({\bf 1}, {\bf 1}, {\bf 3}) & {\bf 45}\\
&({\bf 15}, {\bf 1}, {\bf 1}) & {\bf 45}, {\bf 210}\\
&({\bf 10}, {\bf 1}, {\bf 3}) & {\bf 126}\\
&({\bf 15}, {\bf 1}, {\bf 3}) & {\bf 210}\\
\hline
$\text{SU}(4)_C\otimes \text{SU}(2)_L \otimes \text{U}(1)_R$ 
&({\bf 15}, {\bf 1}, 0)& {\bf 45}, {\bf 210}\\
&({\bf 10}, {\bf 1}, $1$) & {\bf 126}\\
\hline
$\text{SU(3)}_C\otimes \text{SU}(2)_L \otimes \text{SU}(2)_R
 \otimes \text{U}(1)_{B-L}$
& ({\bf 1}, {\bf 1}, {\bf 3}, 0)&{\bf 45}, {\bf 210}\\
& ({\bf 1}, {\bf 1}, {\bf 3}, $-2$)&{\bf 126}\\
\hline
$\text{SU(3)}_C\otimes \text{SU}(2)_L \otimes \text{U}(1)_R 
 \otimes \text{U}(1)_{B-L}$ 
& ({\bf 1}, {\bf 1}, $1$, $-2$)&{\bf 126}\\
\hline
$\text{SU}(5) \otimes \text{U}(1)$ 
&({\bf 24}, 0)&{\bf 45}, {\bf 54}, {\bf 210}\\
&({\bf 1}, $-10$)&{\bf 126}\\
&({\bf 75}, 0)&{\bf 210}\\
\hline
$\text{Flipped}~ \text{SU}(5) \otimes \text{U}(1)$ 
&({\bf 24}, 0)&{\bf 45}, {\bf 54}, {\bf 210}\\
&($\overline{\bf 50}$, $-2$)&{\bf 126}\\
&({\bf 75}, 0)&{\bf 210}\\
\hline
\hline
\end{tabular}
 \end{center}
\end{table}

Below the GUT scale, components in an SO(10) multiplet can obtain
different masses. We assume that only a part of an SO(10) multiplet which
contains the DM candidate and forms a representation under 
$G_{\text{int}}$ has a mass much lighter than the GUT
scale. We denote this representation by $R_{\text{DM}}$. Such a mass
splitting can be realized by the Yukawa coupling of the DM multiplet
with the $R_1$ Higgs field. After the $R_1$ Higgs obtains a VEV, the
Yukawa coupling leads to an additional mass term for the SO(10)
multiplet, which gives different masses among the components. By
carefully choosing the parameters in the Lagrangian, we can make only
$R_{\text{DM}}$ light. This will be discussed in detail in
Sec.~\ref{sec:models}.  

As will be seen in Sec.~\ref{sec:gcuandint}, without $R_{\text{DM}}$, SO(10)
GUTs often predict a low value of either $M_{\text{GUT}}$ or
$M_{\text{int}}$, which could be problematic for proton decay or the
explanation of light neutrino 
masses, respectively. In order to affect the RGE running of the gauge
couplings and possibly increase the mass scales for both
$M_{\text{int}}$ and $M_{\text{GUT}}$, the DM should be charged under
$G_{\text{int}}$. In Table~\ref{tab:NETDM}, we summarize possible
candidates for $R_{\text{DM}}$ for each intermediate gauge group. 
Above the intermediate scale, all of the components have an identical
mass. In fact, it turns out that the degeneracy is not resolved at tree
level even after the intermediate gauge symmetry is broken. This is because
the SO(10) multiplets which contain $R_{\text{DM}}$ displayed in the
table cannot have Yukawa couplings with the {\bf 126} Higgs; such a
coupling is forbidden by the SO(10) symmetry. Thus, the effects of
symmetry breaking by the {\bf 126} Higgs VEV cannot be transmitted to
the mass of the $R_{\text{DM}}$ multiplet at tree level, and a simple
realization of DM in $R_{\text{DM}}$ makes its components degenerate
in mass.

\begin{figure}[t]
\begin{center}
\includegraphics[height=45mm]{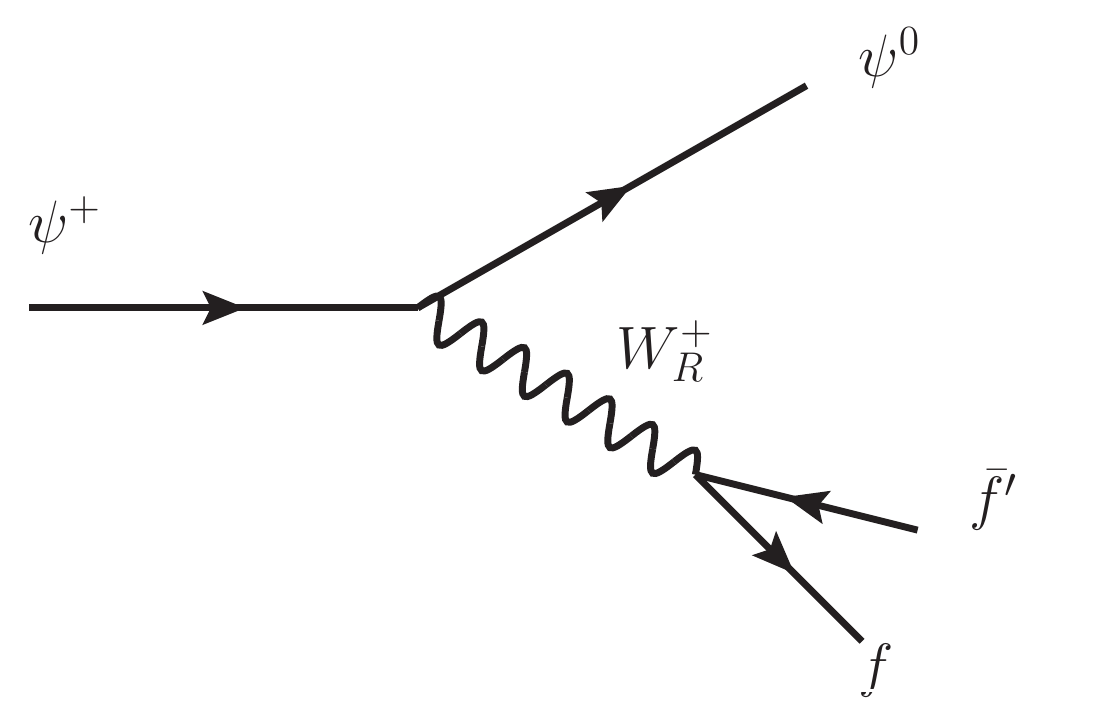}
\caption{{\it Diagram responsible for the decay of $\psi^+$ into the DM
 $\psi^0$. $f$ and $f^\prime$ denote the SM particles.}}
\label{fig:decay}
\end{center}
\end{figure}

Such a degenerate mass spectrum is problematic.
Since the degenerate multiplet contains particles charged under the
SU(3)$_C\otimes\text{U}(1)_{\text{EM}}$ gauge group, they will be
in thermal equilibrium. In general, these components have quite a long
lifetime, and thus their thermal relic density conflicts with various
observations. To see this, let us consider the $({\bf 1}, {\bf 1}, {\bf
3})$ Dirac fermion multiplet $(\psi^0, \psi^\pm)$ in the
$\text{SU}(4)_C\otimes \text{SU}(2)_L \otimes \text{SU}(2)_R$ theory,
which originates from the {\bf 45} representation of SO(10), as an
example. As mentioned above, they have an identical mass $M$ at tree
level, and the mass difference $\Delta M$ induced by the radiative
corrections can be estimated as 
\begin{equation}
\Delta M \simeq \frac{\alpha_1}{4\pi} M\ln
\biggl(\frac{M_{\text{int}}}{M}\biggr)  
\sim 0.01\times M
~,
\end{equation}
where $\alpha_1$ is the $\text{U}(1)$ gauge fine-structure constant. The charged
components $\psi^\pm$ can decay into the neutral DM $\psi^0$ only through the
exchange of the intermediate-scale gauge bosons as shown in
Fig.~\ref{fig:decay}. We estimate the decay width as
\begin{equation}
 \Gamma(\psi^+\to \psi^0 f \bar{f}^\prime) \sim \frac{\alpha_R^2}{\pi}
\frac{(\Delta M)^5}{M_{W_R}^4} ~,
\end{equation}
where $\alpha_R = g_R^2/4\pi$, and $g_R$ and $M_{W_R}$ are the coupling and the mass of the
intermediate gauge boson $W_R$, respectively.
Then, for example, when the DM mass is ${\cal O}(1)$~TeV
and the intermediate scale is ${\cal O}(10^{13})$~GeV, the lifetime of
$\psi^+$ is much longer than the age of the Universe, and thus
cosmologically stable. The abundance of such a stable charged particle
is stringently constrained by the null results of the
search for heavy hydrogen in sea water \cite{Smith:1979rz}. The DM
multiplets in other cases may also be accompanied by stable colored
particles, whose abundance is severely restricted as well. 
If the intermediate scale is relatively low, the charged/colored
particle can have a shorter lifetime. Even in this case, their thermal
relic abundance should be extremely small in order not to spoil the
success of the Big-Bang Nucleosynthesis (BBN). Quite generally, a degenerate
mass spectrum leads to disastrous consequences. We refer to this
problem as the ``degeneracy problem'' in what follows.   

To avoid the degeneracy problem, we need to make the charged/colored
components heavy enough so that they are not in thermal equilibrium and
have very short lifetimes. To that end, it is natural to explore a way
to give them masses of ${\cal O}(M_{\text{int}})$ by using the effects
of the intermediate symmetry breaking. There are several solutions. One
of the simplest ways is to introduce an additional Higgs field that has
a VEV of the order of $M_{\text{int}}$. For this purpose, we can use a
{\bf 45}, {\bf 54}, or {\bf 210} field, as discussed in
Sec.~\ref{sec:discsym}.  The Yukawa coupling between the Higgs
and the DM then yields the desired mass splitting. By fine-tuning the coupling
we can force only the DM to have a mass much below $M_{\text{int}}$ while the
other components remain around the intermediate scale\footnote{This
fine-tuning is similar (though somewhat less severe) to the fine-tuning
associated with the doublet-triplet separation to insure a weak scale
Higgs boson.}. Though other mechanisms are possible, we adopt this
approach in this work. Concrete realizations of the mechanism are
illustrated in Sec.~\ref{sec:models}.

Another solution to the degeneracy problem involves the use of higher-dimensional operators that include
at least two {\bf 126} fields. One would expect that 
such operators suppressed by the Planck scale, $M_{\text{Pl}}$, always exist. These
Planck-suppressed operators can give rise to a mass difference of
${\cal O}(M_{\text{int}}^2/M_{\text{Pl}})$. Another mechanism to
generate higher-dimensional operators is to introduce a vector-like
fermion which has a Yukawa coupling with the DM and the {\bf 126}
Higgs. By integrating out the fermion, we obtain dimension-five
operators which give an ${\cal
O}(M_{\text{int}}^2/M_{\text{fer}})$ mass difference, where
$M_{\text{fer}}$ is the mass of the additional fermion. Moreover, the
higher-dimensional operators can be induced at the loop level, which gives 
rise to a ${\cal O}(\alpha_{\text{GUT}}M_{\text{int}}^2/(4\pi
M_{\text{GUT}}))$ mass difference, where $\alpha_{\text{GUT}}=
g_{\text{GUT}}^2/(4\pi)$ is the fine-structure constant of the unified
gauge coupling $g_{\text{GUT}}$. Realization
of these scenarios will be discussed elsewhere.

\section{Gauge coupling unification}
\label{sec:gcu}

As is well known, gauge coupling unification can be realized in
SO(10) GUTs with an intermediate scale
\cite{Rajpoot:1980xy}.\footnote{For a review, see
Refs.~\cite{Fukugita:1993fr, DiLuzio:2011my}.} Once the intermediate gauge
group as well as the low-energy matter content is given, one can
determine both the intermediate and GUT scales by requiring gauge
coupling unification. In what follows, we reevaluate these scales in the
SO(10) GUT scenarios with different intermediate gauge groups and
up-to-date values for the input parameters. Then, in
Sec.~\ref{sec:NETDMgcu}, we study the effects of the DM and the
intermediate Higgs multiplets on gauge coupling unification. We will
find that the requirement of gauge coupling unification severely
constrains the NETDM models.

\subsection{Gauge coupling unification with the intermediate scale}
\label{sec:gcuandint}

To begin with, let us briefly review SO(10) GUTs with
an intermediate gauge group. In SO(10) GUTs, the SM fermions as well
as three right-handed neutrinos are embedded into three copies of
the ${\bf 16}$ spinor representations, while the SM Higgs boson is
usually included in a ${\bf 10}$ representation. At the GUT scale, the SO(10) GUT group is
spontaneously broken into an intermediate gauge group. Subsequently, the intermediate Higgs
multiplet breaks it into the SM gauge group at the
intermediate scale. In the following analysis, we
work with the so-called extended survival hypothesis
\cite{delAguila:1980at, Mohapatra:1982aq}; that is, we assume that a
minimal set of Higgs multiplets necessary to realize the symmetry
breaking exists in low-energy region. Above the intermediate scale, the
presence of the additional Higgs multiplet and intermediate gauge bosons
change the gauge coupling running from that in the SM. This makes it
possible to realize gauge coupling unification in this scenario.

As displayed in Table~\ref{tab:intgauge}, the intermediate gauge groups
relevant to our discussion are divided into two classes; those which
contain the SU(5) group as a subgroup, and those which do not. The former class
is, however, found to be less promising. In the case of ordinary
$\text{SU}(5) \otimes \text{U}(1)$, the SM gauge couplings should meet
at the intermediate scale, though they do not, as is well known. Failure
of gauge coupling unification is also found in the flipped SU(5)
case. This conclusion cannot be changed even if one adds the DM
and Higgs multiplets in the case of ordinary SU(5). In the flipped SU(5)
case, the addition of the DM and Higgs multiplets may yield gauge
coupling unification. However, it turns out that the intermediate mass
scale is as high as ${\cal O}(10^{17})$~GeV in such cases. Since the
masses of the right-handed neutrinos are expected to be ${\cal
O}(M_{\text{int}})$, if $M_{\text{int}} = {\cal O}(10^{17})$, the simple
seesaw mechanism \cite{Minkowski:1977sc} cannot explain the neutrino
masses required from the observation of the neutrino
oscillations. However, the GUT scale tends to be close to the Planck
scale, and one may need to rely on a double seesaw to explain neutrino
masses \cite{Antoniadis:1987dx,flipped}.  
We do not consider these
possibilities in the following discussion.

The other class of the intermediate gauge groups is related to the
Pati-Salam gauge group \cite{Pati:1974yy}. Therefore, it is useful to 
decompose the SO(10) multiplets into multiplets of the
$\text{SU}(4)_C \otimes \text{SU}(2)_L \otimes \text{SU}(2)_R$ gauge
group. The ${\bf 16}$ spinor representation in SO(10) is decomposed
into a $({\bf 4}, {\bf 2}, {\bf 1})$ and $(\overline{\bf 4}, {\bf 1}, {\bf
2})$ of $\text{SU}(4)_C \otimes \text{SU}(2)_L \otimes
\text{SU}(2)_R$. We denote them by $\Psi_L$ and $\Psi^{\cal C}_R$,
respectively, in which the SM fermions are embedded as follows:
\begin{equation}
 \Psi_L =
\begin{pmatrix}
 u^1_L & u^2_L& u^3_L & \nu_L \\
 d^1_L & d^2_L& d^3_L & e_L
\end{pmatrix}
~, ~~~~~~
\Psi_R^{\cal C} =
\begin{pmatrix} 
 d^{\cal C}_{R1}&  d^{\cal C}_{R2}&  d^{\cal C}_{R3}& e^{\cal C}_R \\
 - u^{\cal C}_{R1} &- u^{\cal C}_{R2} &- u^{\cal C}_{R3} & -\nu _R^{\cal C}
\end{pmatrix}
~,
\label{eq:4spinordef}
\end{equation}
where the indices represent the SU(3)$_C$ color and ${\cal C}$ indicates
charge conjugation. The SM Higgs field is, on the other hand, embedded
in the $({\bf 1}, {\bf 2}, \overline{\bf 2})$ component of the
ten-dimensional representation. As discussed in Ref.~\cite{Bajc:2005zf},
to obtain the viable Yukawa sector,\footnote{For a general discussion on
the Yukawa sector in SO(10) GUTs, see
Refs.~\cite{Bajc:2005zf,Fukuyama:2004xs}. } we need to consider a complex scalar
${\bf 10}_C$ for the representation, not a real one. Thus, $({\bf 1},
{\bf 2}, \overline{\bf 2})$ is also a complex scalar multiplet and
includes the two Higgs doublets. In the following calculation, we regard
one of these doublets as the SM Higgs boson, and the other is assumed to
have a mass around the intermediate scale. The $\text{SU}(4)_C \otimes
\text{SU}(2)_L \otimes \text{SU}(2)_R$ gauge group is broken by the VEV
of the $({\bf 10}, {\bf 1}, {\bf 3})$ component in the ${\bf 126}_C$. In the
presence of the left-right symmetry, we also have a $(\overline{\bf 10},
{\bf 3}, {\bf 1})$ above the intermediate scale. We assume that the
$(\overline{\bf 10}, {\bf 3}, {\bf 1})$ field does not acquire a VEV,
with which the constraint coming from the $\rho$-parameter is avoided.  
From these charge assignments, one can readily obtain the quantum numbers
for the corresponding fields in the other intermediate gauge groups, since
they are subgroups of the $\text{SU}(4)_C \otimes \text{SU}(2)_L \otimes
\text{SU}(2)_R$.

With this field content, we study whether the gauge coupling unification
is actually achieved or not for the first six intermediate gauge groups
listed in Table~\ref{tab:intgauge}. We perform the analysis by using the
two-loop RGEs, which are given in Appendix~\ref{sec:RGEs}. We
will work in the $\overline{\text{DR}}$ scheme \cite{Siegel:1979wq}, 
as there is no constant term in the intermediate and GUT scale matching
conditions. The
input parameters we use in our analysis are listed in
Table~\ref{table:inputparameters} in Appendix~\ref{sec:input}. By
solving the RGEs and assuming gauge coupling unification, we
determine the intermediate scale $M_{\text{int}}$, the 
GUT scale $M_{\text{GUT}}$, and the unified gauge coupling constant
$g_{\text{GUT}}$. If we fail to find the appropriate values for
these quantities, we will conclude that gauge coupling unification is
not realized in this case. To determine their central values as well as the
error coming from the input parameters, we form a $\chi^2$ statistic as 
\begin{equation}
 \chi^2
  =\sum_{a=1}^{3}\frac{(g_a^2-g_{a,\text{exp}}^2)^2}
{\sigma^2(g_{a,\text{exp}}^2)} ~,
\end{equation}
where $g_a$ are the gauge couplings at the electroweak scale obtained by
solving the RGEs on the above assumption, and $g_{a,\text{exp}}$ are the
experimental values of the corresponding gauge couplings, with
$\sigma(g_{a,\text{exp}}^2)$ denoting their error. The central values of
$M_{\text{int}}$, $M_{\text{GUT}}$, and $g_{\text{GUT}}$ are
corresponding to a point at which $\chi^2$ is minimized.\footnote{We
also use the $\chi^2$ statistics to determine the value of the input
Yukawa coupling in a similar manner, though it scarcely affects the
error estimation of $M_{\text{int}}$, $M_{\text{GUT}}$, and
$g_{\text{GUT}}$.  }

\begin{table}[t!]
 \begin{center}
\caption{\it $\log_{10}(M_{\text{int}})$, $\log_{10} (M_{\text{GUT}})$, and
  $g_{\text{GUT}}$. For each $G_{\text{int}}$, the upper shaded (lower)
  row shows the 2-loop (1-loop) result. $M_{\text{int}}$ and
  $M_{\text{GUT}}$ are given in GeV. The blank entries indicate that gauge
  coupling unification is not achieved.  }
\label{tab:resultssm}
\vspace{5pt}
\begin{tabular}{llll}
\hline
\hline
$G_{\text{int}}$ & $\log_{10} (M_{\text{int}})$ & $\log_{10} (M_{\text{GUT}})$ &
 $g_{\text{GUT}}$  \\
\hline
\rowcolor{LightGray}
$\text{SU}(4)_C\otimes \text{SU}(2)_L \otimes \text{SU}(2)_R$
& 11.17(1) & 15.929(4) & 0.52738(4) \\
& 11.740(8)& 16.07(2)& 0.5241(1) \\
\rowcolor{LightGray}
$\text{SU}(4)_C\otimes \text{SU}(2)_L \otimes \text{SU}(2)_R\otimes {D}$
&13.664(3) & 14.95(1)& 0.5559(1)\\
&13.708(7) &15.23(3) & 0.5520(1) \\
\rowcolor{LightGray}
$\text{SU}(4)_C\otimes \text{SU}(2)_L \otimes \text{U}(1)_R$ 
&11.35(2) &14.42(1)& 0.5359(1) \\
&11.23(1) &14.638(8) &0.53227(7) \\
\rowcolor{LightGray}
$\text{SU(3)}_C\otimes \text{SU}(2)_L \otimes \text{SU}(2)_R
 \otimes \text{U}(1)_{B-L}$ 
&9.46(2)& 16.20(2)& 0.52612(8)\\
&8.993(3)&16.68(4)& 0.52124(3) \\
\rowcolor{LightGray}
$\text{SU(3)}_C\otimes \text{SU}(2)_L \otimes \text{SU}(2)_R
 \otimes \text{U}(1)_{B-L} \otimes D$ 
&10.51(1) & 15.38(2)& 0.53880(3)\\
&10.090(9)&15.77(1)&0.53478(6) \\
\rowcolor{LightGray}
$\text{SU(3)}_C\otimes \text{SU}(2)_L \otimes \text{U}(1)_R
 \otimes \text{U}(1)_{B-L}$ &  &  & \\
&&& \\
\hline
\hline
\end{tabular}
 \end{center}
\end{table}

\begin{figure}[t!]
\vspace*{-2mm}
\begin{center} 
\begin{tabular}{cc}
\hspace{0cm} \includegraphics[scale=0.43]{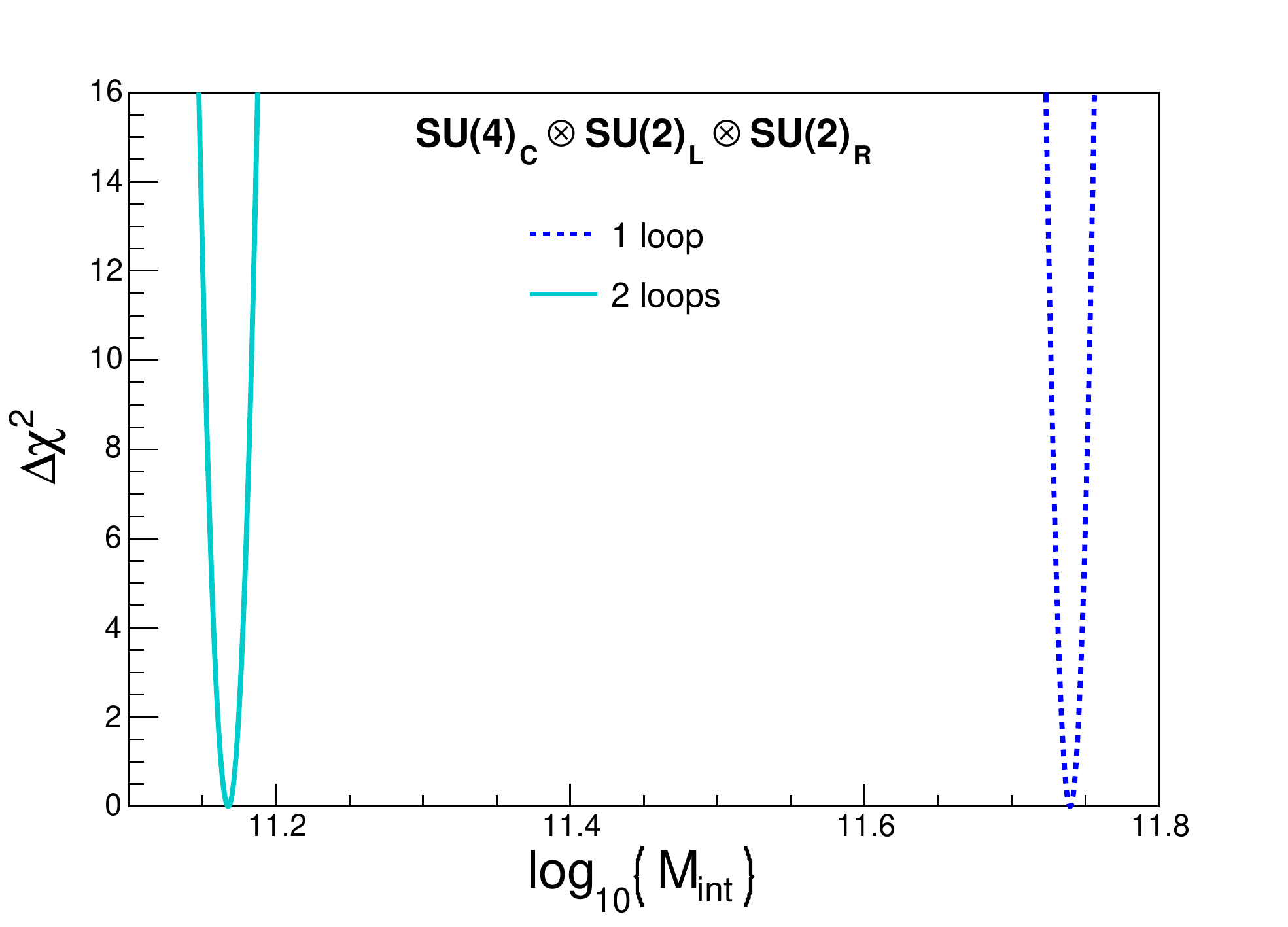} & 
\hspace{-1.3cm}
 \includegraphics[scale=0.43]{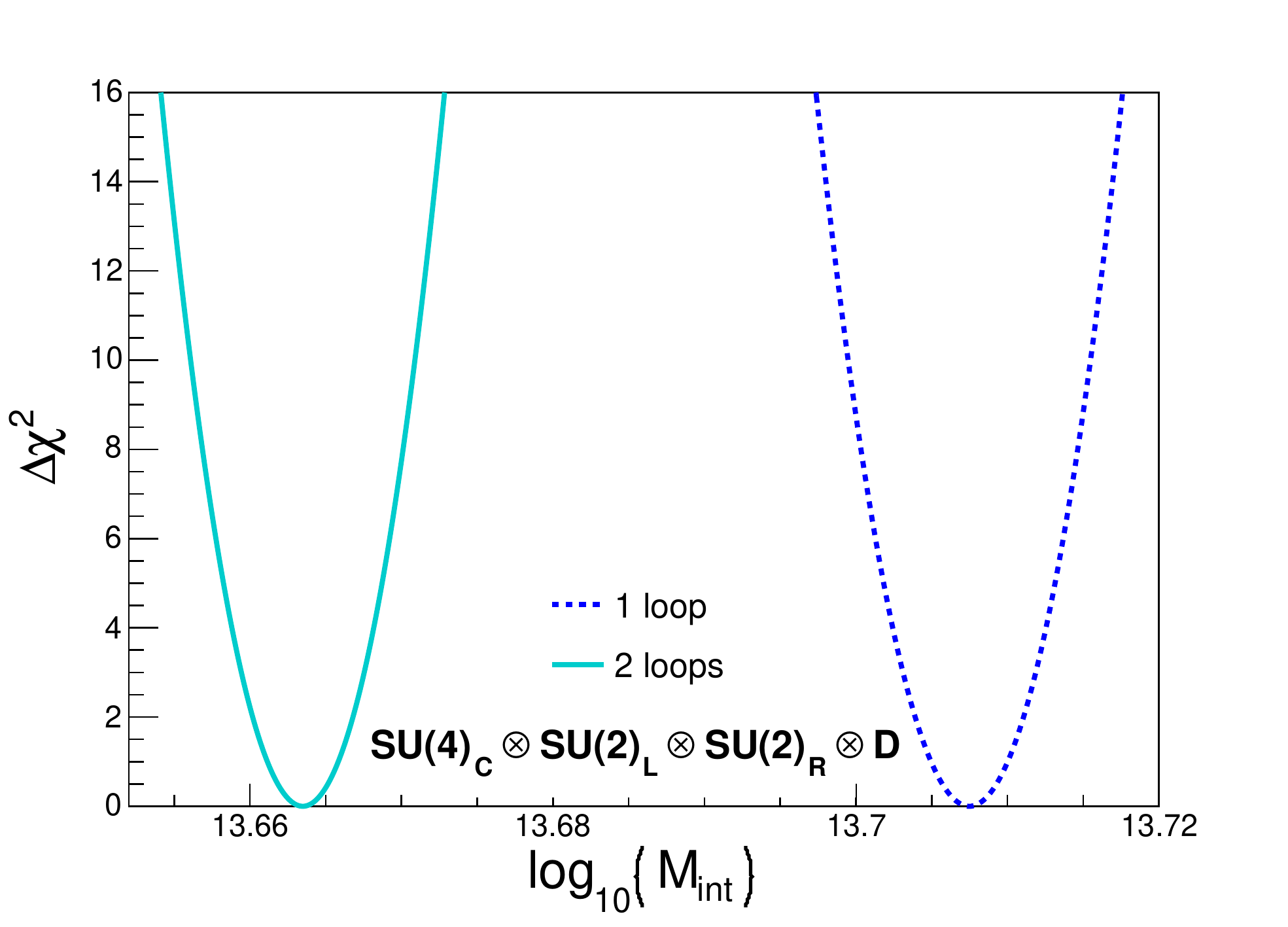} \\
\vspace*{-2mm}
\hspace{0cm} \includegraphics[scale=0.43]{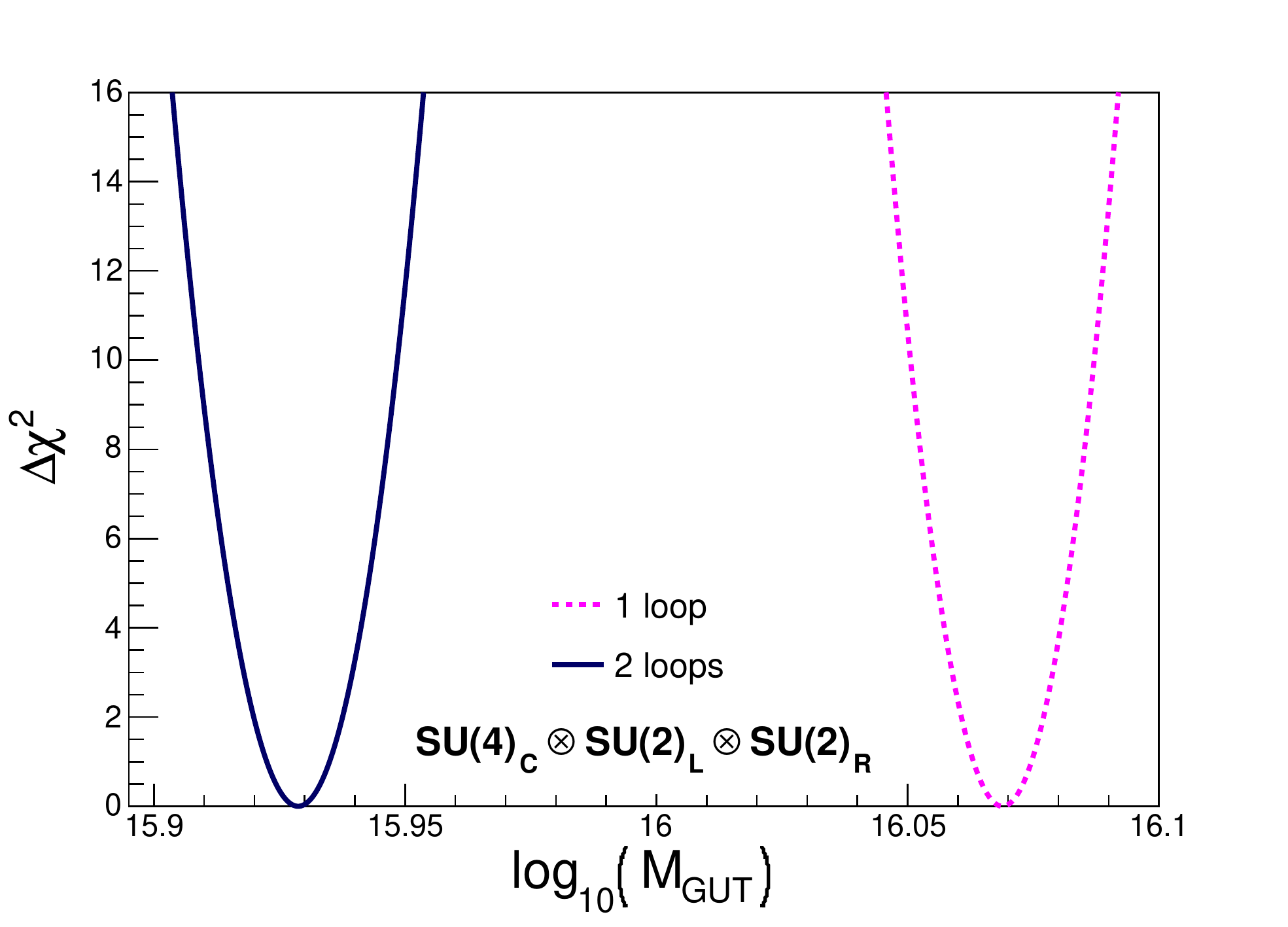} & 
\hspace{-1.3cm} \includegraphics[scale=0.43]{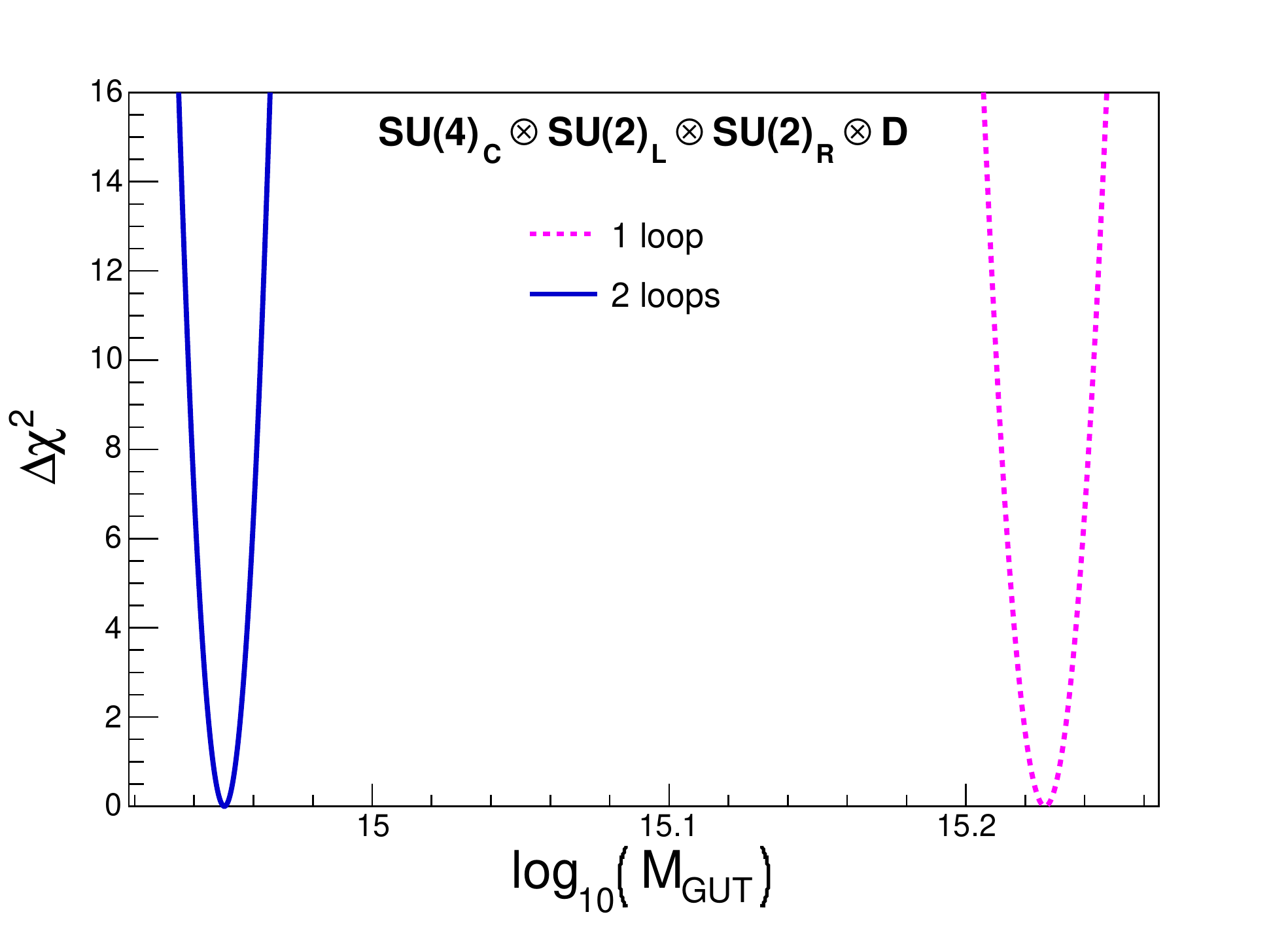} \\
\vspace*{-2mm}
\hspace{0cm} \includegraphics[scale=0.43]{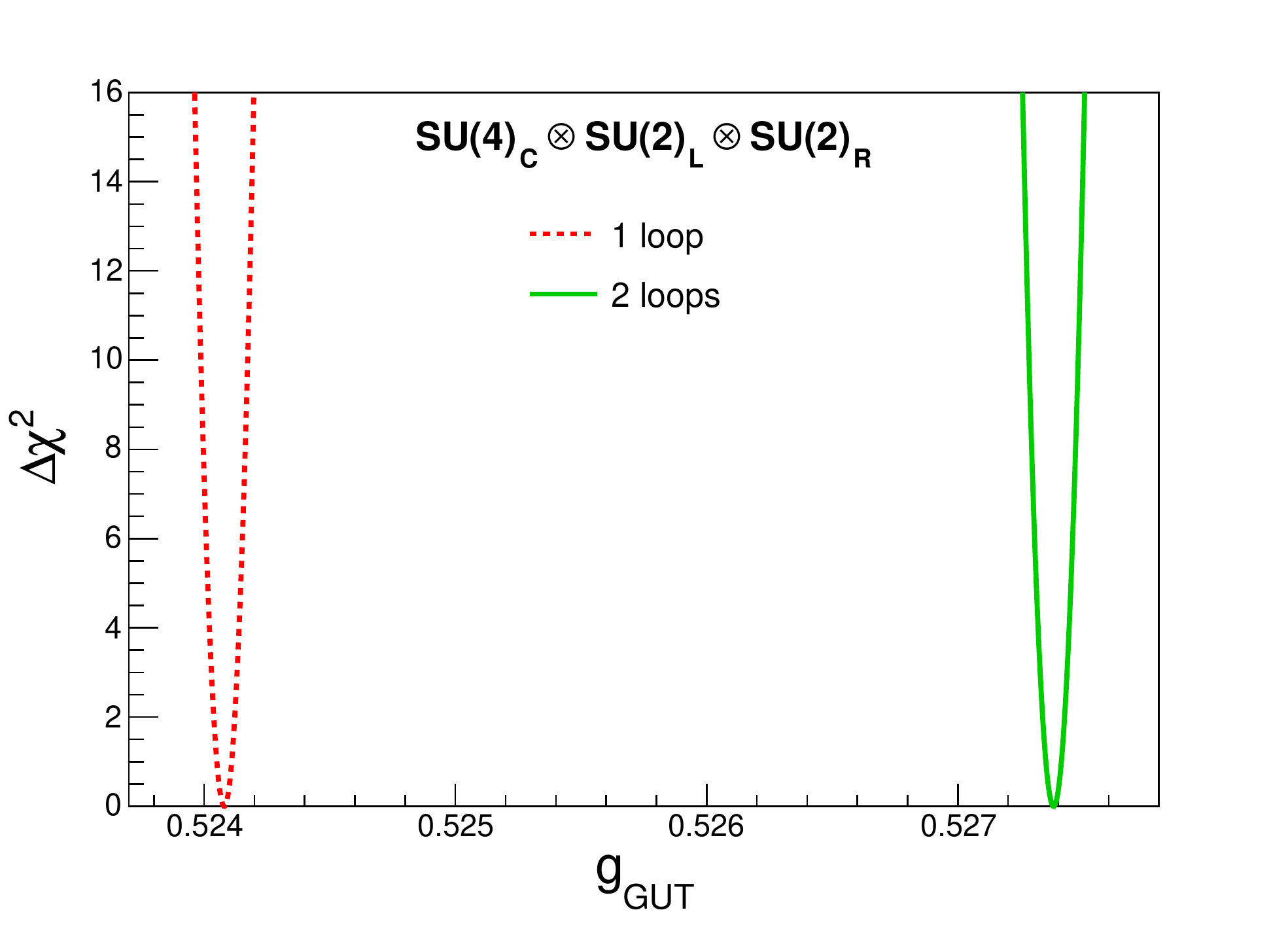} & 
\hspace{-1.3cm} \includegraphics[scale=0.43]{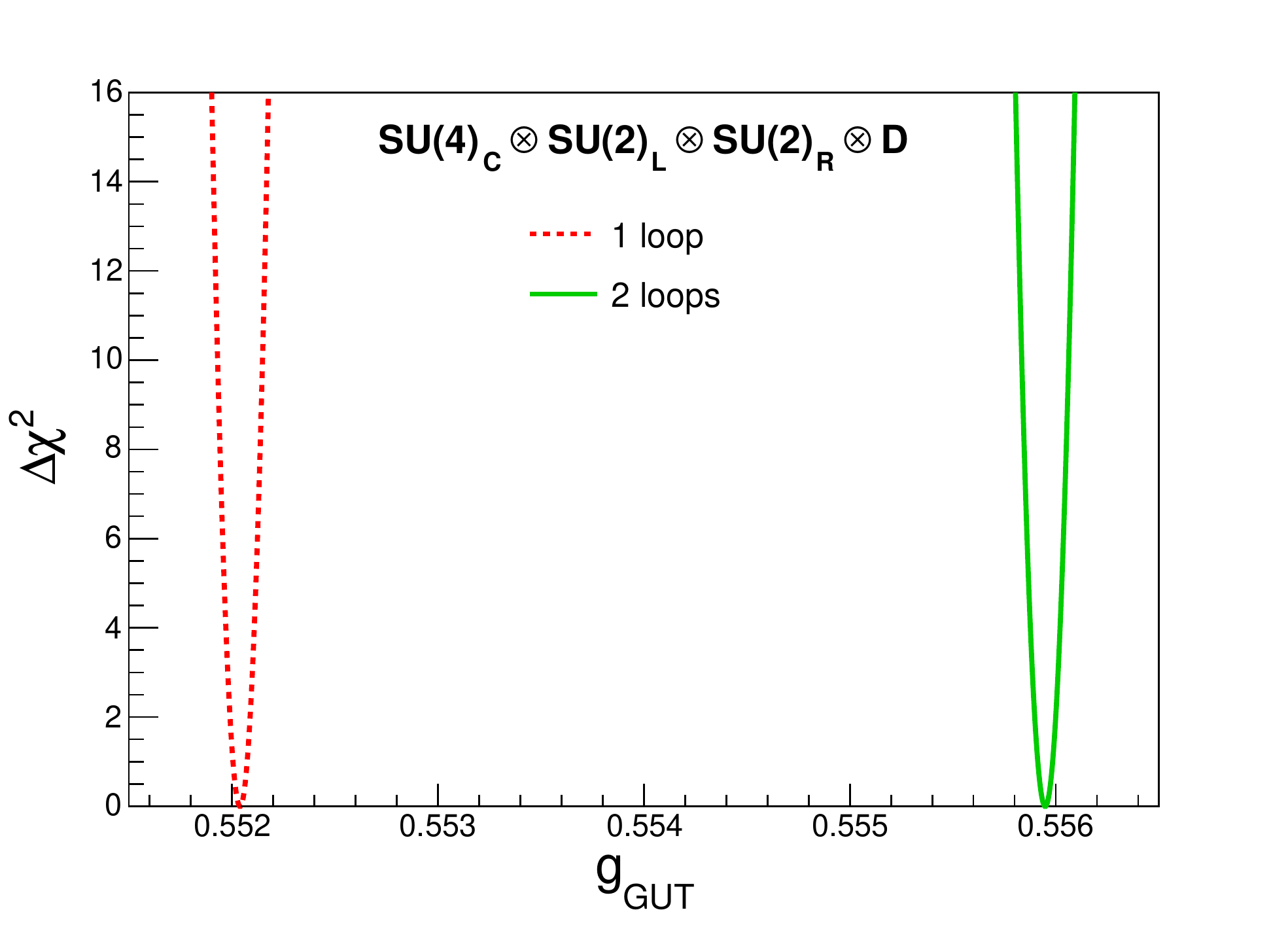} 
\end{tabular}
\caption{\it $\chi^2$ as functions of $\log_{10} (M_{\text{int}})$ (top),
 $\log_{10} (M_{\text{GUT}})$ (middle), and $g_{\text{GUT}}$ (bottom). 
Left and right panels
are for $G_{\text{int}}= \text{SU}(4)_C\otimes \text{SU}(2)_L \otimes
\text{SU}(2)_R$ and $G_{\text{int}}=
\text{SU}(4)_C\otimes \text{SU}(2)_L \otimes \text{SU}(2)_R\otimes
D$, respectively. $M_{\text{int}}$ and $M_{\text{GUT}}$ are given in
GeV.} 
\label{linear_plots}
\end{center}
\end{figure}

\begin{figure}[ht!]
\vspace*{-2mm}
\begin{center} 
\begin{tabular}{cc}
\hspace{-1.0cm}
 \includegraphics[scale=0.41]{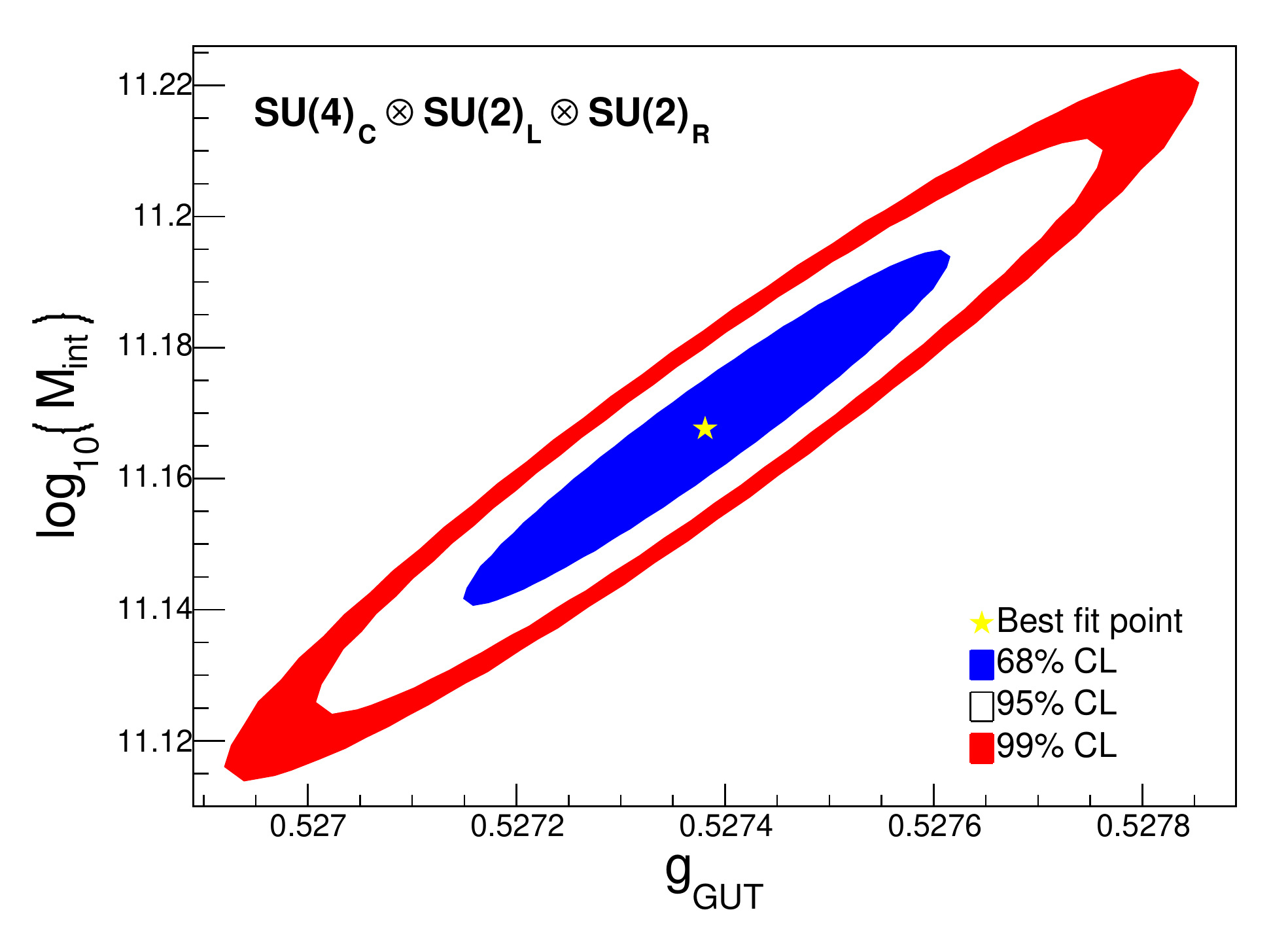} & 
\hspace{-.8cm}
 \includegraphics[scale=0.41]{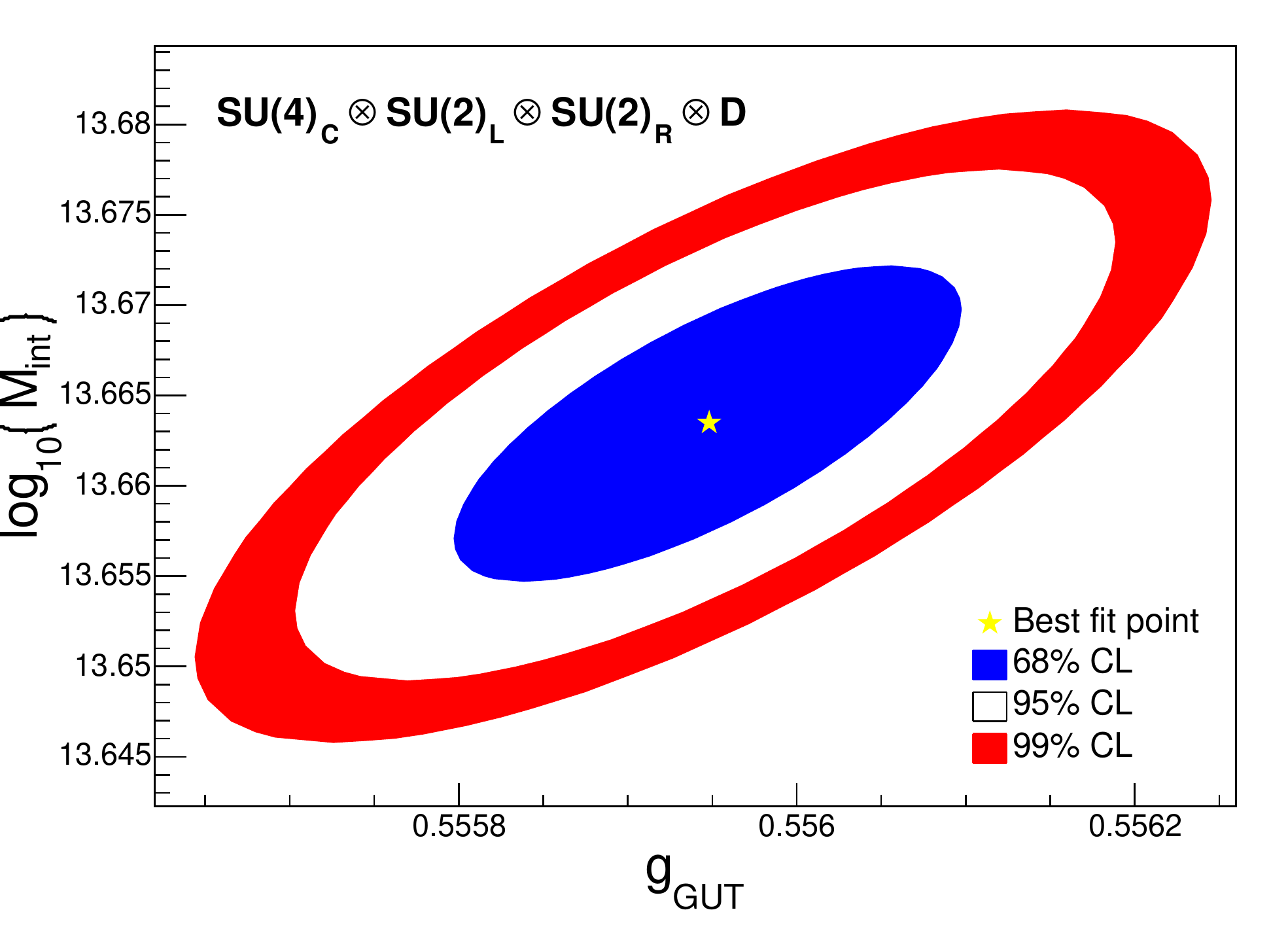} \\
\vspace{-3mm}
\hspace{-1.0cm} \includegraphics[scale=0.41]{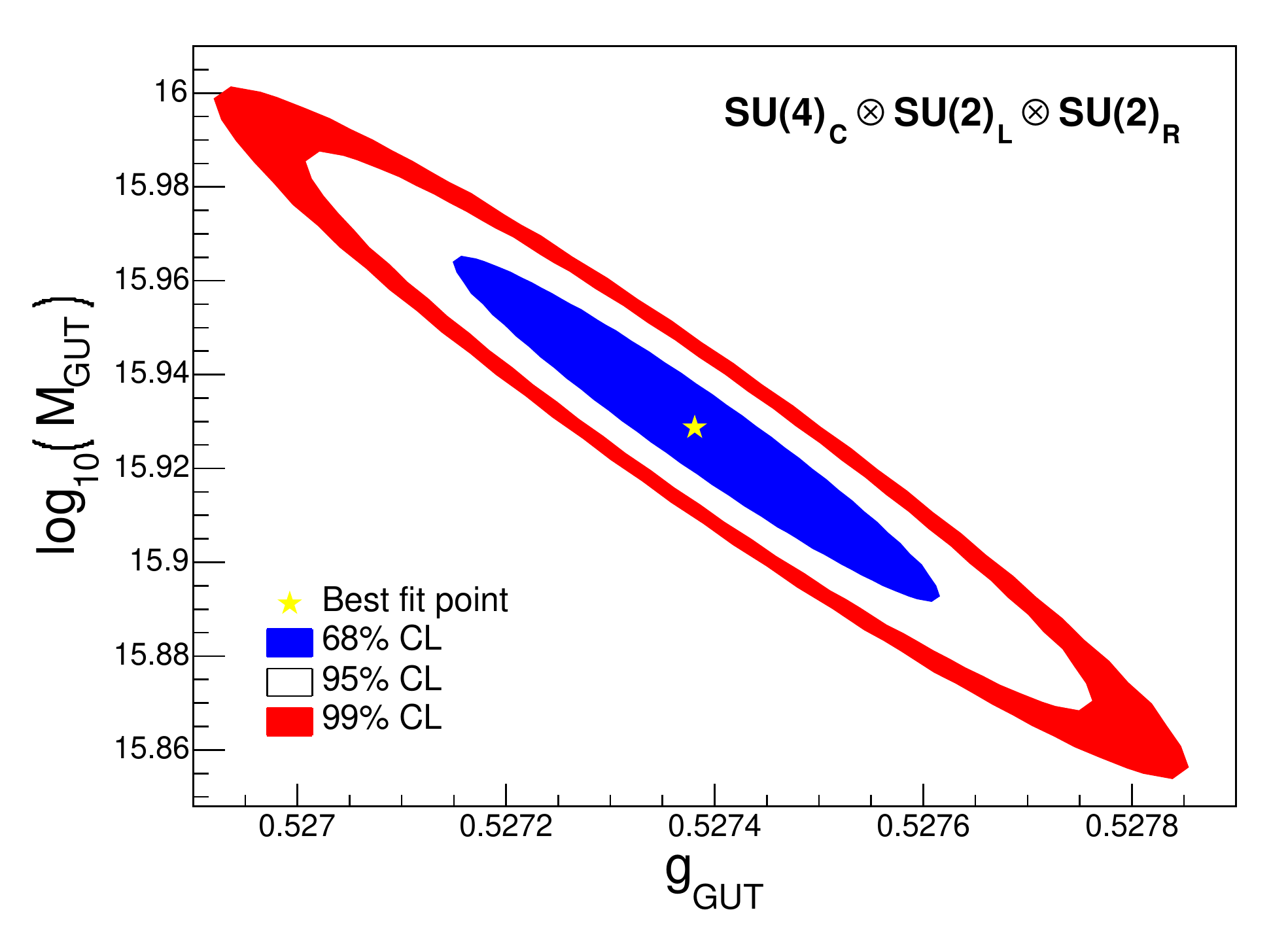}
 & 
\hspace{-0.8cm}
     \includegraphics[scale=0.41]{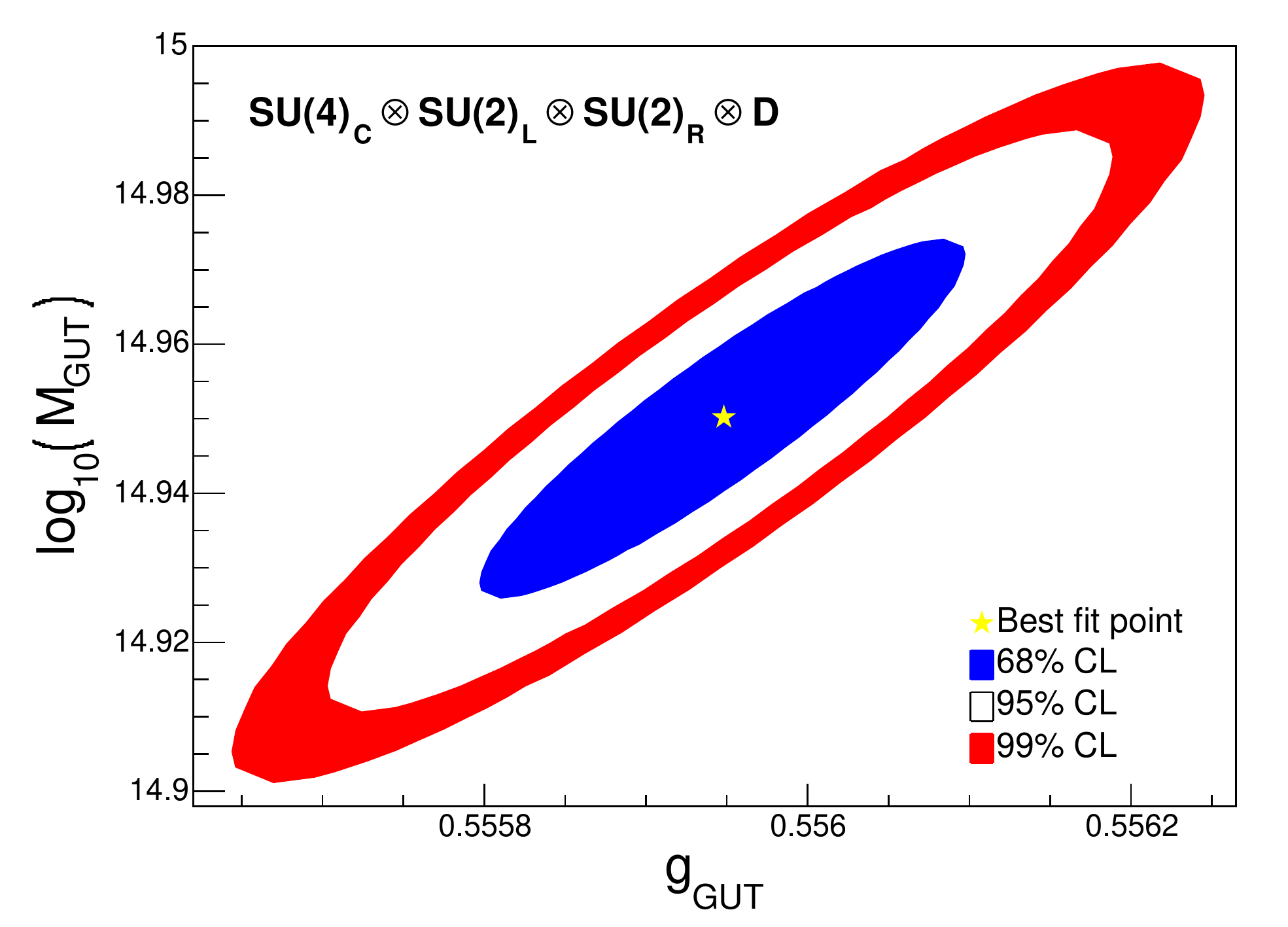} \\
\vspace*{-3mm}
\hspace{-1.0cm}
 \includegraphics[scale=0.41]{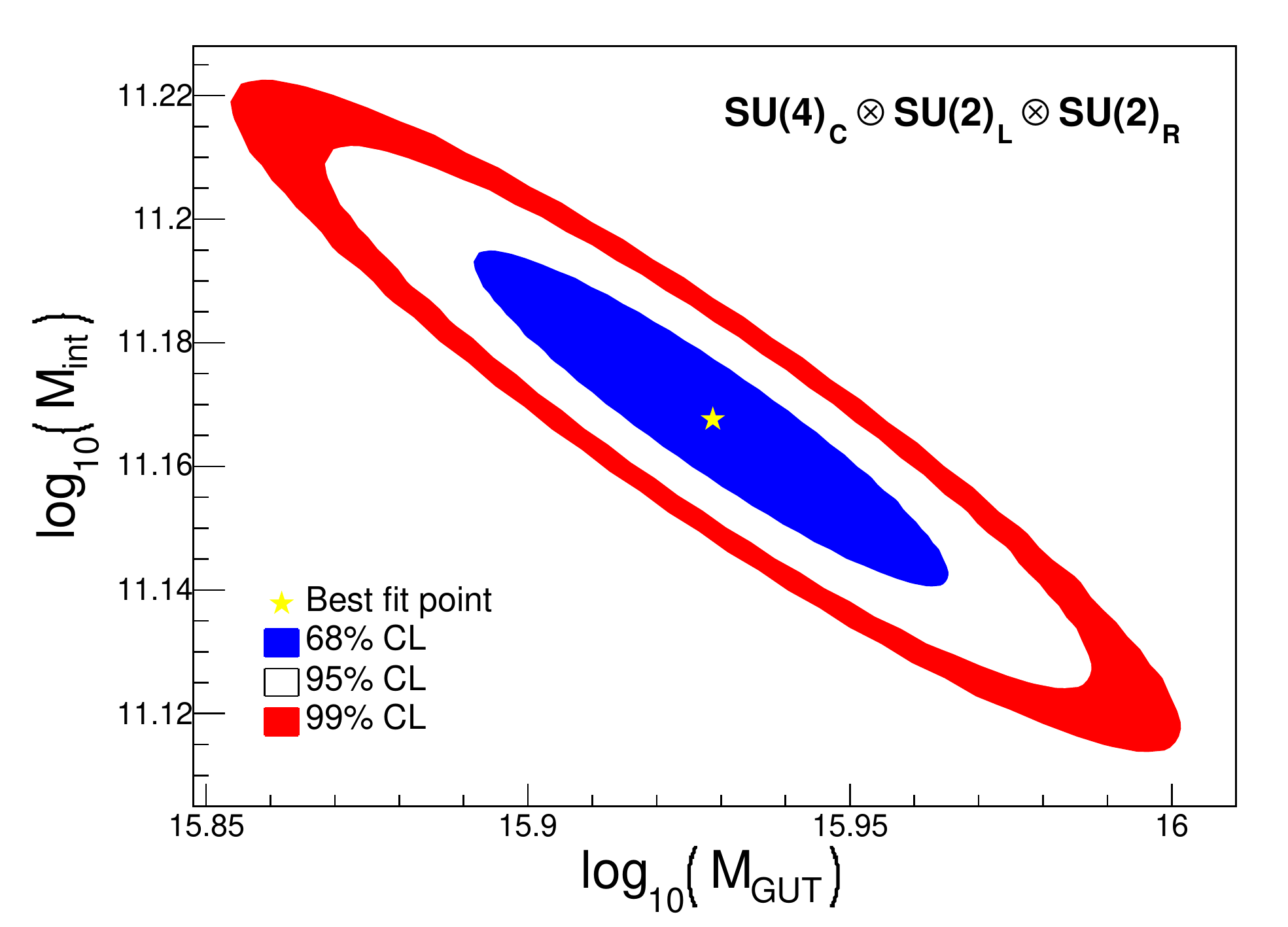} & 
\hspace{-0.8cm} \includegraphics[scale=0.41]{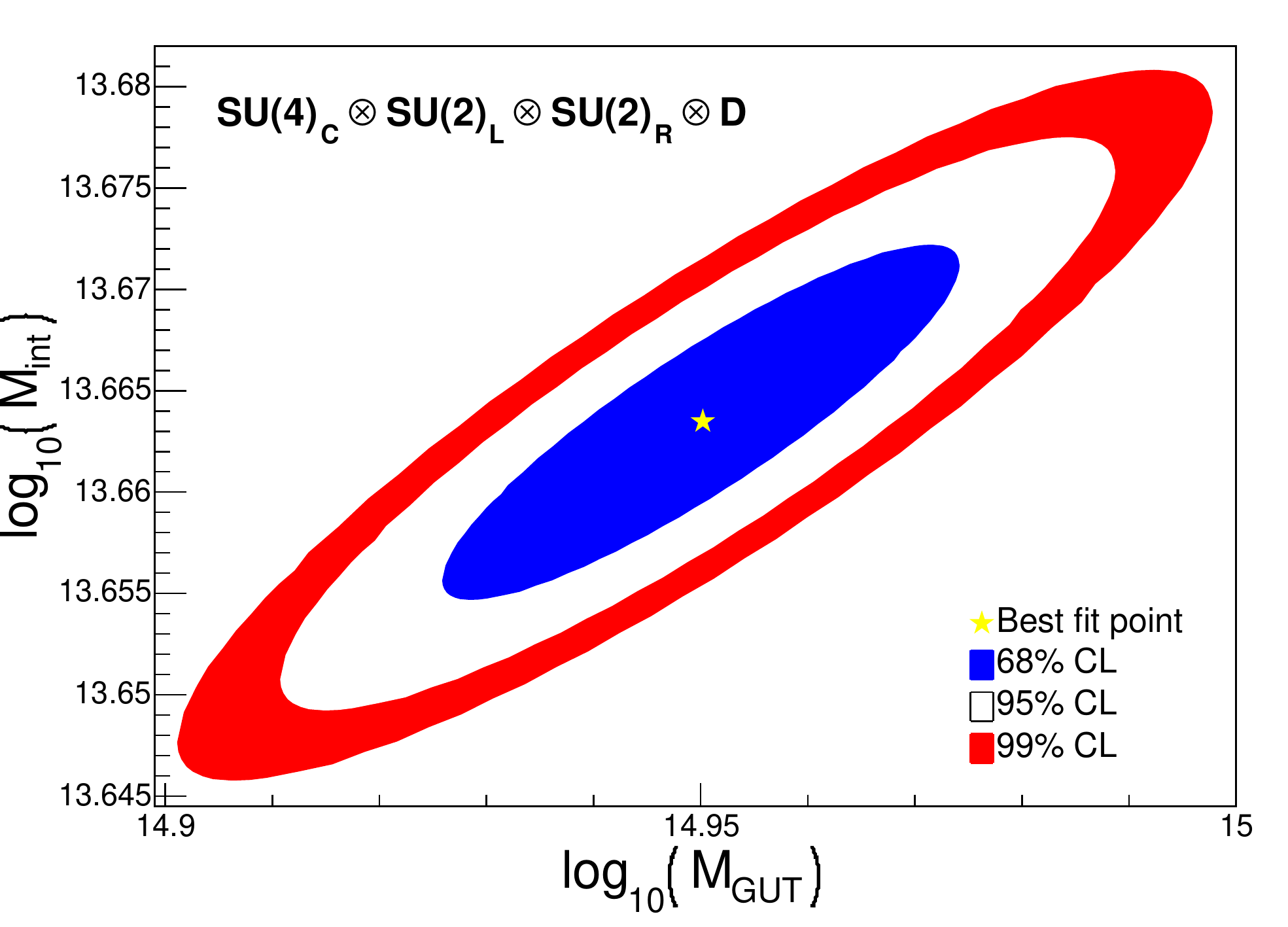}
\end{tabular}
\caption{\it Contour plots for the allowed region in the
$g_{\text{GUT}}$-$\log_{10}(M_{\text{int}})$,
 $g_{\text{GUT}}$-$\log_{10}(M_{\text{GUT}})$, and
 $\log_{10}(M_{\text{GUT}})$-$\log_{10}(M_{\text{int}})$ parameter planes
 in the top, middle, and bottom panels, respectively. Left panels
are for $G_{\text{int}}= \text{SU}(4)_C\otimes \text{SU}(2)_L \otimes
\text{SU}(2)_R$, while right ones are for $G_{\text{int}}=
\text{SU}(4)_C\otimes \text{SU}(2)_L \otimes \text{SU}(2)_R\otimes
D$. Stars represent the best-fit point. The colored regions correspond to
68, 95, and 99 \% C.L. limits determined from $\Delta \chi^2 \simeq  2.30, 5.99, 9.21$.
}
\label{2D_plots}
\end{center}
\end{figure}

By using the method discussed above, we carry out the analysis and 
summarize the results in Table~\ref{tab:resultssm}. Here, we show
$\log_{10} (M_{\text{int}})$, $\log_{10} (M_{\text{GUT}})$, and
$g_{\text{GUT}}$. For each intermediate gauge group, the
upper shaded (lower) row shows the 2-loop (1-loop)
result. $M_{\text{int}}$ and $M_{\text{GUT}}$ are given in GeV. The
blank entries indicate that gauge coupling unification is not achieved. The
uncertainties resulting from the input error are also shown in the
parentheses. To illustrate our procedure more clearly, we show $\chi^2$ as
functions of $\log_{10} (M_{\text{int}})$ (top), $\log (M_{\text{GUT}})$
(middle), and $g_{\text{GUT}}$ (bottom) in Fig.~\ref{linear_plots} for two examples of 
intermediate gauge groups. The
left panels are for $G_{\text{int}}= \text{SU}(4)_C\otimes
\text{SU}(2)_L \otimes \text{SU}(2)_R$, while the right ones are for
$G_{\text{int}}= \text{SU}(4)_C\otimes \text{SU}(2)_L \otimes
\text{SU}(2)_R\otimes D$. The $\chi^2$ functions for the other choices of intermediate scale
gauge groups will be qualitatively similar. Here again, $M_{\text{int}}$ and
$M_{\text{GUT}}$ are given in GeV. In each plot, the other two free
parameters are fixed to their best-fit values. We also plot the one-loop
results (shown as dotted curves) to show the significance of the two-loop effects. In
Fig.~\ref{2D_plots}, we show the $\chi^2$ functions projected down
onto 2D planes
corresponding to
$g_{\text{GUT}}$-$\log_{10}(M_{\text{int}})$,
$g_{\text{GUT}}$-$\log_{10}(M_{\text{GUT}})$, and
$\log_{10}(M_{\text{GUT}})$-$\log_{10}(M_{\text{int}})$ 
in the top, middle, and bottom panels, respectively. Again, the left
panels are for $G_{\text{int}}= \text{SU}(4)_C\otimes \text{SU}(2)_L
\otimes \text{SU}(2)_R$, while the ones on the right are for $G_{\text{int}}=
\text{SU}(4)_C\otimes \text{SU}(2)_L \otimes \text{SU}(2)_R\otimes
D$. The stars represent the best-fit point. 
The uncertainty ellipses represent 68, 95, and 99 \% C.L. uncertainties corresponding
to $\Delta \chi^2 = 2.30$, $5.99$, and 9.21, respectively. Threshold
corrections at $M_{\text{int}}$ and $M_{\text{GUT}}$ \cite{Hall:1980kf}
due to the non-degeneracy of the particles  that have masses of the
order of these scales contribute to the uncertainties.\footnote{ 
Note that the intermediate
scale in the left-right symmetric theories does not depend on physics
beyond $M_{\text{int}}$, as discussed in Appendix~\ref{sec:1loopform}.} 
For a recent discussion of threshold corrections, see Ref.~\cite{Ellis:2015jwa}.
In addition, we neglect the contribution of
Yukawa couplings above the intermediate scale, which causes additional
error. These are expected to give ${\cal O}(1)$\% uncertainty to the
results.

From Table~\ref{tab:resultssm}, it is found that gauge coupling
unification is not achieved in the case of $G_{\text{int}} =
\text{SU(3)}_C\otimes \text{SU}(2)_L \otimes \text{U}(1)_R \otimes
\text{U}(1)_{B-L}$. Moreover, we find that relatively low GUT scales are
predicted for $G_{\text{int}} = \text{SU}(4)_C\otimes \text{SU}(2)_L
\otimes \text{SU}(2)_R\otimes {D}$ and $\text{SU}(4)_C\otimes
\text{SU}(2)_L \otimes \text{U}(1)_R$, and thus the proton decay
constraints may be severe in these cases, as discussed in
Sec.~\ref{sec:protondecay}. Furthermore, except for $G_{\text{int}} =
\text{SU}(4)_C\otimes \text{SU}(2)_L \otimes \text{SU}(2)_R\otimes {D}$,
we obtain low intermediate scales, with which it may be difficult to
account for the neutrino masses, as explained in
Sec.~\ref{sec:neutrinomass}. As we will see below, this situation can be
improved in the NETDM models.

\subsection{NETDM and gauge coupling unification}
\label{sec:NETDMgcu}

Next, we look for the NETDM models in which gauge coupling unification
is realized with an appropriate intermediate unification scale. Here, we
require $10^{15}\lesssim M_{\text{GUT}} \lesssim 10^{18}$~GeV; if
$M_{\text{GUT}} < 10^{15}$~GeV, then proton decays are too rapid to be
consistent with proton decay experiments, while if $M_{\text{GUT}} >
10^{18}$~GeV, then gravitational effects cannot be neglected anymore
and a calculation based on quantum field theories may be invalid
around the GUT scale. To search for promising candidates, we assume the
following conditions. Firstly, a model should contain a NETDM candidate
shown in Table~\ref{tab:NETDM}, where only a singlet component has a
mass much below the intermediate scale. This component does not affect
the running of the gauge couplings. Secondly, the rest of the components
in $R_{\text{DM}}$ are assumed to be around $M_{\text{int}}$ due to the mass
splitting mechanism with an additional Higgs multiplet, discussed in
Sec.~\ref{sec:degenprob}. At this point, we only assume that there
exists an extra Higgs multiplet from either the {\bf 45}, {\bf 54} or {\bf 210}
whose mass is around the intermediate scale. Whether the VEV of the
extra Higgs actually gives rise to the mass splitting or not will be
discussed in the subsequent section. Thirdly, we require that only the
SM fields, the intermediate gauge bosons, $R_{\text{DM}}$, and $R_2$ are
present below the GUT scale.  
For example, if we consider the ({\bf 1}, {\bf 1}, {\bf 3}) DM of the {\bf
45} given in the first column in Table~\ref{tab:NETDM}, then we suppose
that all of the components of the {\bf 45} except $R_{\text{DM}} = ({\bf 1},
{\bf 1}, {\bf 3})$ should have masses around the GUT scale. 
This condition is corresponding to the requirement of the minimal
fine-tunings in the scalar potential to realize an adequate mass
spectrum.

\begin{table}[t!]
 \begin{center}
\caption{\it Models that realize the gauge coupling
  unification. $M_{\text{int}}$ and $M_{\text{GUT}}$ are given in
  GeV. All of the values listed here are evaluated at one-loop level.
}
\label{tab:promisingmodels}
\vspace{5pt}
\begin{tabular}{llccc}
\hline
\hline
\rowcolor{LightGray}
\multicolumn{5}{l}
{SU(4)$_C\otimes$SU(2)$_L\otimes$SU(2)$_R$} \\
\hline
\rowcolor{LightGray}
$R_{\text{DM}}$&$R_2$
&$\log_{10} (M_{\text{int}})$& $\log_{10} (M_{\text{GUT}})$& $g_{\text{GUT}}$\\ 
\hline
\shortstack{$({\bf 1}, {\bf 1}, {\bf 3})_W$\\{}}&
\shortstack{$({\bf 10}, {\bf 1}, {\bf 3})_C$ \\ 
$({\bf 1}, {\bf 1}, {\bf 3})_R$}&
\shortstack{10.8 \\ {}} &
\shortstack{15.9\\ {}} &
\shortstack{0.53\\ {}} \\
\hline
\shortstack{$({\bf 1}, {\bf 1}, {\bf 3})_D$\\{}}&
\shortstack{$({\bf 10}, {\bf 1}, {\bf 3})_C$ \\ 
$({\bf 1}, {\bf 1}, {\bf 3})_R$}&
\shortstack{$9.8$\\ {}} &
\shortstack{$15.7$\\ {}} &
\shortstack{$0.53$\\ {}} \\
\hline
\rowcolor{LightGray}
\multicolumn{5}{l}
{SU(4)$_C\otimes$SU(2)$_L\otimes$SU(2)$_R\otimes D$} \\
\hline
\rowcolor{LightGray}
$R_{\text{DM}}$&$R_2$
&$\log_{10} (M_{\text{int}})$& $\log_{10} (M_{\text{GUT}})$& $g_{\text{GUT}}$\\ 
\hline
\shortstack{$({\bf 15}, {\bf 1}, {\bf 1})_W$\\ {} \\ {} }&
\shortstack{$({\bf 10}, {\bf 1}, {\bf 3})_C$ \\ 
$(\overline{\bf 10}, {\bf 3}, {\bf 1})_C$ \\$({\bf 15}, {\bf 1}, {\bf 1})_R$ }&
\shortstack{13.7\\ {} \\ {} } &
\shortstack{16.2\\ {} \\ {} } &
\shortstack{0.56\\ {} \\ {} } \\
\hline
\shortstack{$({\bf 15}, {\bf 1}, {\bf 1})_W$\\ {} \\ {} \\ {}}&
\shortstack{$({\bf 10}, {\bf 1}, {\bf 3})_C$ \\ 
$(\overline{\bf 10}, {\bf 3}, {\bf 1})_C$ \\$({\bf 15}, {\bf 1}, {\bf 3})_R$ \\ $({\bf 15}, {\bf 3}, {\bf 1})_R$}&
\shortstack{14.2\\ {} \\ {} \\ {}} &
\shortstack{15.5\\ {} \\ {} \\ {}} &
\shortstack{0.56\\ {} \\ {} \\ {}} \\
\hline
\shortstack{$({\bf 15}, {\bf 1}, {\bf 1})_D$\\ {} \\ {} \\ {}}&
\shortstack{$({\bf 10}, {\bf 1}, {\bf 3})_C$ \\ 
$(\overline{\bf 10}, {\bf 3}, {\bf 1})_C$ \\$({\bf 15}, {\bf 1}, {\bf 3})_R$ \\
 $({\bf 15}, {\bf 3}, {\bf 1})_R$}& 
\shortstack{14.4\\ {} \\ {} \\ {}} &
\shortstack{16.3\\ {} \\ {} \\ {}} &
\shortstack{0.58\\ {} \\ {} \\ {}} \\
\hline
\rowcolor{LightGray}
\multicolumn{5}{l}
{SU(3)$_C\otimes$SU(2)$_L\otimes$SU(2)$_R\otimes$U(1)$_{B-L}$} \\
\hline
\rowcolor{LightGray}
$R_{\text{DM}}$&$R_2$
&$\log_{10} (M_{\text{int}})$& $\log_{10} (M_{\text{GUT}})$& $g_{\text{GUT}}$\\ 
\hline
\shortstack{$({\bf 1}, {\bf 1}, {\bf 3}, 0)_W$\\ {} }&
\shortstack{$({\bf 1}, {\bf 1}, {\bf 3}, -2)_C$ \\ 
$({\bf 1}, {\bf 1}, {\bf 3}, 0)_R$}&
\shortstack{6.1 \\ {}} &
\shortstack{16.6 \\ {}} &
\shortstack{0.52 \\ {}} \\
\hline
\hline
\end{tabular}
 \end{center}
\end{table}

With these conditions, we then search for possible candidates by using the
one-loop analytic formula given in Appendix~\ref{sec:1loopform}. In
Table~\ref{tab:promisingmodels}, we summarize the field contents that satisfy
the above requirements, as well as the values of $\log_{10}
(M_{\text{int}})$, $\log_{10} (M_{\text{GUT}})$, and $g_{\text{GUT}}$,
with $M_{\text{int}}$ and $M_{\text{GUT}}$ in GeV. All of the values are
evaluated at one-loop level. Here the subscript
$R$, $C$, $W$, or $D$ of each multiplet indicates that it is a real
scalar, a complex scalar, a Weyl fermion, or a Dirac fermion,
respectively. As for the intermediate Higgs fields, $R_2$, listed in 
Table~\ref{tab:promisingmodels}, $({\bf 10}, {\bf 1},
{\bf 3})_C$ and $({\bf 1},{\bf 1}, {\bf 3}, -2)_C$ are from the ${\bf
126}$ Higgs field, while all other representations included in $R_2$ are
extra Higgs fields introduced to resolve the degeneracy problem. For the
additional Higgs fields, we only show the real scalar cases for
brevity. Indeed, we can also consider complex scalars for the Higgs
fields and find that gauge coupling unification is also realized in
these cases, where both the intermediate and GUT scales are only
slightly modified.

\section{Models}
\label{sec:models}

In the previous section, we have reduced the possible candidates to those
presented in Table~\ref{tab:promisingmodels}. In this section, we study if any of
those models are viable; {\it i.e.}, we check if they actually offer an
appropriate mass spectrum to realize the NETDM scenario, with
the charged/colored components in $R_{\text{DM}}$ acquiring masses of
${\cal O}(M_{\text{int}})$.

First, let us consider the $({\bf 1}, {\bf 1}, {\bf 3})_{W/D}$ DM representation in the
$\text{SU}(4)_C\otimes \text{SU}(2)_L\otimes \text{SU}(2)_R$ gauge
theory. To split the masses in the $({\bf 1}, {\bf 1}, {\bf 3})$
multiplet $\psi^r$, we need to couple the DM with the $({\bf 1}, {\bf 1}, {\bf
3})_{R}$ Higgs $\phi^r$, with $r$ denoting the SU(2)$_R$ index. Since
the fields transform as triplets under the SU(2)$_R$ transformations,
to construct an invariant term from the fields, the indices should be
contracted anti-symmetrically; \textit{i.e.}, the coupling should have a form
like
\begin{equation}
 \epsilon_{pqr}(\overline{\psi})^p\psi^q\phi^r ~.
\label{eq:333}
\end{equation}
Then, if $\psi^r$ is a Majorana fermion, the above term always
vanishes. Thus, $\psi^r$ should be a Dirac fermion, that is, $({\bf
1}, {\bf 1}, {\bf 3})_D$ is the unique candidate for  NETDM in this
case. 

Next, we study the terms in the SO(10) Lagrangian relevant to the masses
of the fields much lighter than the GUT scale. 
In SO(10), $({\bf 1}, {\bf 1}, {\bf 3})_D$, $({\bf 1}, {\bf 1},
{\bf 3})_R$, and $({\bf 10}, {\bf 1}, {\bf 3})_C$ are included in the
${\bf 45}_D$, ${\bf 45}_R$, and ${\bf 126}_C$, respectively. The SO(10)
gauge group is spontaneously broken by the ${\bf 210}_R$ Higgs field
$(R_1)$ into the $\text{SU}(4)_C\otimes \text{SU}(2)_L\otimes
\text{SU}(2)_R$ intermediate gauge group. As is usually done in the
intermediate scale scenario, we fine-tune the Higgs potential so that
the $({\bf 1}, {\bf 1}, {\bf 3})_R$ and $({\bf 10}, {\bf 1}, {\bf 3})_C$
Higgs fields have masses around the intermediate scale. This can always be
performed by using the couplings of the ${\bf 45}_R$ and ${\bf 126}_C$
fields with the ${\bf 210}_R$ Higgs field, which acquires a VEV of the
order of the GUT scale. Similarly, we give desirable masses
to the fields in $({\bf 1}, {\bf 1}, {\bf 3})_D$ by carefully
choosing the couplings of the ${\bf 45}_D$ fermion with the ${\bf 45}_R$ and
${\bf 126}_C$ Higgs fields. Here, the relevant interactions are
\begin{equation}
 {\cal L}_{\text{int}} =
-M_{45_D}\overline{{\bf 45}_D}{\bf 45}_D 
-iy_{45}\overline{{\bf 45}_D}{\bf 45}_D {\bf 45}_R
-y_{210}\overline{{\bf 45}_D}{\bf 45}_D {\bf 210}_R ~.
\label{eq:model1so10lag}
\end{equation}
Notice that ${\bf 45}_D$ does not couple to the ${\bf 126}_C$ field, as
already mentioned in Sec.~\ref{sec:degenprob}. After the $R_1 = {\bf
210}_R$ Higgs field gets a VEV $\langle {\bf
210}_R \rangle = v_{210}$, the interactions in
Eq.~\eqref{eq:model1so10lag} lead to the following terms:\footnote{For the
computation of the Clebsch-Gordan coefficients, we have used the results
given in Ref.~\cite{Fukuyama:2004ps}. }
\begin{equation}
 {\cal L}_{\text{int}}\to
-M_{\text{DM}}(\overline{\psi})^r\psi^r 
-iy_{45} \epsilon_{rst}(\overline{\psi})^r\psi^s\phi^t ~,
\label{eq:model1422lag}
\end{equation}
with $ M_{\text{DM}}= M_{45_D}+y_{210}{v_{210}}/{\sqrt{6}}$. 
Here, $\psi^r$ and $\phi^r$ denote the $({\bf 1}, {\bf 1}, {\bf 3})_D$ and
$({\bf 1}, {\bf 1}, {\bf 3})_R$ components in ${\bf 45}_D$ and ${\bf
45}_R$, respectively. We find that although
$M_{45_D}$ and $v_{210}$ are expected to be ${\cal
O}(M_{\text{GUT}})$, we can let $M_{\text{DM}}$ be much lighter than the
GUT scale by carefully choosing the above parameters so that they cancel
each other. In addition, it 
turns out that the mass term of the $({\bf 1}, {\bf 3}, {\bf 1})_D$
component in ${\bf 45}_D$ is given by
$M_{45_D}-y_{210}{v_{210}}/{\sqrt{6}}$. Thus, even if we fine-tune $M_{45_D}$ 
and $y_{210}$ to realize $M_{\text{DM}}\ll M_{\text{GUT}}$, the mass of
$({\bf 1}, {\bf 3}, {\bf 1})_D$ is still around the GUT scale. This
observation reflects the violation of the $D$-parity in this
model.  
At this point, all of the components in $\psi^r$ have identical masses
(the ``degeneracy problem''). Once the neutral component of $\phi^r$
acquires a VEV $\langle \phi^3 \rangle =v_{45}$, which is assumed to
be ${\cal O}(M_{\text{int}})$, the 
second term in Eq.~\eqref{eq:model1422lag} gives rise to additional mass
terms for $\psi^r$. These are	
\begin{equation}
 {\cal L}_{\text{int}} \to
-M_{\text{DM}}\overline{\psi^0}\psi^0 -M_+ \overline{\psi^+}\psi^+ 
-M_-\overline{\psi^-}\psi^-~,
\end{equation}
where $M_\pm = M_{\text{DM}}\mp y_{45}v_{45}$, and 
$\psi^0$ and $\psi^\pm$ are the neutral and charged components,
respectively.\footnote{Note that since $\psi^r$ are Dirac fermions,
$(\psi^0)^{\cal C}\neq \psi^0$ and $(\psi^\pm)^{\cal C}\neq \psi^\mp$ } 
The above expression shows that the VEV of the ${\bf 45}_R$
Higgs field indeed solves the degeneracy problem; if $M_{\text{DM}} \ll
M_{\text{int}}$ and $y_{45}v_{45} ={\cal O}(M_{\text{int}})$, then the
charged components acquire masses of ${\cal O}(M_{\text{int}})$, while
the neutral component has a mass much lighter than
$M_{\text{int}}$. Thus, we obtain the mass spectrum we have assumed in
the previous section.

In the next example, we consider the DM representation $R_{\text{DM}} = ({\bf 15}, {\bf 1}, {\bf
1})_W$ with $R_2 =({\bf 10}, {\bf 1}, {\bf 3})_C\oplus (\overline{\bf
10}, {\bf 3}, {\bf 1})_C 
\oplus ({\bf 15}, {\bf 1}, {\bf 1})_R$ in the left-right symmetric 
$\text{SU}(4)_C\otimes \text{SU}(2)_L\otimes \text{SU}(2)_R$ gauge theory.
In this case, $R_1 = {\bf 54}_R$. We assume that the $({\bf 15}, {\bf 1},
{\bf 1})_W$ is a part of the ${\bf 45}_W$, while both $({\bf 10},
{\bf 1}, {\bf 3})_C$ and $(\overline{\bf 10}, {\bf 3}, {\bf 1})_C$ are part of the 
${\bf 126}_C$. The couplings of the DM with the Higgs fields, as well
as its mass term, are then given by
\begin{equation}
 {\cal L}_{\text{int}} =
-\frac{M_{45_W}}{2}{{\bf 45}_W}{\bf 45}_W 
-\frac{y_{54}}{2}{{\bf 45}_W}{\bf 45}_W {\bf 54}_R
-\frac{y_{210}}{2}{{\bf 45}_W}{\bf 45}_W {\bf 210}_R 
+\text{h.c.}~,
\label{eq:model2so10lag}
\end{equation}
Here, $({\bf 15}, {\bf 1}, {\bf 1})_R$ is included in the ${\bf 210}_R$
field; we cannot use a ${\bf 45}_R$ in this case since the Weyl fermion ${\bf
45}_W$ has no coupling to the ${\bf 45}_R$.\footnote{It is also possible
to embed $({\bf 15}, {\bf 1}, {\bf 1})_W$ into ${\bf 210}_W$ and $({\bf
15}, {\bf 1}, {\bf 1})_R$ into ${\bf 45}_R$. The phenomenology in this
case is the same as that discussed in the text. } As before, below the GUT
scale, the VEV of ${\bf 54}_R$, $v_{54}$, gives a common mass
$M$ to the $({\bf 15}, {\bf 1}, {\bf 1})_W$ multiplet with
$M =
M_{45_W}-y_{54}v_{54}/\sqrt{15}$. 
We can take $M={\cal O}(M_{\text{int}})$ by fine-tuning $M_{45_W}$ and
$y_{54}v_{54}$. The above Lagrangian then reduces to 
\begin{equation}
 {\cal L}_{\text{int}} \to 
- \frac{M}{2} {\psi^A}\psi^A 
+\frac{2y_{210}}{\sqrt{3}}
\text{Tr}({\psi}\phi\psi) +\text{h.c.} ~,
\label{eq:model2422lag}
\end{equation}
where $\psi^A$ and $\phi^A$ denote the $({\bf 15}, {\bf 1}, {\bf 1})_W$
and $({\bf 15}, {\bf 1}, {\bf 1})_R$ fields, respectively, with
$\psi\equiv \psi^AT^A$ and $\phi\equiv \phi^AT^A$; $A, B, C =
1,\dots 15$ are the SU(4) adjoint indices and $T^A$ are
the SU(4) generators.  
The mass degeneracy in this case is resolved by the VEV of the ${\bf
210}_R$ field,
\begin{equation}
 \langle \phi \rangle =\frac{v_{210}}{2\sqrt{6}}~\text{diag}(1,1,1,-3)
  ,\label{eq:vev1511r}
\end{equation}
with which Eq.~\eqref{eq:model2422lag} leads to
\begin{equation}
 {\cal L}_{\text{int}} \to 
-\frac{M_{\text{DM}}}{2} \psi^0 \psi^0
-\frac{M_{\tilde{g}}}{2}\widetilde{g}^A\widetilde{g}^A
-M_\xi\overline{\xi}_a \xi^a +\text{h.c.} ~,
\end{equation}
where $\psi^0$, $\widetilde{g}^A$, $\xi^a$, and $\overline{\xi}_a$ are
the color singlet, octet, triplet, and anti-triplet components in $({\bf
15}, {\bf 1}, {\bf 1})_W$, respectively, with $a$ denoting the color
index, and
\begin{align}
 M_{\text{DM}} &= M + \frac{\sqrt{2}}{3}y_{210}v_{210} ~, \\
 M_{\tilde{g}} &= M -\frac{1}{3\sqrt{2}}y_{210}v_{210} ~, \\
 M_\xi &= M +\frac{1}{3\sqrt{2}}y_{210}v_{210} ~.
\end{align}
Therefore, by carefully adjusting $y_{210}v_{210}$, we can make the DM
$\psi^0$ much lighter than $M_{\text{int}}$ while keeping the
other components around the intermediate scale.

There are two more possible representations for $R_{\text DM}$ for the left-right symmetric 
$\text{SU}(4)_C\otimes \text{SU}(2)_L\otimes \text{SU}(2)_R$ intermediate
gauge group given in Table~\ref{tab:promisingmodels},
namely $({\bf 15}, {\bf 1}, {\bf 1})_{W/D}$. In this case,
however, one can readily conclude that the degeneracy problem cannot be
solved by the $({\bf 15}, {\bf 1}, {\bf 3})_{R}$ and $({\bf 15}, {\bf
3}, {\bf 1})_{R}$ Higgs fields. This is because the Yukawa couplings
between the DM and these Higgs fields are forbidden by the intermediate
gauge symmetry. As a consequence, we can safely neglect these
possibilities. 

Finally, we discuss the model presented in the last column in
Table~\ref{tab:promisingmodels}. We again find that the $({\bf 1}, {\bf
1}, {\bf 3}, 0)_R$ Higgs field does not yield a mass difference among
the components in the $({\bf 1}, {\bf 1}, {\bf 3}, 0)_W$ DM multiplet,
since the operator in Eq.~\eqref{eq:333} vanishes when the DM is a Weyl
fermion. Thus, we do not consider this model in the following
discussion.

\begin{figure}[ht!]
\begin{center}
\subfigure[Model I]
 {\includegraphics[clip, width = 0.48 \textwidth]{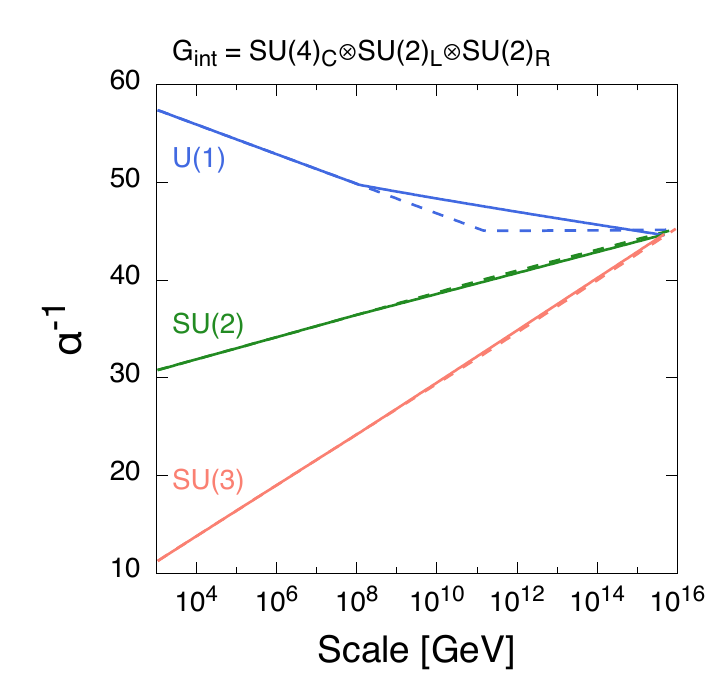}
 \label{fig:mod1}}
\subfigure[Model II]
 {\includegraphics[clip, width = 0.48 \textwidth]{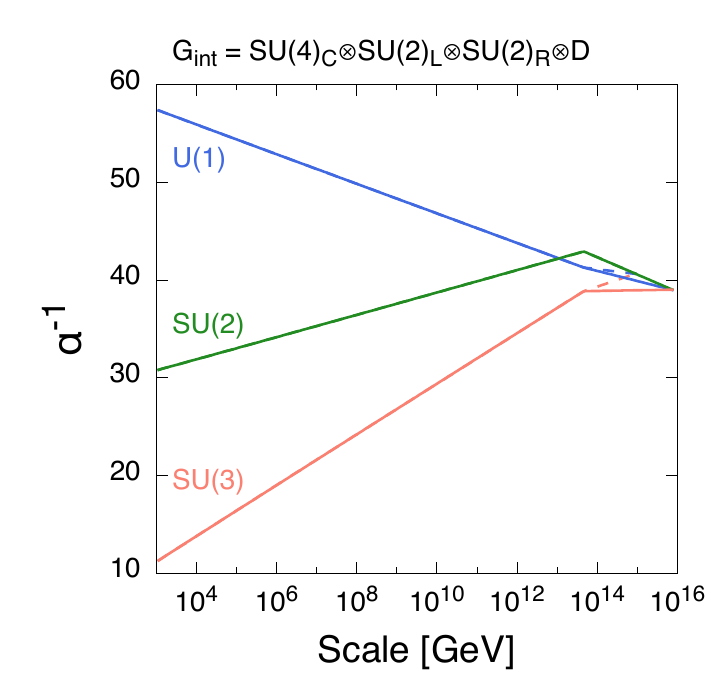}
 \label{fig:mod2}}
\caption{\it Running of gauge couplings. Solid (dashed) lines show the case
with (without) DM and additional Higgs bosons. Blue, green, and red lines
represent the running of the U(1), SU(2) and SU(3) gauge couplings,
respectively.}
\label{fig:gaugerun}
\end{center}
\end{figure}

As a result, we obtain two distinct models for NETDM within SO(10). We summarize these two
models in Table~\ref{tab:model1and2}. We call them Model I and II
in what follows. Here, $M_{\text{int}}$ and $M_{\text{GUT}}$ are given
in GeV, and all of the values are evaluated with two-loop RGEs
and differ somewhat from the one-loop values given in Table~\ref{tab:promisingmodels}. The
errors shown in the parentheses arise from uncertainties in the input
parameters. In addition, we again expect threshold corrections at $M_{\text{int}}$
and $M_{\text{GUT}}$. Furthermore, we neglect the contribution of Yukawa couplings to
the RGEs above the intermediate scale, and this also will contribute to the
theoretical error. We estimate that these two sources cause ${\cal O}(1)$\%
uncertainties in the values displayed in Table~\ref{tab:model1and2}. 
From these results, we find that the presence of the DM component as
well as the extra Higgs bosons can significantly alter the intermediate
and GUT scales,\footnote{However, their existence hardly changes the
intermediate scale in Model II, which is clarified in
Appendix~\ref{sec:1loopform}.  } because of their effects on the gauge
coupling running. To illustrate this more clearly, in
Fig.~\ref{fig:gaugerun} we show the running of gauge couplings in each
theory. The left and right panels of Fig.~\ref{fig:gaugerun} correspond to Model I
and II, respectively. In each figure, solid (dashed) lines show the case
with (without) DM and additional Higgs bosons. The blue, green, and red lines
represent the running of the U(1), SU(2) and SU(3) gauge couplings,
respectively, where the U(1) fine-structure constant $\alpha_1$ is defined by
\begin{equation}
 \frac{1}{\alpha_1} \equiv \frac{3}{5}\frac{1}{\alpha_{2R}}
+\frac{2}{5}\frac{1}{\alpha_{4}} ~,
\end{equation}
while the SU(3)$_C$ coupling $\alpha_3$ is defined by $\alpha_3 \equiv
\alpha_4$ above the intermediate scale. 
These figures clearly show the effects of the extra particles on the
gauge coupling running. In particular, the GUT scale in Model II is now well
above $10^{15}$~GeV, which allows this model to evade the proton decay
constraints, as will be seen in the subsequent section. 

\begin{table}[ht]
 \begin{center}
\caption{\it NETDM models. $M_{\text{int}}$ and $M_{\text{GUT}}$ are given in
  GeV. All of the values are evaluated with the two-loop RGEs.}
\label{tab:model1and2}
\vspace{5pt}
\begin{tabular}{l|cc}
\hline
\hline
 & Model I& Model II \\
\hline
$G_{\text{int}}$ &$\text{SU}(4)_C\otimes \text{SU}(2)_L\otimes
     \text{SU}(2)_R$ &$\text{SU}(4)_C\otimes \text{SU}(2)_L\otimes
	 \text{SU}(2)_R\otimes D$\\ 
$R_{\text{DM}}$ & $({\bf 1}, {\bf 1}, {\bf 3})_D$ in ${\bf 45}_D$
& $({\bf 15}, {\bf 1}, {\bf
	 1})_W$ in ${\bf 45}_W$ \\
$R_1$ & ${\bf 210}_R$ & ${\bf 54}_R$ \\
$R_2$ & $({\bf 10}, {\bf 1}, {\bf 3})_C\oplus ({\bf 1}, {\bf 1}, {\bf
     3})_R$ & $({\bf 10}, {\bf 1}, {\bf 3})_C\oplus ({\bf 10}, {\bf 3},
	 {\bf 1})_C\oplus ({\bf 15}, {\bf 1}, {\bf
     1})_R$ \\
$\log_{10}(M_{\text{int}})$& $8.08(1)$ & $13.664(5)$ \\
$\log_{10}(M_{\text{GUT}})$& $15.645(7)$ & $15.87(2)$ \\ 
$g_{\text{GUT}}$ & $0.53055(3)$ & $0.5675(2)$ \\
\hline
\hline
\end{tabular}
 \end{center}
\end{table}

\section{Phenomenological aspects}
\label{sec:phenomenology}

Now that we have obtained the NETDM models, we can study their
phenomenological aspects and possible implications in future
experiments. In Sec.~\ref{sec:neutrinomass}, we first consider whether
these models can give appropriate masses for light
neutrinos. Next,  in
Sec.~\ref{sec:protondecay}, we evaluate proton lifetimes in each model and discuss the testability in future proton
decay experiments. Finally, we compute the abundance of DM produced by
the NETDM mechanism in Sec.~\ref{sec:NETDMcalc}, and predict the
reheating temperature after inflation.

\subsection{Neutrino mass}
\label{sec:neutrinomass}

In SO(10) GUTs, the Majorana mass terms of the right-handed
neutrinos are induced after the $B-L$ symmetry is broken. These mass
terms are generated from the Yukawa couplings of the ${\bf 16}$ spinors
with the ${\bf 126}_C$ Higgs field. If the Yukawa couplings are ${\cal
O}(1)$, then the Majorana mass terms are ${\cal O}(M_{\text{int}})$. On
the other hand, in these models, the Dirac masses of neutrinos are equal
to the up-type quark masses, $m_u$, at the unification scale. Therefore, via
the seesaw mechanism \cite{Minkowski:1977sc}, light neutrino
masses are given by
\begin{equation}
 m_\nu \simeq \frac{m_u^2}{M_{\text{int}}} ~.  
\label{eq:mnumu}
\end{equation}
In Model II, $M_{\text{int}}={\cal O}(10^{13})$~GeV indeed
gives proper values for neutrino masses.\footnote{Note that in a
left-right symmetric model such as Model II there is in general also a
type-II seesaw contribution to $m_\nu$ from the VEV of the SU(2)$_L$
triplet in the ${\bf 126}_C$. However, we know from
constraints on the $\rho$-parameter that the VEV must be quite small and
definitely much smaller than the VEV of the SU(2)$_R$ triplet. For
example, if the mixing between the SU(2)$_L$ and SU(2)$_R$ triplets with
the Higgs doublets is small, it is safe to assume that the SU(2)$_L$
triplet VEV is small and thus the type-II seesaw contribution is
subdominant \cite{Mohapatra:1980yp}. } However, in Model I, a low 
intermediate scale of ${\cal O}(10^8)$~GeV yields neutrino
masses which are too heavy using the standard seesaw expression \eqref{eq:mnumu}. 
Thus, Model I is disfavored on the basis of small
neutrino masses. 

The defect in Model I may be evaded if the $({\bf 15}, {\bf 2}, {\bf 2})$
component in  ${\bf 126}_C$ has a sizable mixing with the $({\bf 1},
{\bf 2}, \overline{\bf 2})$ Higgs boson and  acquires a VEV of the order
of the electroweak scale. In this case, the neutrino Yukawa couplings
can differ from those of the up quark, and thus the relation
\eqref{eq:mnumu} does not hold any more. For sizable mixing to occur, the 
$({\bf 15}, {\bf 2}, {\bf 2})$ field should lie around the intermediate
scale. One might think that the presence of additional fields below the GUT scale would modify the
running of the gauge couplings and spoil the above discussion based on
gauge coupling unification. However, it turns out that both the
intermediate and GUT scales are hardly affected by the existence of this
field, though the unified gauge coupling constant becomes slightly
larger. This is because its contribution to the one-loop beta function
coefficients is $\Delta b_4 =16/3$ and $\Delta b_{2L} = \Delta b_{2R}
=5$, and thus their difference is very tiny (see the discussion given in
Appendix~\ref{sec:1loopform}). Therefore, we can take the
$({\bf 15}, {\bf 2}, {\bf 2})$ to be at the intermediate scale with
little change in the values of $M_{\text{int}}$ and $M_{\text{GUT}}$.
The presence of the $({\bf 15}, {\bf 2}, {\bf 2})$ is also desirable
to account for the down-type quark and charged lepton Yukawa couplings 
\cite{Bajc:2005zf, Lazarides:1980nt, Babu:1992ia, Matsuda:2000zp}. In addition, the
higher-dimensional operators induced above the GUT scale may also affect
the Yukawa couplings. Constructing a realistic
Yukawa sector in these models is saved for future work.

\subsection{Proton decay}
\label{sec:protondecay}

Proton decay is a smoking-gun signature of GUTs, and thus a powerful
tool for testing them. In non-SUSY GUTs, $p\to e^+ \pi^0$ is
the dominant decay mode, which is caused by the exchange of
GUT-scale gauge bosons. This could be compared with the case of the SUSY
GUTs; in SUSY GUTs, the color-triplet Higgs exchange usually yields the
dominant contribution to proton decay, which gives rise to the $p\to K^+
\bar{\nu}$ decay mode \cite{Sakai:1981pk}.\footnote{For recent analyses
on proton decay in SUSY GUTs, see Ref.~\cite{Liu:2013ula}.}

\begin{figure}[t]
\begin{center}
\includegraphics[height=75mm]{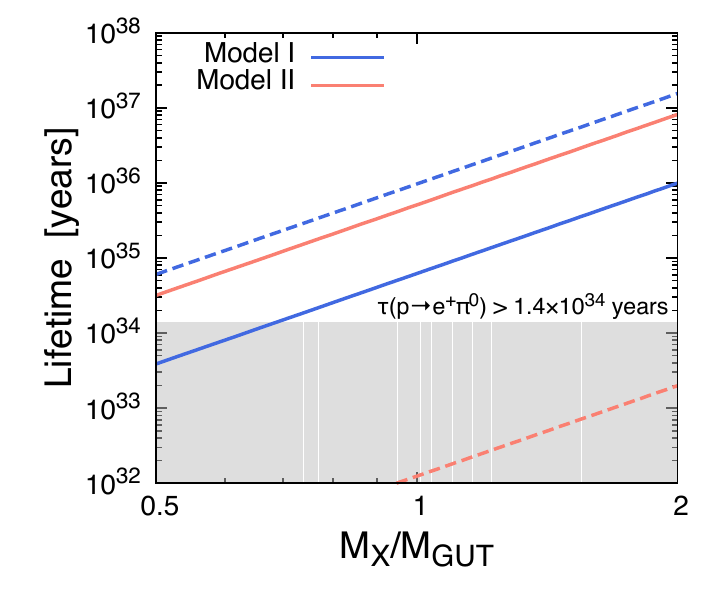}
\caption{{\it Proton lifetimes as functions of $M_X/M_{\text{GUT}}$. Blue
 solid and red solid  lines represent
 Model I and Model II, respectively.  Blue dashed and red dashed lines represent the cases 
 for  $G_{\text{int}} = \text{SU}(4)_C\otimes
 \text{SU}(2)_L \otimes \text{SU}(2)_R$ and $G_{\text{int}} =
 \text{SU}(4)_C\otimes \text{SU}(2)_L \otimes \text{SU}(2)_R \otimes D$
when the DM and extra Higgs multiplets are not included. 
The shaded area shows the region which is excluded by the current experimental
 bound, $\tau (p\to e^+\pi^0)> 1.4 \times 10^{34}~{\rm years}$
 \cite{Shiozawa, Babu:2013jba}. }}
\label{fig:protdec}
\end{center}
\end{figure}

Since the $p\to e^+ \pi^0$ decay mode is induced by gauge interactions, we
can make a robust prediction for the partial decay lifetime of this
mode. Details of the calculation are given in
Appendix~\ref{sec:protondecay422}. By using the results given there, we
evaluate the partial decay lifetime of the $p\to e^+ \pi^0$ mode in each
theory, and plot it as a function of $M_X/M_{\text{GUT}}$ ($M_X$ denotes the
mass of the GUT-scale gauge boson) in Fig.~\ref{fig:protdec}. 
Here, the blue and red solid lines
represent  Models I and II, while the blue and red dashed lines
represent the models without the DM and extra Higgs multiplets 
as given in Table \ref{tab:resultssm}, namely $G_{\text{int}} = \text{SU}(4)_C\otimes
\text{SU}(2)_L \otimes \text{SU}(2)_R$ and $G_{\text{int}} =
\text{SU}(4)_C\otimes \text{SU}(2)_L \otimes \text{SU}(2)_R \otimes D$, 
respectively. The shaded area shows the region which is excluded by the
current experimental bound, $\tau (p\to e^+\pi^0)> 1.4 \times
10^{34}~{\rm years}$ \cite{Shiozawa, Babu:2013jba}. We have varied the
heavy gauge boson mass between 
$M_{\text{GUT}}/2 \leq M_X \leq 2M_{\text{GUT}}$, which reflects our
ignorance of the GUT-scale mass spectrum. From this figure, we see that
the existence of DM and Higgs multiplets produces a large effect on the
proton decay lifetime. In particular, in the case of
$\text{SU}(4)_C\otimes \text{SU}(2)_L \otimes \text{SU}(2)_R \otimes D$,
the predicted lifetime is so small that the present bound has already
excluded the possibility. This conclusion can be evaded, however, once
the DM and $R_2$ Higgs multiplets are included in the theory as they
raise the value of $M_{\text{GUT}}$. Moreover, Model~I
is now being constrained by the proton decay experiments. In this case,
the inclusion of the DM and Higgs multiplets decreases  $M_{\text{GUT}}$.
Future proton
decay experiments, such as the Hyper-Kamiokande experiment
\cite{Abe:2011ts}, may offer much improved sensitivities (by about an
order of magnitude), with which we can probe a wide range of parameter
space in both models.

\subsection{Non-equilibrium thermal dark matter}
\label{sec:NETDMcalc}

Finally, we evaluate the relic abundance of DM produced by the NETDM
mechanism \cite{Mambrini:2013iaa} in Models I and II. In both of these models,
the DM $\psi^0$ is produced in the early Universe via the exchange of
the intermediate-scale particles. Therefore, the
production rate is extremely small and their self-annihilation can be
neglected. In addition, the produced DM cannot be in the thermal bath
since they have no renormalizable interactions with the SM
particles. These two features characterize the NETDM mechanism;
the DM is produced by SM particles in the thermal bath via the
intermediate boson exchange, while they do not annihilate with
each other nor attain thermal equilibrium. In what follows, we
estimate the density of the DM produced via this mechanism and determine
the reheating temperature which realizes the observed DM density.

The Boltzmann equation for the DM $\psi^0$ is given by
\begin{equation}
 \frac{dY_{\text{DM}}}{dx}
=\sqrt{\frac{\pi}{45}}\frac{g_{*s}}{\sqrt{g_{*\rho}}}
 M_{\text{DM}} M_{\text{Pl}} \frac{\langle \sigma v
\rangle}{x^2}Y^2_{\text{eq}} ~,
\label{eq:boltzeq}
\end{equation}
with $Y_{\text{DM}}\equiv n_{\text{DM}}/s$ and $Y_{\text{eq}}\equiv
n_{\text{eq}}/s$, where $n_{\text{DM}}$ is the DM number density,
$n_{\text{eq}}$ is the equilibrium number density of each individual
initial state SM particle, and $s$ is the entropy of the Universe; $x\equiv
M_{\text{DM}}/T$, with $T$ being the temperature of the Universe; $g_{*s}$
and $g_{*\rho}$ are the effective degrees of freedom for the entropy and
energy density in the thermal bath, respectively; $M_{\text{Pl}}\equiv
1/\sqrt{G_N} = 1.22\times 10^{19}$~GeV is the Planck mass; $\langle
\sigma v \rangle$ is the thermally averaged total annihilation cross
section of the initial SM particles, $f$, into the DM pair. When we derive
Eq.~\eqref{eq:boltzeq}, we neglect the DM self-annihilation contribution
as discussed above. From now on, we assume $g_{*s} = g_{*\rho}\equiv
g_*$ for brevity. 

We evaluate the thermal averaged cross section $\langle\sigma v\rangle$
multiplied by the equilibrium number density squared $n_{\text{eq}}^2$ as
\begin{equation}
 \langle \sigma v\rangle n_{\text{eq}}^2 \simeq \frac{T}{512\pi^5}
\int_{4M_{\text{DM}}^2}^\infty d\hat{s} \sqrt{\hat{s}-4M_{\text{DM}}^2} 
K_1 (\sqrt{\hat{s}}/T) \sum |{\cal M}|^2 ~, 
\label{eq:sigvn2}
\end{equation}
where $\sqrt{\hat{s}}$ denotes the center-of-mass energy, and $K_n(x)$
is the modified Bessel function of the second kind. Here, we have used
the approximation $m_f \ll \sqrt{\hat{s}}$, with $m_f$ being the
masses of the SM particles, since the particle production predominantly occurs at
high temperature, and we have neglected the angular dependence of
${\cal M}$ for simplicity. In addition, we have assumed the initial
particles follow a Maxwell-Boltzmann distribution and ignored
statistical mechanical factors which may result from the Fermi-Dirac or
Bose-Einstein distribution. $\sum |{\cal M}|^2 $ indicates the sum of the
squared amplitude over all possible incoming
SM particles, as well as the spin of the final state.

\begin{figure}[t]
\begin{center}
\includegraphics[height=45mm]{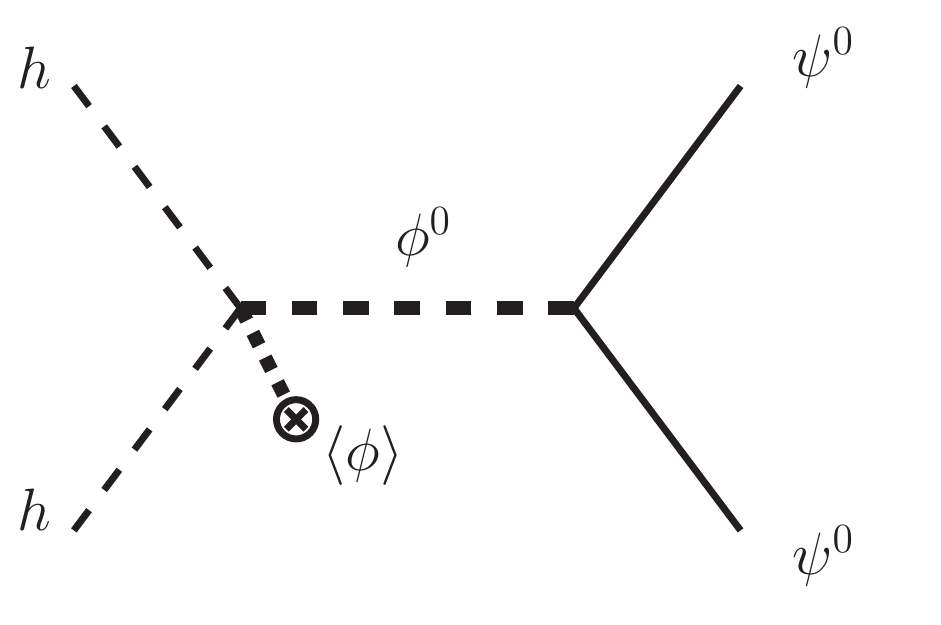}
\caption{{\it Diagram responsible for the DM production in Model II.}}
\label{fig:annmod2}
\end{center}
\end{figure}

Next, we evaluate the amplitude ${\cal M}$ in each model. First, we
consider the case of Model II. In this case, the dominant contribution
comes from the tree-level Higgs-boson annihilation process displayed in
Fig.~\ref{fig:annmod2}. Here, $\psi^0$, $h$, and $\phi^0$ denote the DM,
the SM Higgs boson, and the singlet component of the $({\bf 15}, {\bf
1}, {\bf 1})_R$, respectively, and the VEV $\langle \phi \rangle$ is
given in Eq.~\eqref{eq:vev1511r}. From the dimensional analysis, we
estimate the contribution as 
\begin{equation}
 \sum |{\cal M}|^2 \simeq c~
  \frac{\hat{s}-4M_{\text{DM}}^2}{M_{\text{int}}^2} ~,
\label{eq:summsq}
\end{equation}
where $c$ is a numerical factor which includes the unknown
couplings appearing in the diagram.
By substituting Eqs.~\eqref{eq:sigvn2} and \eqref{eq:summsq} into
Eq.~\eqref{eq:boltzeq}, we have
\begin{equation}
 \frac{d Y_{\text{DM}}}{dx} \simeq
\frac{c }{1024 \pi^7}
\biggl(\frac{45}{\pi g_*}\biggr)^{\frac{3}{2}}
\frac{M_{\text{Pl}}M_{\text{DM}}}{M_{\text{int}}^2} 
\frac{1}{x^2}
\int^\infty_{2x}  t(t^2-4x^2)^{\frac{3}{2}} K_1(t) dt
~.
\label{eq:dydx}
\end{equation}
When $M_{\text{DM}}\ll T_{\text{RH}}$ with  $T_{\text{RH}}$ being the
reheating temperature, the above equation is easily integrated to give 
\begin{equation}
 Y^{(0)}_{\text{DM}} \simeq \frac{c }{64 \pi^7}
\biggl(\frac{45}{\pi g_*}\biggr)^{\frac{3}{2}}
\frac{M_{\text{Pl}}T_{\text{RH}}}{M_{\text{int}}^2} 
~,
\end{equation}
where the superscript ``(0)'' implies the present-day value. On the
other hand, the current value of $Y^{(0)}_{\text{DM}}$ is given by
\begin{equation}
 Y^{(0)}_{\text{DM}} = \frac{\Omega_{\text{DM}}
  \rho_{\text{crit}}^{(0)}}{M_{\text{DM}} s^{(0)}}~,
\end{equation}
where $\Omega_{\text{DM}}$ is the DM density parameter and
$\rho_{\text{crit}}^{(0)}$ is the critical density of the Universe. In
the following calculation, we use $\Omega_{\text{DM}}h^2 =0.12$,
$\rho_{\text{crit}}^{(0)} = 1.05\times
10^{-5}h^2~\text{GeV}\cdot\text{cm}^{-3}$, and $s^{(0)}=2.89\times
10^{3}~\text{cm}^{-3}$, with $h$ the Hubble parameter. As a result, we
obtain
\begin{equation}
 T_{\text{RH}}\simeq 2.7\times 10^4~\text{GeV}\times
\biggl(\frac{\Omega_{\text{DM}}h^2}{0.12}\biggr)
\biggl(\frac{g_*^{\frac{3}{2}}c^{-1}}{10^4}\biggr)
\biggl(\frac{M_{\text{DM}}}{100~\text{GeV}}\biggr)^{-1} ~,
\end{equation}
where we have set the value of $M_{\text{int}} = 10^{13.66}$ GeV from the 
result in Table \ref{tab:model1and2}. This approximate formula is valid
when $M_{\text{DM}}\ll T_{\text{RH}}$. 
Here, $g_*^{\frac{3}{2}}c^{-1}$ is an unknown factor and thus
causes an uncertainty in the computation. For instance, if $g_*={\cal O}(100)$
and the quartic coupling of $h$ and $\phi$ is $\sim 0.3$, then
$g_*^{\frac{3}{2}}c^{-1} = {\cal O}(10^4)$. Note that the
perturbativity of the quartic coupling ensures that this factor cannot
become too small. On the other hand, it also has an upper bound; if
$c$ is extremely small, then the one-loop gauge-boson exchange
contribution dominates over the tree level. Taking this consideration
into account, we vary the value of $g_*^{\frac{3}{2}}c^{-1}$ by a
factor of ten to estimate the uncertainty in the analysis given below.

\begin{figure}[t]
\begin{center}
\includegraphics[height=45mm]{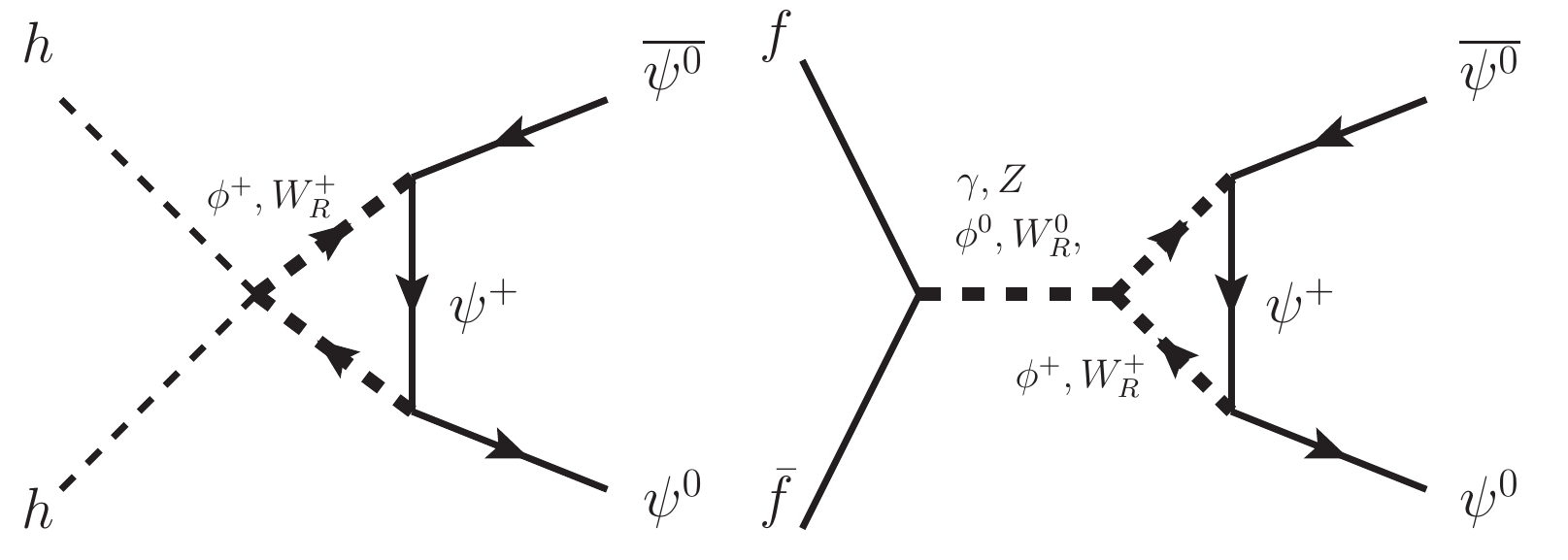}
\caption{{\it Examples of diagrams responsible for the DM production in
 Model I.}} 
\label{fig:annmod1}
\end{center}
\end{figure}

Next, we evaluate the relic abundance of the DM in Model I. In this
case, there is no tree-level process for the DM production, since the DM
does not couple to the singlet component $\phi^0$ in the $({\bf 1}, {\bf
1}, {\bf 3})_R$. Therefore, the DM is produced at the loop level. In
Fig.~\ref{fig:annmod1}, we show examples of one-loop diagrams which give
the dominant contribution to the DM production. The amplitude is then
estimated as
\begin{equation}
 \sum |{\cal M}|^2 \simeq \frac{c^\prime}{(16\pi^2)^2} ~
  \frac{\hat{s}-4M_{\text{DM}}^2}{M_{\text{int}}^2} ~,
\label{eq:summsqmod1}
\end{equation}
where we have included the one-loop factor. After a similar computation,
we obtain
\begin{equation}
 Y^{(0)}_{\text{DM}} \simeq \frac{c^\prime }{64 \pi^7(16\pi^2)^2}
\biggl(\frac{45}{\pi g_*}\biggr)^{\frac{3}{2}}
\frac{M_{\text{Pl}}T_{\text{RH}}}{M_{\text{int}}^2} 
~,
\end{equation}
and 
\begin{equation}
 T_{\text{RH}}\simeq 4.6~\text{GeV}\times
\biggl(\frac{\Omega_{\text{DM}}h^2}{0.12}\biggr)
\biggl(\frac{g_*^{\frac{3}{2}}c^{\prime -1}}{10^5}\biggr)
\biggl(\frac{M_{\text{DM}}}{\text{GeV}}\biggr)^{-1} ~,
\end{equation}
on the assumption of $M_{\text{DM}}\ll T_{\text{RH}}$.
Here, we have set $M_{\text{int}} = 10^{8.08}$ GeV. 

\begin{figure}[t]
\begin{center}
\subfigure[Model I]
 {\includegraphics[clip, width = 0.48 \textwidth]{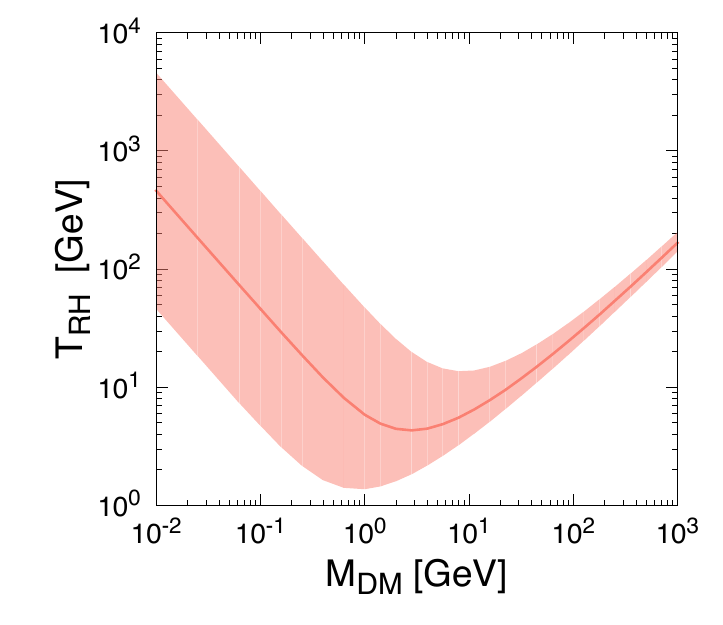}
 \label{fig:trhmod1}}
\subfigure[Model II]
 {\includegraphics[clip, width = 0.48 \textwidth]{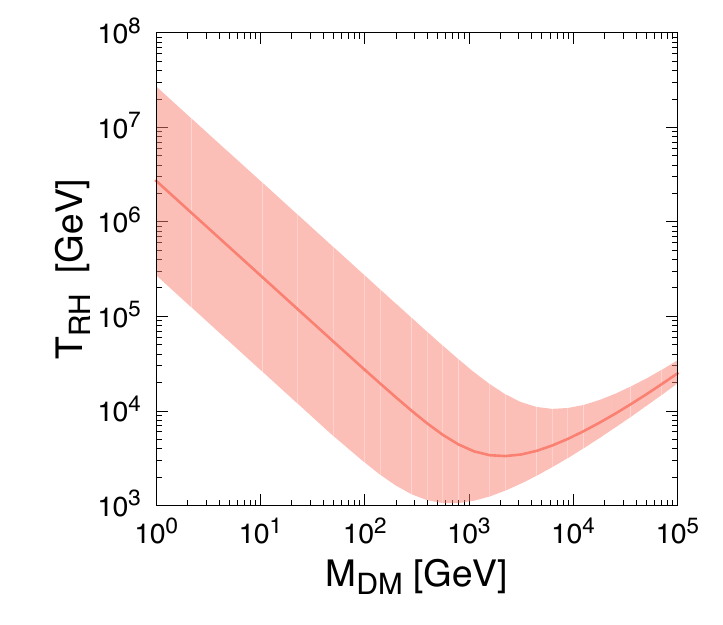}
 \label{fig:trhmod2}}
\caption{\it Reheating temperature as a function of DM mass. Pink band
 shows the theoretical uncertainty.}
\label{fig:trh}
\end{center}
\end{figure}

In Fig.~\ref{fig:trh}, we plot the predicted reheating temperature as a
function of the DM mass after numerically integrating Eq.~\eqref{eq:dydx}. The left and right panels show the cases of
Model I and II, respectively. The pink band shows the uncertainty of the
calculation, which we estimate by varying the unknown factor by a factor
of ten. 
It turns out that when $M_{\text{DM}}\ll T_{\text{RH}}$, in the
case of Model I, only a small DM mass is allowed and the reheating
temperature must be quite low. In Model II, on the other hand, DM with a
mass of around the electroweak scale accounts for the observed DM
density with an acceptably high reheating temperature. For a larger
$M_{\text{DM}}$, in both models, the DM relic abundance can only be
explained in the narrow strip region where $M_{\text{DM}}\simeq
T_{\text{RH}}$. 

\section{Lonely Singlet Fermion Dark Matter}
\label{sec:singletDM}

In the above discussion, we have assumed that there exists a DM
multiplet (as well as extra Higgs multiplets) above the intermediate
scale, and studied how the presence of
the additional fields affect the gauge coupling running in such models. As seen in
Sec.~\ref{sec:NETDMgcu}, these fields can indeed improve the solutions
for both the intermediate and GUT scales, which allow the models to
evade the limit from the proton decay experiment and to explain
light neutrino masses via the seesaw mechanism. Before concluding our
discussion, we briefly consider another possibility in this
section; that is, we have only a singlet DM fermion on top of the
standard SO(10) setup discussed in Sec.~\ref{sec:gcuandint}. In this
case, the DM, of course, cannot affect the gauge coupling running, and
thus it does not solve the problems regarding the low intermediate/GUT
scales in the ordinary SO(10) GUT models. Since there may be another
solution to these problems, it is worthwhile studying this possibility
as well. 

In fact, we can easily construct such a model by exploiting an
appropriate Higgs field at the GUT scale and fine-tuning its VEV so that
only the singlet fermion DM has a mass much lighter than the GUT
scale. For example, let us consider the case of $\text{SU}(4)_C\otimes
\text{SU}(2)_L \otimes \text{SU}(2)_R\otimes D$. In this case, the
singlet field under the intermediate gauge interactions, 
$({\bf 1}, {\bf 1}, {\bf 1})$, is contained in a
${\bf 54}$ or ${\bf 210}$ of SO(10). Since only the ${\bf 210}$ can have
a Yukawa coupling to the $R_1={\bf 54}_R$ Higgs field, we focus on the
case where the singlet DM fermion is a component of the ${\bf 210}$
field. In this case, both Majorana and Dirac fermions can couple to
the $R_1$ Higgs. Then, by fine-tuning the Yukawa coupling, we can make
only the singlet component have a light mass, as is done in
Sec.~\ref{sec:models}.  Similarly, we can obtain other models with
different intermediate gauge groups by using appropriate multiplets for
the fields which contain the singlet DM.  

The NETDM mechanism again works for this singlet DM through the $R_1$ Higgs
exchange process at tree level, with a diagram similar to that
illustrated in Fig.~\ref{fig:annmod2}. Following the discussion given in
Sec.~\ref{sec:NETDMcalc}, we can readily evaluate the reheating
temperature required to produce the right amount of DM. When
$M_{\text{DM}} \ll T_{\text{RH}}$, we have
\begin{equation}
 T_{\text{RH}}\simeq 1.3\times 10^9~\text{GeV}\times
\biggl(\frac{\Omega_{\text{DM}}h^2}{0.12}\biggr)
\biggl(\frac{g_*^{\frac{3}{2}}c^{-1}}{10^4}\biggr)
\biggl(\frac{M_{\text{DM}}}{100~\text{GeV}}\biggr)^{-1}
\biggl(\frac{M_{\text{GUT}}}{10^{16}~\text{GeV}}\biggr)^2 ~.
\end{equation}
Compared with Model I and II, the present scenario in general predicts a
high reheating temperature, as the production occurs via the GUT-scale
particle exchange. Such a high reheating temperature may be consistent
with thermal leptogenesis \cite{Fukugita:1986hr}.

As for proton decay and neutrino masses, the consequence of the singlet
DM models is the same as that without DM. Thus, for  $G_{\text{int}} =
\text{SU}(4)_C\otimes \text{SU}(2)_L \otimes \text{SU}(2)_R\otimes {D}$
and $\text{SU}(4)_C\otimes \text{SU}(2)_L \otimes \text{U}(1)_R$, the
proton decay constraints are still problematic, and thus it may be
required that we assume a relatively heavy GUT-scale gauge boson when
compared to the GUT scale. Other intermediate groups are not suitable for the
explanation of neutrino masses. The solution discussed in
Sec.~\ref{sec:neutrinomass} can again be exploited in these cases.

\section{Conclusion and discussion}
\label{sec:conclusion}

For over 40 years now, we have wondered whether grand unification is
actually realized in nature. Its simplicity, its capacity for an explanation of charge 
quantization and the apparent focusing of the gauge couplings as they 
run to high energy has kept grand unification (supersymmetric or not) 
at the center of most ultra-violet completions of the SM, though
experimental verification is still lacking.

On the other hand, we know from the existence of neutrino masses, the baryon asymmetry of the Universe
and the existence of DM that there must be new physics beyond the SM.
The presence of a natural DM candidate in SUSY extensions
of the SM (with conserved $R$-parity) is often taken to be one of the motivations for low-energy SUSY.
The ingredients for the baryon asymmetry are contained in most grand unified theories
(supersymmetric or not) including SU(5) and SO(10), and while a neutrino seesaw can be 
accomplished in SU(5) (by including the right-handed neutrino as a SU(5) singlet), it is more natural
in SO(10). 

We have, here, examined several breaking schemes of SO(10)
which lead to gauge coupling unification (by altering the SM
running of the gauge couplings at an intermediate scale), and contain a remnant 
$\mathbb{Z}_N$ symmetry which can account for the stability of DM. 
Having established the possible intermediate-scale gauge groups capable of
both gauge coupling unification and of supporting a stable DM candidate,
we considered specific possible representations (of dimension no larger than {\bf 210} for
simplicity) which contain a suitable non-degenerate SM singlet DM candidate.
If the DM candidate couples to the SM only through intermediate scale fields,
it may never equilibrate in the early Universe after reheating, and its production from the thermal bath is
an example of the NETDM scenario. Despite the fact that there are several possible
intermediate-scale gauge groups to consider and many possible representations for the 
DM candidate and intermediate-scale Higgs fields needed to break the degeneracy in the
DM multiplet, we found only two surviving models: one each based on 
$\text{SU}(4)_C\otimes
 \text{SU}(2)_L \otimes \text{SU}(2)_R$ and $
 \text{SU}(4)_C\otimes \text{SU}(2)_L \otimes \text{SU}(2)_R \otimes D$,
 with DM contained in a $({\bf 1}, {\bf 1}, {\bf 3})_D \in {\bf 45}_D$
 and $({\bf 15}, {\bf 1}, {\bf	 1})_W \in {\bf 45}_W$, respectively.

Both of the surviving models are capable of producing light neutrino masses
(though it is more difficult in Model I due to its relatively low intermediate scale).
We also showed that while the proton decay lifetime (to $e^+ \pi^0$) is at least
a factor of two longer than the current experimental bound for $M_X/M_{\text {GUT}} > 1/2$ in Model I,
the current bound excludes masses $M_X/M_{\text{GUT}} \lesssim 0.7$, and higher masses may be probed in
future proton decay experiments.  Finally, within the NETDM production scenario,
we have related our two models to a specific reheat temperature after inflation
needed to obtain the current relic density. While Model II predicts a
reheat temperature which easily allows for (non thermal) leptogenesis
\cite{Fukugita:1986hr, Lazarides:1991wu}, the reheat temperature in
Model I is rather low and presents a challenge for baryogenesis.

\section*{Acknowledgments}

Florian Lyonnet is gratefully acknowledged for useful discussions. 
This work was supported by the Spanish MICINN's Consolider-Ingenio 2010
Programme under grant Multi-Dark CSD2009-00064, and the contract
FPA2010-17747. Y.M. is grateful to the Mainz Institute for Theoretical
Physics (MITP) for its hospitality and its partial support during the
completion of this work. Y.M. acknowledges partial support from the
European Union FP7 ITN INVISIBLES (Marie Curie Actions, PITN- GA-2011-
289442) and the ERC advanced grants MassTeV and Higgs@LHC.
The work of N.N. is supported by Research
Fellowships of the Japan Society for the Promotion of Science for Young
Scientists. The work of K.A.O. and J.Z. was supported in part
by DOE grant DE-SC0011842 at the University of Minnesota.
The work of J.Q. was supported by the STFC Grant
ST/J002798/1.

\section*{Appendix}
\appendix

\section{Input parameters}
\label{sec:input}

The values for the input parameters we have used in this paper are
summarized in Table~\ref{table:inputparameters}. They are taken from
Ref.~\cite{Agashe:2014kda} except for the top-quark pole mass and the
Higgs mass, for which we use the values given in
Refs.~\cite{ATLAS:2014wva} and \cite{CMS:ril}, respectively. In this
table, the gauge coupling constants are defined in the
$\overline{\text{MS}}$ scheme, and thus we convert them to the
$\overline{\text{DR}}$ scheme at the electroweak scale using the
one-loop relation \cite{Yamada:1993uh}: 
\begin{equation}
 g_a(m_Z)_{\overline{\text{DR}}} =
g_a(m_Z)_{\overline{\text{MS}}}
\biggl(1+\frac{C(G_a)\alpha_a(m_Z)_{\overline{\text{MS}}}}
{24\pi}\biggr) ~,
\end{equation}
where $C(G_a)$ is the quadratic Casimir invariant. For the mass of the top
quark, we convert the pole mass to its $\overline{\text{MS}}$ mass by
using \cite{Agashe:2014kda}
\begin{equation}
 m_t^{\overline{\text{MS}}}(m_t^{\overline{\text{MS}}}) = m_t
\biggl(1-\frac{4\alpha_s(m_t^{\overline{\text{MS}}})}{3\pi}\biggr) ~,
\end{equation}
from which we obtain the $\overline{\text{MS}}$ top Yukawa coupling. The 
$\overline{\text{DR}}$ Yukawa coupling is then given by
\begin{equation}
 y_t^{\overline{\text{DR}}} = 
 y_t^{\overline{\text{MS}}}
\biggl[1
+\frac{\alpha_1}{480\pi}
+\frac{3\alpha_2}{32\pi}
-\frac{\alpha_3}{3\pi}\biggr] ~.
\end{equation}

\begin{table}[ht]
\caption{\it Input parameters \cite{Agashe:2014kda, ATLAS:2014wva, CMS:ril}.}
\label{table:inputparameters}
\begin{center}
\begin{tabular}{ll|l}
\hline
\hline
Strong coupling constant &$\alpha_s(m_Z)$ &
 $0.1185(6)$ \\
QED coupling constant& $\alpha(m_Z)$ & $1/127.944(14)$ \\
Fermi coupling constant &$G_F$& $1.1663787(6)\times 10^{-5}~\text{GeV}^{-2}$ \\
Weak-mixing angle &$\sin^2\theta_W (m_Z)$ & $0.23126(5)$ \\
$Z$-boson mass& $m_Z$ & $91.1876(21)$~GeV \\
Top pole mass & $m_t$ & $173.34(82)$~GeV \\
Higgs mass & $m_h$ & $125.15(24)$~GeV \\
\hline
\hline
\end{tabular}
\end{center}
\end{table}

\section{Renormalization group equations}
\label{sec:RGEs}

In this section, we summarize the RGEs and the matching conditions used
in text. The two-loop RGEs \cite{Machacek:1983tz} of the gauge coupling
constants $g_a$ are written as 
\begin{equation}
 \mu \frac{d g_a}{d\mu} = \frac{b_a^{(1)}}{16\pi^2} g_a^3 
+\frac{g_a^3}{(16\pi^2)^2}\biggl[ \sum_{b=1}^{3}b_{ab}^{(2)}g_b^2 
-c_{a} y_t^2
\biggr]~.
\end{equation}
Below, we will give the coefficients in each theory discussed in this
paper. For the contribution of Yukawa couplings, we include them only in the SM
running, as unknown Yukawa couplings appear above the intermediate
scale. Their effects should be taken into account as theoretical
uncertainties. All of the one-loop RGEs have been checked with the code
PyR@TE \cite{Lyonnet:2013dna}, and more importantly, the two-loop RGEs have
been computed with this code.

\subsection{Standard Model}

In the SM, we have
\begin{equation}
 b_a^{(1)}=
\begin{pmatrix}
 41/10 \\ -19/6 \\ -7
\end{pmatrix}
,~~~~b^{(2)}_{ab}=
\begin{pmatrix}
 199/50 & 27/10 & 44/5 \\
 9/10 & 35/6 & 12 \\
 11/10 & 9/2 & -26
\end{pmatrix}
,~~~~
 c_{a}=
\begin{pmatrix}
 17/10 \\
 3/2 \\
 2 
\end{pmatrix}
~.
\end{equation}
Here, $a =1,2,3$ correspond to U(1), SU(2)$_L$, and SU(3)$_C$,
respectively, with the U(1) gauge coupling constant normalized as $g_1
\equiv  \sqrt{5/3} g^\prime$. Since the top Yukawa
coupling contributes to the running of the gauge couplings at the two-loop
level, it is sufficient to consider the one-loop RGE for the top Yukawa
coupling. Furthermore, we can safely neglect the contribution of the other
Yukawa couplings. Thus, the relevant RGE is
\begin{align}
 \mu\frac{d}{d \mu}y_t=\frac{1}{16\pi^2}y_t
\biggl[\frac{9}{2}y_t^2 
 -\frac{17}{20}g^2_1-\frac{9}{4}g_2^2-8
 g_3^2\biggr] ~ .
\end{align}

\subsection{$\text{SU}(4)_C\otimes \text{SU}(2)_L \otimes
  \text{SU}(2)_R$}
\label{sec:beta422}

As discussed in
Sec.~\ref{sec:gcuandint}, above the intermediate mass scale, the theory
contains the SM fermions, the gauge bosons, the $({\bf 10}, {\bf 1},
{\bf 3})_C$ field, and the $({\bf 1}, {\bf 2}, \overline{\bf 2})_C$ Higgs
field. The beta-function coefficients in this case are given by
\begin{equation}
 b_a^{(1)}=
\begin{pmatrix}
 -3 \\ 11/3 \\ -{23}/{3}
\end{pmatrix}
,~~~~~~b^{(2)}_{ab}=
\begin{pmatrix}
 8 & 3 & 45/2 \\
 3 & 584/3 & 765/2 \\
 9/2 & 153/2 & 643/6
\end{pmatrix}
~,
\end{equation}
where $a=2L,2R,4$ correspond to SU(2)$_L$, SU(2)$_R$, and SU(4)$_C$,
respectively. The matching conditions at the intermediate mass scale are
\begin{align}
 \frac{1}{g_1^2(M_{\text{int}})} &= 
\frac{3}{5}\frac{1}{g^2_{2R}(M_{\text{int}})}
+\frac{2}{5}\frac{1}{g^2_{4}(M_{\text{int}})} ~, \nonumber \\
 g_{2}(M_{\text{int}}) &= g_{2L}(M_{\text{int}}) ~, \nonumber \\
 g_{3}(M_{\text{int}}) &= g_{4}(M_{\text{int}}) ~.
\end{align}

\subsection{$\text{SU}(4)_C\otimes \text{SU}(2)_L \otimes
  \text{SU}(2)_R\otimes D$}
\label{sec:beta422D}

In this case, the $(\overline{\bf 10}, {\bf 3}, {\bf 1})_C$ field is
added to the previous theory. The beta-function coefficients then become
\begin{equation}
 b_a^{(1)}=
\begin{pmatrix}
 11/3 \\ 11/3 \\ -{14}/{3}
\end{pmatrix}
,~~~~~~b^{(2)}_{ab}=
\begin{pmatrix}
 584/3 & 3 & 765/2 \\
 3 & 584/3 & 765/2 \\
 153/2 & 153/2 & 1759/6
\end{pmatrix}
~,
\end{equation}
where $a=2L,2R,4$ correspond to SU(2)$_L$, SU(2)$_R$, and SU(4)$_C$,
respectively.

\subsection{$\text{SU}(4)_C\otimes \text{SU}(2)_L \otimes \text{U}(1)_R$}

This theory contains the SM fermions, the gauge bosons, the $({\bf 10}, {\bf 1},
1)_C$ field, and the $({\bf 1}, {\bf 2}, \frac{1}{2})$ Higgs
field. The beta-function coefficients in this case are given by
\begin{equation}
 b_a^{(1)}=
\begin{pmatrix}
 -19/6 \\ 15/2 \\ -{29}/{3}
\end{pmatrix}
,~~~~~~b^{(2)}_{ab}=
\begin{pmatrix}
 35/6 & 1/2 & 45/2 \\
 3/2 & 87/2 & 405/2 \\
 9/2 & 27/2 & -101/6
\end{pmatrix}
~,
\end{equation}
where $a=2L,1R,4$ correspond to SU(2)$_L$, U(1)$_R$, and SU(4)$_C$,
respectively. The matching conditions at the intermediate mass scale are
\begin{align}
 \frac{1}{g_1^2(M_{\text{int}})} &= 
\frac{3}{5}\frac{1}{g^2_{1R}(M_{\text{int}})}
+\frac{2}{5}\frac{1}{g^2_{4}(M_{\text{int}})} ~, \nonumber \\
 g_{2}(M_{\text{int}}) &= g_{2L}(M_{\text{int}}) ~, \nonumber \\
 g_{3}(M_{\text{int}}) &= g_{4}(M_{\text{int}}) ~.
\end{align}

\subsection{$\text{SU}(3)_C\otimes \text{SU}(2)_L \otimes \text{SU}(2)_R
  \otimes \text{U}(1)_{B-L}$}

This theory contains the SM fermions, the gauge bosons, the $({\bf 1},
{\bf 1}, {\bf 3}, -2)_C$ field, and the $({\bf 1}, {\bf 2}, {\bf 2}, 0)$
Higgs field. The beta-function coefficients in this case are given by
\begin{equation}
 b_a^{(1)}=
\begin{pmatrix}
 -3 \\ -7/3 \\ 11/2 \\ -7
\end{pmatrix}
,~~~~~~b^{(2)}_{ab}=
\begin{pmatrix}
 8 & 3 & 3/2& 12 \\
 3 & 80/3 & 27/2 &12 \\
 9/2 & 81/2 & 61/2 &4 \\
 9/2 & 9/2 & 1/2 & -26
\end{pmatrix}
~,
\end{equation}
where $a=2L,2R, BL, 3$ correspond to SU(2)$_L$, SU(2)$_R$, U(1)$_{B-L}$
and SU(3)$_C$, respectively. The U(1)$_{B-L}$ charge is normalized such that
it satisfies the normalization condition of the SO(10) generators:
$T_{B-L} = \sqrt{3/8} (B-L)$. 
The matching conditions at the intermediate mass scale are
\begin{align}
 \frac{1}{g_1^2(M_{\text{int}})} &= 
\frac{3}{5}\frac{1}{g^2_{2R}(M_{\text{int}})}
+\frac{2}{5}\frac{1}{g^2_{BL}(M_{\text{int}})} ~, \nonumber \\
 g_{2}(M_{\text{int}}) &= g_{2L}(M_{\text{int}}) ~, \nonumber \\
 g_{3}(M_{\text{int}}) &= g_{3}(M_{\text{int}}) ~.
\end{align}

\subsection{$\text{SU}(3)_C\otimes \text{SU}(2)_L \otimes \text{SU}(2)_R
  \otimes \text{U}(1)_{B-L}\otimes D$}

For this left-right symmetric theory, the $({\bf 1}, {\bf 3}, {\bf 1}, 2)_C$ field is added to the previous
case. The beta-function coefficients are then modified to
\begin{equation}
 b_a^{(1)}=
\begin{pmatrix}
 -7/3 \\ -7/3 \\ 7 \\ -7
\end{pmatrix}
,~~~~~~b^{(2)}_{ab}=
\begin{pmatrix}
 80/3 & 3 & 27/2& 12 \\
 3 & 80/3 & 27/2 &12 \\
 81/2 & 81/2 & 115/2 &4 \\
 9/2 & 9/2 & 1/2 & -26
\end{pmatrix}
~,
\end{equation}
where $a=2L,2R, BL, 3$ correspond to SU(2)$_L$, SU(2)$_R$, U(1)$_{B-L}$
and SU(3)$_C$, respectively.

\subsection{$\text{SU}(3)_C\otimes \text{SU}(2)_L \otimes \text{U}(1)_R
  \otimes \text{U}(1)_{B-L}$}

This theory contains the SM fermions, the gauge bosons, the $({\bf 1},
{\bf 1}, {\bf 1}, -2)_C$ field, and the $({\bf 1}, {\bf 2}, 1/2, 0)$
Higgs field. The beta-function coefficients in this case are given by
\begin{equation}
 b_a^{(1)}=
\begin{pmatrix}
 -19/6 \\ 9/2 \\ 9/2 \\ -7
\end{pmatrix}
,~~~~~~b^{(2)}_{ab}=
\begin{pmatrix}
 35/6 & 1/2 & 3/2& 12 \\
 3/2 & 15/2 & 15/2 &12 \\
 9/2 & 15/2 & 25/2 &4 \\
 9/2 & 3/2 & 1/2 & -26
\end{pmatrix}
~,
\end{equation}
where $a=2L,1R, BL, 3$ correspond to SU(2)$_L$, U(1)$_R$, U(1)$_{B-L}$
and SU(3)$_C$, respectively. 
The matching conditions at the intermediate mass scale are
\begin{align}
 \frac{1}{g_1^2(M_{\text{int}})} &= 
\frac{3}{5}\frac{1}{g^2_{1R}(M_{\text{int}})}
+\frac{2}{5}\frac{1}{g^2_{BL}(M_{\text{int}})} ~, \nonumber \\
 g_{2}(M_{\text{int}}) &= g_{2L}(M_{\text{int}}) ~, \nonumber \\
 g_{3}(M_{\text{int}}) &= g_{3}(M_{\text{int}}) ~.
\end{align}

\subsection{Model I}
\label{sec:betamodel1}

For DM model I, a $({\bf 1}, {\bf 1}, {\bf 3})_D$ Dirac fermion and a
$({\bf 1}, {\bf 1}, {\bf 3})_R$ real scalar field are added to the
theory described in Appendix~\ref{sec:beta422}. 
The beta-function coefficients are then computed as
\begin{equation}
 b_a^{(1)}=
\begin{pmatrix}
 -3 \\ 20/3 \\ -{23}/{3}
\end{pmatrix}
,~~~~~~b^{(2)}_{ab}=
\begin{pmatrix}
 8 & 3 & 45/2 \\
 3 & 740/3 & 765/2 \\
 9/2 & 153/2 & 643/6
\end{pmatrix}
~,
\end{equation}
where $a=2L,2R,4$ correspond to SU(2)$_L$, SU(2)$_R$, and SU(4)$_C$,
respectively. 

\subsection{Model II}
\label{sec:betamodel2}

For DM model II,  a $({\bf 15}, {\bf 1}, {\bf 1})_W$ Weyl fermion and a
$({\bf 15}, {\bf 1}, {\bf 1})_R$ real scalar field are added to the
theory described in Appendix~\ref{sec:beta422D}. 
The beta-function coefficients are then computed as
\begin{equation}
 b_a^{(1)}=
\begin{pmatrix}
 11/3 \\ 11/3 \\ -{4}/{3}
\end{pmatrix}
,~~~~~~b^{(2)}_{ab}=
\begin{pmatrix}
 584/3 & 3 & 765/2 \\
 3 & 584/3 & 765/2 \\
 153/2 & 153/2 & 2495/6
\end{pmatrix}
~,
\end{equation}
where $a=2L,2R,4$ correspond to SU(2)$_L$, SU(2)$_R$, and SU(4)$_C$,
respectively.

\section{One-loop formulae for gauge coupling unification}
\label{sec:1loopform}

At the one-loop level, the gauge coupling RGEs are easily solved
analytically. By using the solutions, we can  obtain analytic
expressions for $M_{\text{int}}$, $M_{\text{GUT}}$, and
$\alpha_{\text{GUT}}$ as follows:
\begin{align}
 M_{\text{int}}&= m_Z\exp \biggl[\frac{2\pi(\bm{\tilde{b}}\times  \bm{n}
)\cdot \bm{\alpha_{-1}}}{(\bm{\tilde{b}}\times \bm{n} )\cdot
\bm{b}}\biggr] 
~, \label{eq:mintanaly} \\[3pt]
 M_{\text{GUT}}&= m_Z\exp \biggl[\frac{2\pi(\Delta\bm{{b}}\times
 \bm{n} )\cdot\bm{\alpha_{-1}}}
{(\bm{\tilde{b}}\times \bm{n} )\cdot \bm{b}}\biggr]
~, \\[3pt]
 \alpha_{\text{GUT}}^{-1} &=
\frac{(\bm{\tilde{b}}\times \bm{\alpha_{-1}})\cdot   \bm{b}}
{(\bm{\tilde{b}}\times \bm{n} )\cdot \bm{b}} ~,
\end{align}
with
\begin{align}
 \bm{\alpha_{-1}}&\equiv 
\begin{pmatrix}
 \alpha_1^{-1}(m_Z) \\ \alpha_2^{-1}(m_Z) \\ \alpha_3^{-1}(m_Z)
\end{pmatrix}
~,~~~~~~ 
\bm{b}\equiv
\begin{pmatrix}
 b_1 \\ b_2 \\ b_3
\end{pmatrix}
~,~~~~~~ 
\bm{\tilde{b}}\equiv
\begin{pmatrix}
 \tilde{b}_1 \\ \tilde{b}_2 \\ \tilde{b}_3
\end{pmatrix}
~,~~~~~~
\bm{n}\equiv
\begin{pmatrix}
 1\\ 1\\ 1
\end{pmatrix}
~, 
\end{align}
where $\Delta \bm{b}\equiv \bm{\tilde{b}} - \bm{b}$, and $b_a$ and
$\tilde{b}_a$ denote the beta-function coefficients below and above the
intermediate scale, respectively. The U(1) beta function above the
intermediate scale is given by a linear combination of the beta
functions of the intermediate gauge group. For instance, in the case of 
$\text{SU}(4)_C\otimes \text{SU}(2)_L \otimes \text{SU}(2)_R$, we have 
\begin{equation}
 \tilde{b}_1 =\frac{2}{5}b_4 + \frac{3}{5}b_{2R} ~.
\end{equation}
Similar expressions are obtained for other intermediate groups. 
Notice that the components of the beta-function
coefficients which are proportional to $\bm{n}$ do not affect
$M_{\text{GUT}}$ and $M_{\text{int}}$, as one can see from the
formulae. Therefore, if one adds a multiplet to, {\it e.g.}, the
$\text{SU}(4)_C\otimes \text{SU}(2)_L \otimes \text{SU}(2)_R$ theory
whose contribution to the beta-function coefficients is $\Delta b_4 =
\Delta b_{2L} = \Delta b_{2R}$, then the multiplet does not alter 
$M_{\text{GUT}}$ and $M_{\text{int}}$ at the one-loop level. 

We also note that physics above the intermediate scale gives negligible
effects on the determination of $M_{\text{int}}$ in the presence of the
left-right symmetry. We can see this feature by using
Eq.~\eqref{eq:mintanaly}. Let us consider the case of $\text{SU}(4)_C\otimes
\text{SU}(2)_L \otimes \text{SU}(2)_R\otimes D$. In the left-right
symmetric theories, the beta functions of the SU(2)$_L$
and SU(2)$_R$ gauge couplings should be the same. Therefore, we have $b_{2L} =
b_{2R}$, and 
\begin{equation}
 \bm{\tilde{b}} \times \bm{n} =(b_{2L}-b_4)
\bm{c} ~,
\end{equation}
with 
\begin{equation}
 \bm{c} = \begin{pmatrix}
 1 \\ -\frac{3}{5} \\ -\frac{2}{5}
\end{pmatrix} ~.
\end{equation}
Therefore, Eq.~\eqref{eq:mintanaly} reads
\begin{equation}
 M_{\text{int}}
= m_Z\exp \biggl[\frac{2\pi\bm{c}\cdot \bm{\alpha_{-1}}}{\bm{c}\cdot
\bm{b}}\biggr] 
~,
\end{equation}
and thus, the intermediate scale does not depend on the beta function
above $M_{\text{int}}$. One can also see this feature by noting that
above the intermediate scale $g_{2L} = g_{2R}$ holds at any
scale. Hence, the intermediate scale corresponds to a point at
which $g_{2L}$ becomes equivalent to $g_{2R}$, which is determined only
by the running below $M_{\text{int}}$.
A similar argument holds in the case of
$\text{SU}(3)_C\otimes \text{SU}(2)_L \otimes \text{SU}(2)_R \otimes
\text{U}(1)_{B-L}\otimes D$.

\section{Proton decay in $\text{SO}(10)\to \text{SU}(4)\otimes
 \text{SU}(2) \otimes \text{SU}(2)$}
\label{sec:protondecay422}

Here, we give details of the calculation for the proton decay lifetime in
the intermediate-scale scenario. We consider the case of $\text{SO}(10)\to
\text{SU}(4)\otimes  \text{SU}(2) \otimes \text{SU}(2)$, which was
discussed in Sec.~\ref{sec:protondecay}. 

In non-SUSY GUTs, proton decay is induced by gauge
interactions. The relevant interactions are written as
\begin{align}
 {\cal L}_{\text{int}} =\frac{g_{\text{GUT}}}{\sqrt{2}}
\bigl[
(\overline{Q})_{ar}\Slash{X}^{air}P_R (L^{\cal C})_i
+(\overline{Q})_{ai}\Slash{X}^{air}P_L (L^{\cal C})_r
+\epsilon_{ij}\epsilon_{rs}\epsilon_{abc}
(\overline{Q^{\cal C}})^{ar}\Slash{X}^{bis}P_L Q^{cj}
+\text{h.c.}\bigr] ~,
\end{align}
where
\begin{equation}
 Q=
\begin{pmatrix}
    u \\ d
\end{pmatrix}
~, ~~~~~~
 L=
\begin{pmatrix}
    \nu \\ e^-
\end{pmatrix}
~,
\end{equation}
and $X$ denotes the superheavy gauge bosons which induce the
baryon-number violating interactions; $g_{\text{GUT}}$ is the unified gauge
coupling constant; $a,b,c$,
$i,j$, and $r,s$ are the SU(3)$_C$, SU(2)$_L$, and SU(2)$_R$ indices,
respectively; $P_{R/L}\equiv (1\pm \gamma_5)/2$ are the chirality
projection operators.

After integrating out the SO(10) gauge fields $X$, we obtain the
dimension-six proton decay operator. The operator is expressed in a
form that respects the intermediate gauge symmetry, $\text{SU}(4)\otimes
\text{SU}(2) \otimes \text{SU}(2)$:
\begin{equation}
 {\cal L}_{\text{eff}} = C(M_{\text{GUT}})\cdot 
\epsilon_{ij}\epsilon_{rs}\epsilon_{\alpha\beta\gamma\delta}
(\overline{\Psi^{\cal C}})^{\alpha i} P_L \Psi^{\beta j}
(\overline{\Psi^{\cal C}})^{\gamma r} P_R \Psi^{\delta s}~,
\label{eq:effop422}
\end{equation} 
where $\alpha, \beta,\dots$ denote the SU(4) indices, and $\Psi$ is
given in Eq.~\eqref{eq:4spinordef}. Notice that 
\begin{equation}
 \epsilon_{ij}\epsilon_{kl}\epsilon_{\alpha\beta\gamma\delta}
(\overline{\Psi^{\cal C}})^{\alpha i} P_L \Psi^{\beta j}
(\overline{\Psi^{\cal C}})^{\gamma k} P_L \Psi^{\delta l}
=
\epsilon_{rs}\epsilon_{tu}\epsilon_{\alpha\beta\gamma\delta}
(\overline{\Psi^{\cal C}})^{\alpha r} P_R \Psi^{\beta s}
(\overline{\Psi^{\cal C}})^{\gamma t} P_R \Psi^{\delta u}
= 0 ~,
\end{equation}
and thus the operator in Eq.~\eqref{eq:effop422} is the unique choice. 
At tree level, the coefficient of the effective operator is evaluated as
\begin{equation}
 C(M_{\text{GUT}}) = \frac{g_{\text{GUT}}^2}{2M_X^2} ~,
\end{equation}
with $M_X$ the mass of the heavy gauge field $X$. Here, we have
neglected fermion flavor mixings \cite{FileviezPerez:2004hn} for
simplicity.

The Wilson coefficient is evolved down to the intermediate scale
using the RGE. The renormalization factor is computed to be
\cite{Munoz:1986kq} 
\begin{equation}
 C(M_{\text{int}}) = 
\biggl[\frac{\alpha_4(M_{\text{int}})}{\alpha_{\text{GUT}}}\biggr]
^{-\frac{15}{4b_4}}\biggl[\frac{\alpha_{2L}
(M_{\text{int}})}{\alpha_{\text{GUT}}}\biggr]^{-\frac{9}{4b_{2L}}}
\biggl[\frac{\alpha_{2R}(M_{\text{int}})}{\alpha_\text{GUT}}\biggr]
^{-\frac{9}{4b_{2R}}} C(M_{\text{GUT}})~.
\end{equation}

At the intermediate scale, the $\text{SU}(4)\otimes  \text{SU}(2)
\otimes \text{SU}(2)$ theory is matched onto the SM. The
effective Lagrangian is written as
\begin{equation}
 {\cal L}_{\text{eff}} = \sum_{I=1}^4 C_I {\cal O}_I ~,
\end{equation}
with the effective operators given by \cite{Weinberg:1979sa,
Wilczek:1979hc, Abbott:1980zj}
\begin{align}
 {\cal O}_1&=
\epsilon_{abc}\epsilon_{ij}(u^a_{R}d^b_{R})(Q_{L}^{ci} L_{L}^{j})~,\nonumber \\
 {\cal O}_{2}&=
\epsilon_{abc}\epsilon_{ij}
(Q^{ai}_{L} Q^{bj}_{L})(u_{R}^ce_{R}^{})~,\nonumber \\
{\cal O}_{3}&=
\epsilon_{abc}\epsilon_{ij}\epsilon_{kl}
(Q^{ai}_{L} Q^{bk}_{L})(Q_{L}^{cl}
 L_{L}^{j})~,\nonumber \\ 
 {\cal O}_{4}&=
\epsilon_{abc}(u^a_{R}d^b_{R})(u_{R}^c e_{R}^{})~.
\label{eq:fourfermidef}
\end{align}
We evaluate the coefficients $C_I$ as
\begin{align}
 C_1(M_{\text{int}}) &= 4 C(M_{\text{int}}) ~, \nonumber \\
 C_2(M_{\text{int}}) &= - 4 C(M_{\text{int}}) ~, \nonumber \\
 C_3(M_{\text{int}}) &=  C_4(M_{\text{int}}) = 0.
\end{align}

We then run down the coefficients to the electroweak scale. The
renormalization factors are given by 
\cite{Abbott:1980zj}
\begin{align}
  C_1(m_Z) &= 
\biggl[\frac{\alpha_3(m_Z)}{\alpha_3
 (M_{\text{int}})}\biggr]^{-\frac{2}{b_3}} 
\biggl[\frac{\alpha_2(m_Z)}{\alpha_2
 (M_{\text{int}})}\biggr]^{-\frac{9}{4b_2}} 
\biggl[\frac{\alpha_1(m_Z)}{\alpha_1
 (M_{\text{int}})}\biggr]^{-\frac{11}{20b_1}} 
C_1(M_{\text{int}})
~, \\
  C_2(m_Z) &= 
\biggl[\frac{\alpha_3(m_Z)}{\alpha_3
 (M_{\text{int}})}\biggr]^{-\frac{2}{b_3}} 
\biggl[\frac{\alpha_2(m_Z)}{\alpha_2
 (M_{\text{int}})}\biggr]^{-\frac{9}{4b_2}} 
\biggl[\frac{\alpha_1(m_Z)}{\alpha_1
 (M_{\text{int}})}\biggr]^{-\frac{23}{20b_1}} 
C_2(M_{\text{int}})
~.
\end{align}
Note that the beta-function coefficients should be appropriately
modified when the number of quark flavors changes. Below the electroweak
scale, the QCD corrections are the dominant contribution. By using the
two-loop RGE given in Ref.~\cite{Nihei:1994tx}, we compute the Wilson
coefficients at the hadronic scale $\mu_{\text{had}}$ as
\begin{equation}
C_i(\mu_{\text{had}}) =
 \biggl[
\frac{\alpha_s(\mu_{\text{had}})}{\alpha_s(m_b)}
\biggr]^{\frac{6}{25}}\biggl[
\frac{\alpha_s(m_b)}{\alpha_s(m_Z)}
\biggr]^{\frac{6}{23}}
\biggl[
\frac{\alpha_s(\mu_{\text{had}})+\frac{50\pi}{77}}
{\alpha_s(m_b)+\frac{50\pi}{77}}
\biggr]^{-\frac{173}{825}}
\biggl[
\frac{\alpha_s(m_b)+\frac{23\pi}{29}}
{\alpha_s(m_Z)+\frac{23\pi}{29}}
\biggr]^{-\frac{430}{2001}}C_i(m_Z)~,
\end{equation}
with $i=1,2$. 

In non-SUSY GUTs, the dominant decay mode of proton is
$p\to \pi^0 e^+$. The partial decay width of the mode is computed as
\begin{equation}
 \Gamma (p \to \pi^0 e^+)
=\frac{m_p}{32\pi}\biggl(1-\frac{m_\pi^2}{m_p^2}\biggr)^2
\bigl[|{\cal A}_L|^2+|{\cal A}_R|^2\bigr] ~,
\end{equation}
where $m_p$ and $m_\pi$ are the masses of the proton and the neutral pion,
respectively, and
\begin{align}
 {\cal A}_L&= C_1(\mu_{\text{had}})\langle \pi^0|(ud)_R u_L|p\rangle ~,
 \nonumber \\
 {\cal A}_R&= 2C_2(\mu_{\text{had}})\langle \pi^0|(ud)_L u_R|p\rangle ~.
\end{align}
The hadron matrix elements are evaluated with the lattice QCD
simulations in Ref.~\cite{Aoki:2013yxa}. We have
\begin{align}
 \langle \pi^0|(ud)_R u_L|p\rangle &=
 \langle \pi^0|(ud)_L u_R|p\rangle = -0.103(23)(34)~\text{GeV}^2 ~,
\end{align}
with $\mu_{\text{had}}=2$~GeV. Here, the first and second parentheses
indicate statistical and systematic errors, respectively.




\begin{thebibliography}{99}

\bibitem{susy}
J.~R.~Ellis, S.~Kelley and D.~V.~Nanopoulos,
  Phys.\ Lett.\ B {\bf 249} (1990) 441 and
  Phys.\ Lett.\ B {\bf 260} (1991) 131;
U.~Amaldi, W.~de Boer and H.~Furstenau,
  Phys.\ Lett.\ B {\bf 260} (1991) 447;
  C.~Giunti, C.~W.~Kim and U.~W.~Lee,
  Mod.\ Phys.\ Lett.\ A {\bf 6} (1991) 1745;
  P.~Langacker and M.~x.~Luo,
  Phys.\ Rev.\ D {\bf 44}, 817 (1991).


\bibitem{Rajpoot:1980xy} 
  S.~Rajpoot,
  Phys.\ Rev.\ D {\bf 22}, 2244 (1980);
%
  M.~Yasue,
  Prog.\ Theor.\ Phys.\  {\bf 65}, 708 (1981)
  [Erratum-ibid.\  {\bf 65}, 1480 (1981)];
  J.~M.~Gipson and R.~E.~Marshak,
  Phys.\ Rev.\ D {\bf 31}, 1705 (1985);
  D.~Chang, R.~N.~Mohapatra, J.~Gipson, R.~E.~Marshak and M.~K.~Parida,
  Phys.\ Rev.\ D {\bf 31}, 1718 (1985);
  N.~G.~Deshpande, E.~Keith and P.~B.~Pal,
  Phys.\ Rev.\ D {\bf 46}, 2261 (1993);
  N.~G.~Deshpande, E.~Keith and P.~B.~Pal,
  Phys.\ Rev.\ D {\bf 47}, 2892 (1993)
  [hep-ph/9211232];
  S.~Bertolini, L.~Di Luzio and M.~Malinsky,
  Phys.\ Rev.\ D {\bf 81}, 035015 (2010)
  [arXiv:0912.1796 [hep-ph]].
  
\bibitem{Fukugita:1993fr} 
  M.~Fukugita and T.~Yanagida,
  In *Fukugita, M. (ed.), Suzuki, A. (ed.): Physics and astrophysics of neutrinos* 1-248. and Kyoto Univ. - YITP-K-1050 (93/12,rec.Feb.94) 248 p. C;



\bibitem{Mambrini:2013iaa} 
  Y.~Mambrini, K.~A.~Olive, J.~Quevillon and B.~Zaldivar,
  Phys.\ Rev.\ Lett.\  {\bf 110}, 241306 (2013)
  [arXiv:1302.4438 [hep-ph]].

  \bibitem{EHNOS}
H. Goldberg, Phys. Rev. Lett. {\bf 50} (1983) 1419;
J. Ellis, J.S. Hagelin, D.V. Nanopoulos, K.A. Olive
and M. Srednicki, Nucl. Phys. {\bf B238} (1984) 453.

\bibitem{Kibble:1982ae} 
  T.~W.~B.~Kibble, G.~Lazarides and Q.~Shafi,
  Phys.\ Lett.\ B {\bf 113}, 237 (1982).

\bibitem{Krauss:1988zc} 
  L.~M.~Krauss and F.~Wilczek,
  Phys.\ Rev.\ Lett.\  {\bf 62}, 1221 (1989).

\bibitem{Ibanez:1991hv} 
  L.~E.~Ibanez and G.~G.~Ross,
  Phys.\ Lett.\ B {\bf 260}, 291 (1991);
%
  L.~E.~Ibanez and G.~G.~Ross,
  Nucl.\ Phys.\ B {\bf 368}, 3 (1992).

\bibitem{Martin:1992mq} 
  S.~P.~Martin,
  Phys.\ Rev.\ D {\bf 46}, 2769 (1992)
  [hep-ph/9207218].



\bibitem{Kadastik:2009dj} 
  M.~Kadastik, K.~Kannike and M.~Raidal,
  Phys.\ Rev.\ D {\bf 81}, 015002 (2010)
  [arXiv:0903.2475 [hep-ph]];
%
  M.~Kadastik, K.~Kannike and M.~Raidal,
  Phys.\ Rev.\ D {\bf 80}, 085020 (2009)
  [Erratum-ibid.\ D {\bf 81}, 029903 (2010)]
  [arXiv:0907.1894 [hep-ph]].

\bibitem{Frigerio:2009wf} 
  M.~Frigerio and T.~Hambye,
  Phys.\ Rev.\ D {\bf 81}, 075002 (2010)
  [arXiv:0912.1545 [hep-ph]];
%
  T.~Hambye,
  PoS IDM {\bf 2010}, 098 (2011)
  [arXiv:1012.4587 [hep-ph]].

  \bibitem{freezein}
 L.~J.~Hall, K.~Jedamzik, J.~March-Russell and S.~M.~West,
  JHEP {\bf 1003} (2010) 080
  [arXiv:0911.1120 [hep-ph]].
  J.~McDonald,
  Phys.\ Rev.\ Lett.\  {\bf 88} (2002) 091304
  [hep-ph/0106249].
 X.~Chu, T.~Hambye and M.~H.~G.~Tytgat,
  JCAP {\bf 1205} (2012) 034
  [arXiv:1112.0493 [hep-ph]];
  C.~E.~Yaguna,
  JCAP {\bf 1202} (2012) 006
  [arXiv:1111.6831 [hep-ph]];



\bibitem{DeMontigny:1993gy} 
  M.~De Montigny and M.~Masip,
  Phys.\ Rev.\ D {\bf 49}, 3734 (1994)
  [hep-ph/9309312].

\bibitem{DYNKINE.B.:1957}
  E.~B.~Dynkin,
  Amer.\ Math.\ Soc.\ Transl.\ {\bf 6}, 111 (1957);
  E. Dynkin, 
  Selected Papers of EB Dynkin with Commentary, Amer. Math. Soc.,
  Providence, RI, 37 (2000). 

\bibitem{Slansky:1981yr} 
  R.~Slansky,
  Phys.\ Rept.\  {\bf 79}, 1 (1981).

\bibitem{Georgi:1982jb} 
  H.~Georgi,
  Front.\ Phys.\  {\bf 54}, 1 (1982).

\bibitem{Barr:1981qv} 
  S.~M.~Barr,
  Phys.\ Lett.\ B {\bf 112}, 219 (1982);
%
  S.~M.~Barr,
  Phys.\ Rev.\ D {\bf 40}, 2457 (1989);
%
  J.~P.~Derendinger, J.~E.~Kim and D.~V.~Nanopoulos,
  Phys.\ Lett.\ B {\bf 139}, 170 (1984).
%
\bibitem{Antoniadis:1987dx} 
  I.~Antoniadis, J.~R.~Ellis, J.~S.~Hagelin and D.~V.~Nanopoulos,
  Phys.\ Lett.\ B {\bf 194}, 231 (1987).



\bibitem{Farrar:1978xj} 
  G.~R.~Farrar and P.~Fayet,
  Phys.\ Lett.\ B {\bf 76}, 575 (1978);
%
  S.~Dimopoulos and H.~Georgi,
  Nucl.\ Phys.\ B {\bf 193}, 150 (1981);
%
  S.~Weinberg,
  Phys.\ Rev.\ D {\bf 26}, 287 (1982);
%
  N.~Sakai and T.~Yanagida,
  Nucl.\ Phys.\ B {\bf 197}, 533 (1982);
%
  S.~Dimopoulos, S.~Raby and F.~Wilczek,
  Phys.\ Lett.\ B {\bf 112}, 133 (1982).

\bibitem{Kuzmin:1980yp} 
  V.~A.~Kuzmin and M.~E.~Shaposhnikov,
  Phys.\ Lett.\ B {\bf 92}, 115 (1980);
%
  T.~W.~B.~Kibble, G.~Lazarides and Q.~Shafi,
  Phys.\ Rev.\ D {\bf 26}, 435 (1982);
%
  D.~Chang, R.~N.~Mohapatra and M.~K.~Parida,
  Phys.\ Rev.\ Lett.\  {\bf 52}, 1072 (1984);
%
  D.~Chang, R.~N.~Mohapatra and M.~K.~Parida,
  Phys.\ Rev.\ D {\bf 30}, 1052 (1984);
%
  D.~Chang, R.~N.~Mohapatra, J.~Gipson, R.~E.~Marshak and M.~K.~Parida,
  Phys.\ Rev.\ D {\bf 31}, 1718 (1985).

\bibitem{Smith:1979rz} 
  P.~F.~Smith and J.~R.~J.~Bennett,
  Nucl.\ Phys.\ B {\bf 149}, 525 (1979);
  P.~F.~Smith, J.~R.~J.~Bennett, G.~J.~Homer, J.~D.~Lewin, H.~E.~Walford and W.~A.~Smith,
  Nucl.\ Phys.\ B {\bf 206}, 333 (1982);
  T.~K.~Hemmick, D.~Elmore, T.~Gentile, P.~W.~Kubik, S.~L.~Olsen, D.~Ciampa, D.~Nitz and H.~Kagan {\it et al.},
  Phys.\ Rev.\ D {\bf 41}, 2074 (1990);
  P.~Verkerk, G.~Grynberg, B.~Pichard, M.~Spiro, S.~Zylberajch, M.~E.~Goldberg and P.~Fayet,
  Phys.\ Rev.\ Lett.\  {\bf 68}, 1116 (1992);
  T.~Yamagata, Y.~Takamori and H.~Utsunomiya,
  Phys.\ Rev.\ D {\bf 47}, 1231 (1993).



\bibitem{DiLuzio:2011my} 
  L.~Di Luzio,
  arXiv:1110.3210 [hep-ph].


\bibitem{delAguila:1980at} 
  F.~del Aguila and L.~E.~Ibanez,
  Nucl.\ Phys.\ B {\bf 177}, 60 (1981).

\bibitem{Mohapatra:1982aq} 
  R.~N.~Mohapatra and G.~Senjanovic,
  Phys.\ Rev.\ D {\bf 27}, 1601 (1983).

\bibitem{Minkowski:1977sc} 
  P.~Minkowski,
  Phys.\ Lett.\ B {\bf 67}, 421 (1977);
  T.~Yanagida,
  Conf.\ Proc.\ C {\bf 7902131}, 95 (1979);
  M.~Gell-Mann, P.~Ramond and R.~Slansky,
  Conf.\ Proc.\ C {\bf 790927}, 315 (1979)
  [arXiv:1306.4669 [hep-th]];
  S.~L.~Glashow,
  NATO Sci.\ Ser.\ B {\bf 59}, 687 (1980);
  R.~N.~Mohapatra and G.~Senjanovic,
  Phys.\ Rev.\ Lett.\  {\bf 44}, 912 (1980).
  
  \bibitem{flipped}
   I.~Antoniadis, J.~R.~Ellis, J.~S.~Hagelin and D.~V.~Nanopoulos,
  Phys.\ Lett.\ B {\bf 208}, 209 (1988)
  [Addendum-ibid.\ B {\bf 213}, 562 (1988)];
  J.~R.~Ellis, J.~L.~Lopez and D.~V.~Nanopoulos,
  Phys.\ Lett.\ B {\bf 292}, 189 (1992)
  [hep-ph/9207237];
  J.~R.~Ellis, D.~V.~Nanopoulos and K.~A.~Olive,
  Phys.\ Lett.\ B {\bf 300} (1993) 121
  [hep-ph/9211325];
 J.~R.~Ellis, J.~L.~Lopez, D.~V.~Nanopoulos and K.~A.~Olive,
  Phys.\ Lett.\ B {\bf 308}, 70 (1993)
  [hep-ph/9303307].


\bibitem{Pati:1974yy} 
  J.~C.~Pati and A.~Salam,
  Phys.\ Rev.\ D {\bf 10}, 275 (1974)
  [Erratum-ibid.\ D {\bf 11}, 703 (1975)].

\bibitem{Bajc:2005zf} 
  B.~Bajc, A.~Melfo, G.~Senjanovic and F.~Vissani,
  Phys.\ Rev.\ D {\bf 73}, 055001 (2006)
  [hep-ph/0510139].

\bibitem{Fukuyama:2004xs} 
  T.~Fukuyama, A.~Ilakovac, T.~Kikuchi, S.~Meljanac and N.~Okada,
  Eur.\ Phys.\ J.\ C {\bf 42}, 191 (2005)
  [hep-ph/0401213].

\bibitem{Siegel:1979wq} 
  W.~Siegel,
  Phys.\ Lett.\ B {\bf 84}, 193 (1979).
  
\bibitem{Hall:1980kf}
S.~Weinberg,
  Phys.\ Lett.\ B {\bf 91}, 51 (1980);
 L.~J.~Hall,
 Nucl.\ Phys.\ B {\bf 178} (1981) 75.


\bibitem{Ellis:2015jwa} 
  S.~A.~R.~Ellis and J.~D.~Wells,
  arXiv:1502.01362 [hep-ph].

  

\bibitem{Fukuyama:2004ps} 
  T.~Fukuyama, A.~Ilakovac, T.~Kikuchi, S.~Meljanac and N.~Okada,
  J.\ Math.\ Phys.\  {\bf 46}, 033505 (2005)
  [hep-ph/0405300].

\bibitem{Mohapatra:1980yp} 
  R.~N.~Mohapatra and G.~Senjanovic,
  Phys.\ Rev.\ D {\bf 23}, 165 (1981).

\bibitem{Lazarides:1980nt} 
  G.~Lazarides, Q.~Shafi and C.~Wetterich,
  Nucl.\ Phys.\ B {\bf 181}, 287 (1981).

\bibitem{Babu:1992ia} 
  K.~S.~Babu and R.~N.~Mohapatra,
  Phys.\ Rev.\ Lett.\  {\bf 70}, 2845 (1993)
  [hep-ph/9209215].

\bibitem{Matsuda:2000zp} 
  K.~Matsuda, Y.~Koide and T.~Fukuyama,
  Phys.\ Rev.\ D {\bf 64}, 053015 (2001)
  [hep-ph/0010026].

\bibitem{Sakai:1981pk} 
  N.~Sakai and T.~Yanagida,
  Nucl.\ Phys.\ B {\bf 197}, 533 (1982);
  S.~Weinberg,
  Phys.\ Rev.\ D {\bf 26}, 287 (1982).

\bibitem{Liu:2013ula} 
  M.~Liu and P.~Nath,
  Phys.\ Rev.\ D {\bf 87}, no. 9, 095012 (2013)
  [arXiv:1303.7472 [hep-ph]];
  J.~Hisano, T.~Kuwahara and N.~Nagata,
  Phys.\ Lett.\ B {\bf 723}, 324 (2013)
  [arXiv:1304.0343 [hep-ph]];
  J.~Hisano, D.~Kobayashi, T.~Kuwahara and N.~Nagata,
  JHEP {\bf 1307}, 038 (2013)
  [arXiv:1304.3651 [hep-ph]];
  J.~Hisano, D.~Kobayashi and N.~Nagata,
  Phys.\ Lett.\ B {\bf 716}, 406 (2012)
  [arXiv:1204.6274 [hep-ph]];
  N.~Yamatsu,
  PTEP {\bf 2013}, no. 12, 123B01 (2013)
  [arXiv:1304.5215 [hep-ph]];
  M.~Dine, P.~Draper and W.~Shepherd,
  JHEP {\bf 1402}, 027 (2014)
  [arXiv:1308.0274 [hep-ph]];
  L.~Du, X.~Li and D.~X.~Zhang,
  JHEP {\bf 1404}, 027 (2014)
  [arXiv:1312.1786 [hep-ph]];
  N.~Nagata and S.~Shirai,
  JHEP {\bf 1403}, 049 (2014)
  [arXiv:1312.7854 [hep-ph]];
  L.~J.~Hall, Y.~Nomura and S.~Shirai,
  JHEP {\bf 1406}, 137 (2014)
  [arXiv:1403.8138 [hep-ph]];
  A.~Hebecker and J.~Unwin,
  JHEP {\bf 1409}, 125 (2014)
  [arXiv:1405.2930 [hep-th]];
  J.~L.~Evans, N.~Nagata and K.~A.~Olive,
  Phys.\ Rev.\ D {\bf 91}, 055027 (2015)
  [arXiv:1502.00034 [hep-ph]].

\bibitem{Shiozawa}
M.~Shiozawa, talk presented at TAUP 2013, September 8--13, Asilomar, CA,
	USA.

\bibitem{Babu:2013jba}
  K.~S.~Babu, E.~Kearns, U.~Al-Binni, S.~Banerjee, D.~V.~Baxter, Z.~Berezhiani, M.~Bergevin and S.~Bhattacharya {\it et al.},
  arXiv:1311.5285 [hep-ph].

\bibitem{Abe:2011ts} 
  K.~Abe, T.~Abe, H.~Aihara, Y.~Fukuda, Y.~Hayato, K.~Huang, A.~K.~Ichikawa and M.~Ikeda {\it et al.},
  arXiv:1109.3262 [hep-ex].

\bibitem{Fukugita:1986hr} 
  M.~Fukugita and T.~Yanagida,
  Phys.\ Lett.\ B {\bf 174}, 45 (1986).

\bibitem{Lazarides:1991wu} 
  G.~Lazarides and Q.~Shafi,
  Phys.\ Lett.\ B {\bf 258}, 305 (1991).

\bibitem{Agashe:2014kda} 
  K.~A.~Olive {\it et al.}  [Particle Data Group Collaboration],
  Chin.\ Phys.\ C {\bf 38}, 090001 (2014).

\bibitem{ATLAS:2014wva} 
  [ATLAS and CDF and CMS and D0 Collaborations],
  arXiv:1403.4427 [hep-ex].

\bibitem{CMS:ril} 
  [CMS Collaboration],
  CMS-PAS-HIG-13-001;
  [ATLAS Collaboration],
  ATLAS-CONF-2013-012, ATLAS-COM-CONF-2013-015;
  G.~Aad {\it et al.}  [ATLAS Collaboration],
  Phys.\ Lett.\ B {\bf 726}, 88 (2013)
  [Erratum-ibid.\ B {\bf 734}, 406 (2014)]
  [arXiv:1307.1427 [hep-ex]];
  P.~P.~Giardino, K.~Kannike, I.~Masina, M.~Raidal and A.~Strumia,
  JHEP {\bf 1405}, 046 (2014)
  [arXiv:1303.3570 [hep-ph]].

\bibitem{Yamada:1993uh} 
  Y.~Yamada,
  Phys.\ Lett.\ B {\bf 316}, 109 (1993)
  [hep-ph/9307217].


\bibitem{Machacek:1983tz} 
  M.~E.~Machacek and M.~T.~Vaughn,
  Nucl.\ Phys.\ B {\bf 222}, 83 (1983).
  
\bibitem{Lyonnet:2013dna}
 F.~Lyonnet, I.~Schienbein, F.~Staub and A.~Wingerter,
 Comput.\ Phys.\ Commun.\  {\bf 185} (2014) 1130
 [arXiv:1309.7030 [hep-ph]].


\bibitem{FileviezPerez:2004hn} 
  P.~Fileviez Perez,
  Phys.\ Lett.\ B {\bf 595}, 476 (2004)
  [hep-ph/0403286].

\bibitem{Munoz:1986kq} 
  C.~Munoz,
  Phys.\ Lett.\ B {\bf 177}, 55 (1986).

\bibitem{Weinberg:1979sa} 
  S.~Weinberg,
  Phys.\ Rev.\ Lett.\  {\bf 43}, 1566 (1979).

\bibitem{Wilczek:1979hc} 
  F.~Wilczek and A.~Zee,
  Phys.\ Rev.\ Lett.\  {\bf 43}, 1571 (1979).

\bibitem{Abbott:1980zj} 
  L.~F.~Abbott and M.~B.~Wise,
  Phys.\ Rev.\ D {\bf 22}, 2208 (1980).

\bibitem{Nihei:1994tx} 
  T.~Nihei and J.~Arafune,
  Prog.\ Theor.\ Phys.\  {\bf 93}, 665 (1995)
  [hep-ph/9412325].

\bibitem{Aoki:2013yxa} 
  Y.~Aoki, E.~Shintani and A.~Soni,
  Phys.\ Rev.\ D {\bf 89}, 014505 (2014)
  [arXiv:1304.7424 [hep-lat]].

\end{thebibliography}
\end{document}